\documentclass[epsfig,12pt,onecolumn]{article}
\usepackage{amsfonts}
\usepackage{amssymb}
\usepackage{multicol}
\usepackage{graphicx}
\usepackage{float}
\usepackage{caption}
\usepackage{xcolor}
\usepackage[utf8]{inputenc}
\usepackage{amsmath}

\makeatletter \@addtoreset{equation}{section}
\renewcommand{\theequation}{\thesection.\arabic{equation}}
\def \be{\begin{equation}}
\def \ee{\end{equation}}
\def \bea{\begin{eqnarray}}
\def \eea{\end{eqnarray}}

\newcommand{\nc}{\newcommand}
\nc{\al}{\alpha} \nc{\bib}{\bibitem} \nc{\la}{\lambda}
\nc{\C}{\mbox{\hspace{1.24mm}\rule{0.2mm}{2.5mm}\hspace{-2.7mm} C}}
\nc{\R}{\mbox{\hspace{.04mm}\rule{0.2mm}{2.8mm}\hspace{-1.5mm} R}}
\input{tcilatex}
\begin{document}

\title{\textbf{Higher spin AdS}$_{\mathbf{3}}$\textbf{\ gravity and }\\
\textbf{Tits-Satake diagrams}}
\author{R. Sammani, Y. Boujakhrout, E.H Saidi, R. Ahl Laamara, L.B Drissi \\
%EndAName
{\small 1. LPHE-MS, Science Faculty}, {\small Mohammed V University in
Rabat, Morocco}\\
{\small 2. Centre of Physics and Mathematics, CPM- Morocco}}
\maketitle

\begin{abstract}
We investigate higher spin AdS$_{\mathbf{3}}$ gravity with real split forms
of complex A$_{N}$ B$_{N}$, C$_{N}$ and D$_{N}$ Lie algebras. This is done
by linking $SO(1,2)$ spin multiplets with splitted root systems using
Tits-Satake diagrams of real forms. Unlike $SL(N,R)$, we show that the
orthogonal families have two different higher spin (HS) spectrums: vectorial
and spinorial. We find amongst others that the spinorial spectrum has an
isolated spin $\mathfrak{j}_{\mathcal{N}}$ given by $\mathcal{N}\left(
\mathcal{N}+1\right) /2$ for $SO(\mathcal{N},1+\mathcal{N})$ and $\mathcal{N}%
\left( \mathcal{N}-1\right) /2$ for $SO(\mathcal{N},\mathcal{N})$. We
implement these results into the computation of the HS partition functions
in these gravity theories and identify the individual contributions of the
higher spin fields; valuable to manoeuver the HS-BTZ black hole partition
function.\newline
\textbf{Keywords}: 3D gravity, AdS$_{3}$/CFT$_{2}$ correspondence,
Tits-Satake diagrams, higher spin partition function, BTZ black holes.
\end{abstract}

%\tableofcontents

\section{Introduction}

The correspondence between the three dimensional gravity with a negative
cosmological constant and the conformal field theory in two dimensions has
been well established since the inaugural AdS$_{3}$/CFT$_{2}$ model \textrm{%
\cite{1A}}. Brown and Henneaux showed that the asymptotic AdS$_{3}$ is much
more generous than the bulk theory, the asymptotic symmetry algebra is
\textrm{no longer} the anticipated $SO\left( 2,2,\mathbb{R}\right) $ group
isomorphic to $SO\left( 1,2,\mathbb{R}\right) \times SO\left( 1,2,\mathbb{R}%
\right) $ but it is enhanced to two copies of the conformal Virasoro algebra
\textrm{\cite{AD1}-\cite{AD6}}. And because the AdS$_{3}$ theory is
inherently topological and its dynamic depends on the boundary, it has been
the go-to model to recast all our inquiries about \textrm{3D} gravity,
especially with the Chern-Simons (CS) reformulation \textrm{\cite{2A,3A}}
where general relativity is equivalent to a gauge theory based on $SL\left(
2,\mathbb{R}\right) \times SL\left( 2,\mathbb{R}\right) $ gauge group.

Two of the major inquiries that have been investigated in this context are
the higher spin (HS) gravity \textrm{\cite{HS1}-\cite{HS8} }and the physics
of black holes \textrm{\cite{BL1}-\cite{BLH5}}. The three dimensional
Chern-Simons formalism enables us to couple higher spin fields to 3D gravity
via a process known as the principal\emph{\ }embedding \textrm{\cite{4A}}.
There, the authors argued that the principal embedding of the $SO\left(
1,2\right) \sim sl\left( 2,\mathbb{R}\right) $ in a Lie algebra $\mathfrak{g}
$ gives\textbf{\ }rise to higher spin fields with boundary conformal spins $%
s $ taking values in a set \{$s_{1},...,s_{r_{\mathfrak{g}}}$\} with $r_{%
\mathfrak{g}}$ designating the rank of $\mathfrak{g}.$ For instance,\emph{\ }%
the AdS$_{3}$ theory with $\mathfrak{g}=sl\left( \mathcal{N},\mathbb{R}%
\right) $ gives a HS gauge theory with boundary conformal spin range $%
s=2,3,...,\mathcal{N};$ and the emerging asymptotic symmetry algebra is in
this case given by two copies (left and right) of the $\boldsymbol{WA}_{%
\mathcal{N}}$-algebra. Moreover, the bigger surprise was the introduction of
a dynamical black hole solution as an excitation of the AdS$_{3}$ space
known as the BTZ black hole \textrm{\cite{5A}}; this is because most black
holes were assumed to exist only in spacetime dimensions $d\geq 4.$
Furthermore, BTZ black hole can carry conformal spin $s\geq 2$ charge
\textrm{\cite{6A}} in which case is called the higher spin black hole where
the partition function encompasses for the HS states constructed via the
vacuum character of the asymptotic $\boldsymbol{WA}_{\mathcal{N}}$-symmetry
algebra.

In this work, we study AdS$_{\mathbf{3}}$ gravity with real split forms of
complex A$_{N}$ B$_{N}$, C$_{N}$ and D$_{N}$ Lie algebras to investigate
higher spin gravity and HS-BTZ black holes with various gauge symmetries. We
first exploit the rich structure of real forms of complex Lie algebras to
recast the linear $A_{\mathcal{N}}$- gravity models into a graphic
description using Tits-Satake diagrams. In this construction, we realise the
principal embedding\textrm{\ }by cutting an extremal node in the Tits-Satake
diagram of the Lie algebra of the gauge symmetry; thus leading to left and
right decompositions that happen to be identical for $SL(\mathcal{N},\mathbb{%
R})$. We refer to this graphic procedure as the Extremal Node Decomposition
(END for short). Then, we build the orthogonal $B_{\mathcal{N}}$- and $D_{%
\mathcal{N}}$- theories while focusing on their real split forms $SO(%
\mathcal{N},1+\mathcal{N})$ and $SO(\mathcal{N},\mathcal{N})$. Following the
embedding algorithm, we find that the orthogonal theories have two different
ENDs; and therefore two HS spectrums termed \textrm{below} as vectorial and
spinorial. For these HS gauge symmetries, we determine the higher spin
content of the dual boundary CFT$_{2}$ and calculate their HS partition
functions, valuable for the computation of the HS-BTZ black hole partition
function.

The organisation of this paper is as follows: In section 2, we briefly
review the CS canonical formulation of AdS$_{3}$ gravity with $SL(\mathcal{N}%
,R)$ symmetry before recasting the description in a more suitable basis for
the coupling with HS. In section 3, we study the partition function of $SL(%
\mathcal{N},R)$ and its factorization motivated by the principal embedding
and realised by the extremal node decomposition. In section 4, we introduce
the real forms of the complex $B_{\mathcal{N}}$ series and focus on the real
split form to derive the set of boundary conformal spins by using the END
procedure. Then, we calculate the HS partition function in the asymptotic AdS%
$_{3}$. In section 5, we consider the real forms of the $D_{\mathcal{N}}$
series and use its real split form to determine \textrm{the} vector and
spinorial CFT$_{2}$ spectrums by using the extremal node decomposition.
Section 6 is devoted to conclusions and comments; and in section 7, we give
two appendices where technical details are reported.

\section{Recasting HS-AdS$_{3}$ gravity}

In this section, we first describe the modeling of higher spins in pure AdS$%
_{3}$ gravity with $\boldsymbol{A}_{\mathcal{N}}$ symmetry using the
canonical formulation. We exploit this study to introduce useful tools as a
front matter for an alternative approach to describe higher spin symmetries
in 3D gravity to be developed later on in this investigation. For that, we
first focus on the spins $2$ and $3$ of pure AdS$_{3}$ gravity using the
Chern-Simons fields with $SL\left( 3,\mathbb{R}\right) $ gauge symmetry
given by the diagonal group of,%
\begin{equation}
SL\left( 3,\mathbb{R}\right) _{L}\times SL\left( 3,\mathbb{R}\right) _{R}
\end{equation}%
where the use of the left and right sectors is demanded by the CS
formulation of AdS$_{3}$ in correspondence with the CFT$_{2}$ \textrm{\cite%
{2A,3A}}. Then, we comment on the generalisation of this theory to higher
spins within the $SL\left( \mathcal{N},\mathbb{R}\right) $ family ($\mathcal{%
N}\geq 3$) sitting in $SL\left( \mathcal{N},\mathbb{R}\right) _{L}\times
SL\left( \mathcal{N},\mathbb{R}\right) _{R}.$

\subsection{Canonical formulation of HS-AdS$_{3}$ gravity}

The gravity fields in pure Anti-de Sitter 3D gravity are given by the
Dreibein $e_{\mu }^{a}$ and the $SO\left( 1,2\right) $ spin connection $%
\omega _{\mu }^{a}$ carrying two labels; a curved $\mu $ and a flat $a;$
each taking three values. In differential geometry language, these gravity
fields are described by the 1-forms $e^{a}=e_{\mu }^{a}dx^{\mu }$ and $%
\omega ^{a}=\omega _{\mu }^{a}dx^{\mu }$ transforming as vectors of $%
SO\left( 1,2\right) ;$ the Lorentz group in 3D. As such they have an $%
SO\left( 1,2\right) $ spin $\mathfrak{j}_{so_{1,2}}$ equal to unity; that is
\begin{equation}
\mathfrak{j}\left( e^{a}\right) =\mathfrak{j}\left( \omega ^{a}\right)
=1\qquad ,\qquad \mathfrak{j}\equiv \mathfrak{j}_{so_{1,2}}\qquad ,\qquad
s\equiv s_{cft}  \label{e}
\end{equation}%
We refer below to this integer as the Lorentz spin $\mathfrak{j}$ (L-spin $%
\mathfrak{j}$) to distinguish it from the CFT$_{2}$-spin\footnote{%
\ \ In CFT$_{2}$, the conformal spin $s=h-\bar{h}$ is given by the
difference of conformal weights $h$ and $\bar{h}$; their sum $\Delta =h+\bar{%
h}$ gives the scale dimension.} $s=\mathfrak{j}+1$ living on the boundary of
AdS$_{3}$ with an asymptotic symmetry given by two copies Vir$_{c}\times $Vir%
$_{\bar{c}}$ of the Virasoro algebra; each containing $SO\left( 1,2\right) $
as a non anomalous finite dimensional subalgebra. The pure AdS$_{3}$ gravity
metric is
\begin{equation}
g_{\mu \nu }=e_{\mu }^{a}\eta _{ab}e_{\nu }^{b}
\end{equation}%
with flat $\eta _{ab}=diag(-,+,+).$ In this regard, notice that this curved
3D metric $g_{\mu \nu }$ and the associated flat $\eta _{ab}$ of $\mathbb{R}%
^{1,2}$ can be put in correspondence with the Riemannian metric $g_{\text{%
\textsc{mn}}}^{E}=$\textsc{e}$_{\text{\textsc{m}}}^{\text{\textsc{a}}}\delta
_{\text{\textsc{ab}}}$\textsc{e}$_{\text{\textsc{n}}}^{\text{\textsc{b}}}$
and the flat $\delta _{\text{\textsc{ab}}}$ of the euclidian $\mathbb{R}^{3}$
with the isotropy symmetry $SO(3)$; this metric is useful for application to
BTZ black holes \textrm{\cite{5A}}. The Lie algebras $so\left( 1,2\right) $
and $so\left( 3\right) ,$ of the non compact $SO\left( 1,2\right) $ and the
compact $SO\left( 3\right) $ Lie groups, are also discriminated by the
compacity of their generators including the Cartan ones sitting either in
the abelian $so\left( 1,1\right) $ or the compact $so\left( 2\right) $. The $%
so\left( 1,2\right) $ and $so\left( 3\right) $ will be often handled below
through their homomorphic forms $su\left( 1,1\right) \sim sl\left( 2,\mathbb{%
R}\right) $ and $su\left( 2\right) $ which are the two real forms of $sl(2,%
\mathbb{C})$ as described \textrm{in appendix A}. The homomorphism $so\left(
1,2\right) \sim sl\left( 2,\mathbb{R}\right) $ is a key feature behind
higher spins in the $\boldsymbol{A}_{\mathcal{N}}$ family description of AdS$%
_{3}$ gravity. Similar features are also present when studying the
generalisation of $\boldsymbol{A}_{\mathcal{N}}$ towards $\boldsymbol{B}_{%
\mathcal{N}}$ and $\boldsymbol{D}_{\mathcal{N}}$ Anti-de Sitter gravities;
they will be treated in sections 4 and 5. \newline
The $sl\left( 2,\mathbb{R}\right) $ and $su\left( 2\right) $ real forms are
nicely described by Tits-Satake diagrams, which roughly speaking, are given
by Dynkin diagrams with (un)painted node as depicted by the last column of
the following table.
\begin{table}[tbp]
\begin{center}
$%
\begin{tabular}{|c|c|c|c|}
\hline
$sl(2,C)$ real forms & Cartan subalgebra & \multicolumn{2}{|c|}{Tits-Satake
diagram} \\ \hline
$so\left( 3\right) \simeq su\left( 2\right) $ & $so\left( 2\right) \simeq
u\left( 1\right) $ & black node & \includegraphics[width=0.7cm]{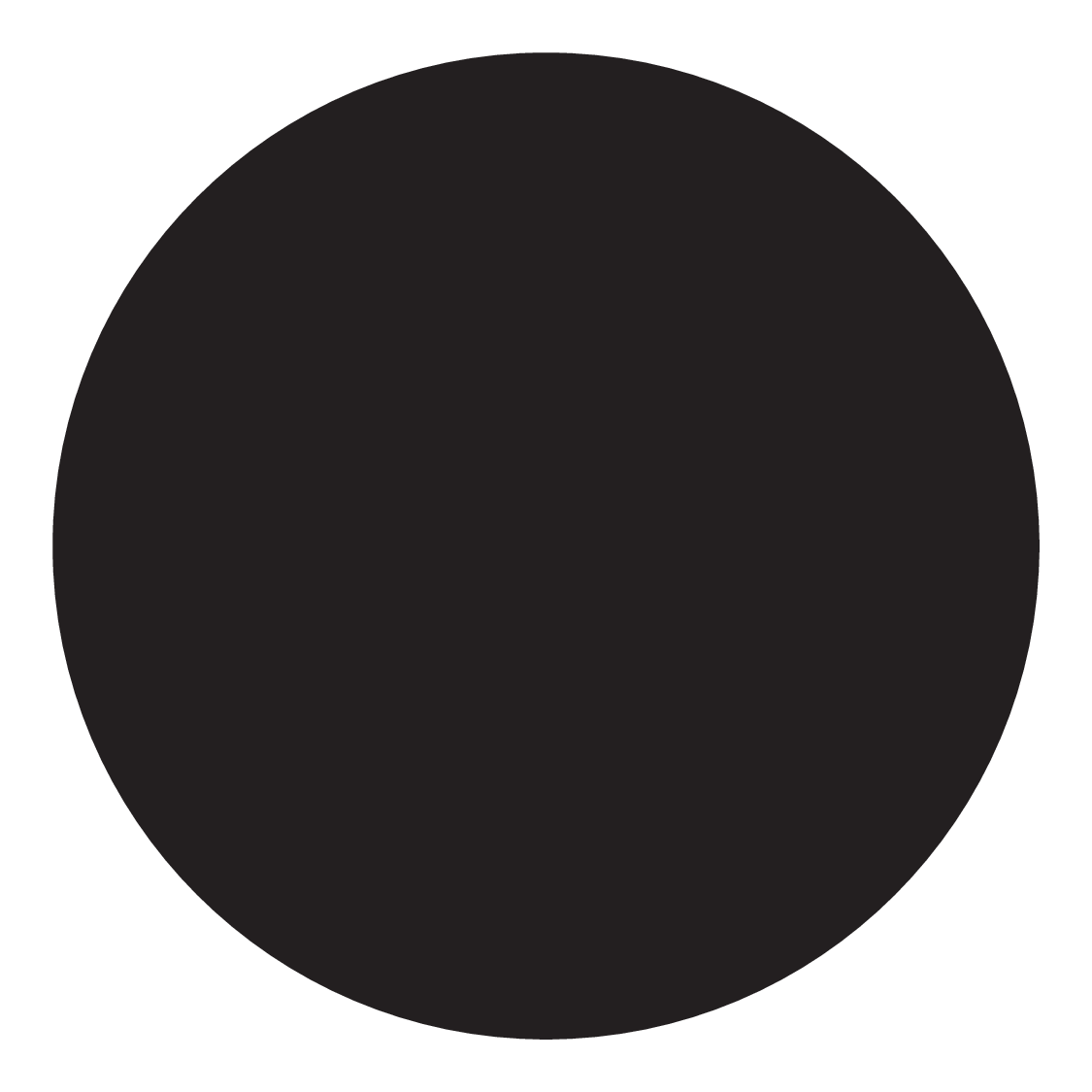} \\
\hline
$so\left( 1,2\right) \simeq sl\left( 2,R\right) $ & $so\left( 1,1\right)
\simeq sl\left( 1,R\right) $ & white node & %
\includegraphics[width=0.7cm]{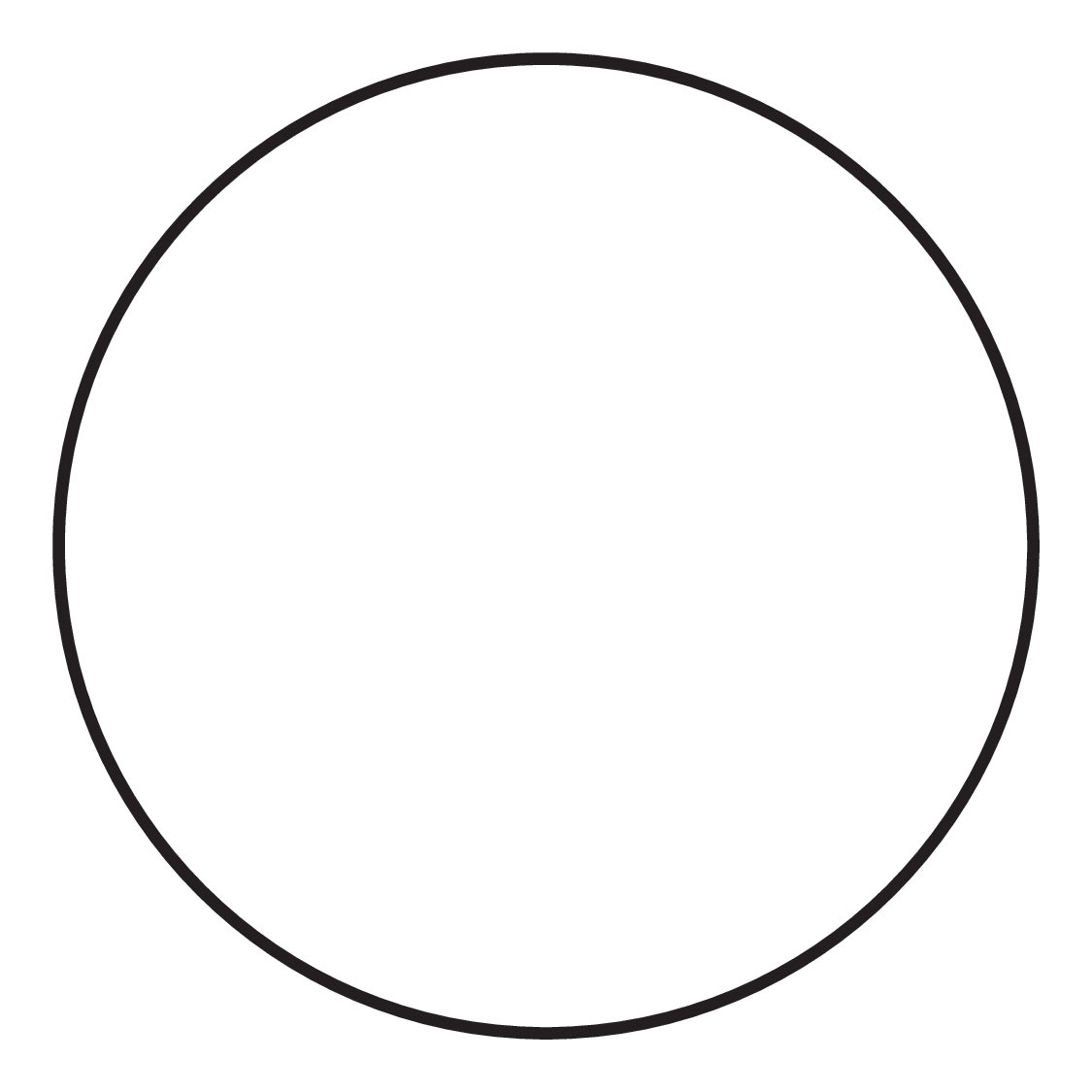} \\ \hline
\end{tabular}%
$%
\end{center}
\caption{The two real forms of the complex Lie algebra $A_{1}.$ }
\label{SA1N}
\end{table}
\ The Tits-Satake diagrammatic description for real forms of complex Lie
algebras constitutes a basic tool in our approach for studying higher spin
3D gravity. To fix ideas about Tits-Satake graphs, see for instance the
Figure \textbf{\ref{STA} }and its homologue given later.\textrm{\newline
}In the 3D Chern-Simons description of AdS$_{3}$ gravity, the field action $%
\mathcal{S}_{grav}$ of pure Anti-de Sitter gravity is given by the
difference of two Chern-Simons field actions $CS\left[ A_{L}\right] -CS\left[
A_{R}\right] $ as follows \textrm{\cite{3A}}%
\begin{equation}
\mathcal{S}_{grav}=\dint\nolimits_{\mathcal{M}_{3D}}Tr\Omega \left[ A_{L}%
\right] -\dint\nolimits_{\mathcal{M}_{3D}}Tr\Omega \left[ A_{R}\right]
\label{1}
\end{equation}%
with Chern-Simons 3-form \textrm{\cite{7A,8A,9A},}
\begin{equation}
\Omega \left[ A\right] =AdA+\frac{2}{3}A^{3}  \label{2}
\end{equation}%
Here, the real 1-form $A_{L}$ (resp. $A_{R}$) is the 3D Chern-Simons gauge
potential valued in the Lie algebra of the $SL\left( 2,\mathbb{R}\right) _{L}
$ gauge symmetry (resp. $SL\left( 2,\mathbb{R}\right) _{R}$). These 1-forms
have the expansions%
\begin{equation}
A=\sum_{a=0}^{2}A^{a}J_{a}=\sum_{\mu =0}^{2}A_{\mu }dx^{\mu }  \label{AA}
\end{equation}%
with $J_{a}$ being the three generators of $SL\left( 2,\mathbb{R}\right) $
satisfying the commutation relations%
\begin{equation}
\left[ J_{a},J_{b}\right] =\varepsilon _{abc}J^{c}  \label{K}
\end{equation}%
with $J^{a}=\eta ^{ab}J_{b}$ \textrm{and }$\varepsilon _{abc}$ \textrm{is
the Levi-Civita tensor in Lorentzian R}$^{1,2}$ \textrm{with metric }$\eta
_{ab}.$\textrm{\ }The trace of the CS 3-form reads as%
\begin{equation}
Tr\Omega \left[ A\right] =q_{ab}A^{a}dA^{b}+\frac{2}{3}q_{abc}A^{a}A^{b}A^{c}
\label{G}
\end{equation}%
where $q_{ab}$ and $q_{abc}$ are given by the intersections
\begin{equation}
q_{ab}=Tr\left( J_{a}J_{b}\right) \qquad ,\qquad q_{abc}=Tr\left(
J_{a}J_{b}J_{c}\right)   \label{int}
\end{equation}%
with $q_{ab}$ related to the flat metric as $\frac{1}{2}\eta _{ab}$ and $%
q_{abc}=\varepsilon _{abc}/4.$ The relationship between the CS gauge fields $%
A_{L},$ $A_{R}$ and the gravity fields $e_{\mu }^{a},$ $\omega _{\mu }^{a}$
respecting scaling dimension is%
\begin{equation}
\begin{tabular}{lll}
$\left( A_{\mu }^{a}\right) _{L}$ & $=$ & $\omega _{\mu }^{a}+\frac{1}{l_{%
{\small AdS}}}e_{\mu }^{a}$ \\
$\left( A_{\mu }^{a}\right) _{R}$ & $=$ & $\omega _{\mu }^{a}-\frac{1}{l_{%
{\small AdS}}}e_{\mu }^{a}$%
\end{tabular}%
\qquad ,\qquad
\begin{tabular}{lll}
$A_{L}^{a}$ & $=$ & $\omega ^{a}+\frac{1}{l_{{\small AdS}}}e^{a}$ \\
$A_{R}^{a}$ & $=$ & $\omega ^{a}-\frac{1}{l_{{\small AdS}}}e^{a}$%
\end{tabular}
\label{14}
\end{equation}%
indicating that%
\begin{equation}
\mathfrak{j}\left( A_{L}^{a}\right) =\mathfrak{j}\left( A_{R}^{a}\right) =1
\end{equation}%
By substituting these expressions into (\ref{1}) and using (\ref{2}), we
re-discover the standard action of $\mathcal{S}_{grav}$ in terms of $e^{a}$
and $\omega ^{a}.$ Below, we shall think of the $SL\left( 2,\mathbb{R}%
\right) $ \textrm{generators as} $J_{0}=L_{0}$ and $J_{1}=(L_{-}+L_{+})/2,$
as well as $J_{2}=(L_{-}-L_{+})/2$ with $L_{n}^{\dagger }=L_{-n}.$ The new
generators satisfy the following commutation relations%
\begin{equation}
\left[ L_{0},L_{\mp }\right] =\pm L_{\mp }\qquad ,\qquad \left[ L_{+},L_{-}%
\right] =2L_{0}  \label{19}
\end{equation}%
The higher spin- 3 extension of AdS$_{3}$ gravity is obtained from the above
description by the principal embedding of \textrm{\cite{4A}}; it relies on
promoting the $SL\left( 2,\mathbb{R}\right) _{L}\times SL\left( 2,\mathbb{R}%
\right) _{R}$ gauge symmetry to the larger group $SL\left( 3,\mathbb{R}%
\right) _{L}\times SL\left( 3,\mathbb{R}\right) _{R}$. In this
generalisation, the CS 1-form $A_{L}$ (resp. $A_{R}$) is valued in the Lie
algebra of the $SL\left( 3,\mathbb{R}\right) _{L}$ gauge symmetry (resp. $%
SL\left( 3,\mathbb{R}\right) _{R}$). So they can be canonically expanded
like
\begin{equation}
A=\sum_{\text{\textsc{a}}=1}^{8}A^{\text{\textsc{a}}}\mathfrak{J}_{\text{%
\textsc{a}}}  \label{3}
\end{equation}%
where $\mathfrak{J}_{\text{\textsc{a}}}$ are the generators of $SL\left( 3,%
\mathbb{R}\right) ,$ the real split form of $SL\left( 3,\mathbb{C}\right) $.
However, to explicitly exhibit the L- spin $\mathfrak{j}=2$ content in AdS$%
_{3}$ gravity, one uses an alternative basis of $SL\left( 3,\mathbb{R}%
\right) $ where the above expansion is presented as follows%
\begin{equation}
A=\sum_{a=0}^{2}A^{a}J_{a}+\sum_{a,b=0}^{2}A^{ab}T_{ab}  \label{4}
\end{equation}%
with $T_{ab}=T_{ba}$ having a vanishing trace $\eta ^{ba}T_{ab}=0.$ The
L-spins $\mathfrak{j}$ of these components read as%
\begin{equation}
\begin{tabular}{lllll}
$\mathfrak{j}\left( A^{a}\right) $ & $=1$ & $\qquad ,\qquad $ & $\mathfrak{j}%
\left( J_{a}\right) $ & $=1$ \\
$\mathfrak{j}\left( A^{ab}\right) $ & $=2$ & $\qquad ,\qquad $ & $\mathfrak{j%
}\left( T_{ab}\right) $ & $=2$%
\end{tabular}
\label{jt}
\end{equation}%
Adopting the representation of \cite{4A}, the L-spin $\mathfrak{j}=2$
generators $T_{\left( ab\right) }$ (\textrm{for short just }$T_{ab}$)
\textrm{obey the following commutation relations}%
\begin{equation}
\begin{tabular}{lll}
$\left[ J_{a},J_{b}\right] $ & $=$ & $\varepsilon _{abc}J^{c}$ \\
$\left[ J_{a},T_{bc}\right] $ & $=$ & $\epsilon _{a(b}^{m}T_{c)m}$ \\
$\left[ T_{ab},T_{cd}\right] $ & $=$ & $-[\eta _{a(c}\epsilon _{d)bm}+\eta
_{b(c}\epsilon _{d)am}]J^{m}$%
\end{tabular}%
\end{equation}%
These commutators are satisfied if one takes the higher spin generators $%
T_{ab}$ as a symmetrised product of $SL\left( 2,\mathbb{R}\right) $
generators. In fact, one can build the fundamental representation of $%
SL\left( 3,\mathbb{R}\right) $ by taking
\begin{equation}
\begin{tabular}{lll}
$T_{ab}$ & $=$ & $\left( J_{a}J_{b}+J_{b}J_{a}\right) -\frac{2}{3}\eta _{ab}%
\boldsymbol{J}^{2}$ \\
$\boldsymbol{J}^{2}$ & $=$ & $J_{a}\eta ^{ab}J_{b}$%
\end{tabular}%
\end{equation}%
for more on the $SL\left( 3,\mathbb{R}\right) $ generators and the
associated commutation relations, report to\ appendix A.

The Dreibein and the spin connections of $SL\left( 3,\mathbb{R}\right) $
therefore expand as%
\begin{equation}
\begin{tabular}{lll}
$\omega _{\mu }$ & $=$ & $\dsum\limits_{a=0}^{2}\omega _{\mu
}^{a}J_{a}+\dsum\limits_{a,b=0}^{2}\omega _{\mu }^{ab}T_{ab}$ \\
$e_{\mu }$ & $=$ & $\dsum\limits_{a=0}^{2}e_{\mu
}^{a}J_{a}+\dsum\limits_{a,b=0}^{2}e_{\mu }^{ab}T_{ab}$%
\end{tabular}%
\end{equation}%
The relations with CS gauge fields are given by%
\begin{equation}
\begin{tabular}{lll}
$\left( A_{\mu }^{a}\right) _{L}$ & $=$ & $\omega _{\mu }^{a}+\frac{1}{l_{%
{\small AdS}}}e_{\mu }^{a}$ \\
$\left( A_{\mu }^{a}\right) _{R}$ & $=$ & $\omega _{\mu }^{a}-\frac{1}{l_{%
{\small AdS}}}e_{\mu }^{a}$%
\end{tabular}%
\qquad ,\qquad
\begin{tabular}{lll}
$\left( A_{\mu }^{ab}\right) _{L}$ & $=$ & $\omega _{\mu }^{ab}+\frac{1}{l_{%
{\small AdS}}}e_{\mu }^{ab}$ \\
$\left( A_{\mu }^{ab}\right) _{R}$ & $=$ & $\omega _{\mu }^{ab}-\frac{1}{l_{%
{\small AdS}}}e_{\mu }^{ab}$%
\end{tabular}
\label{25}
\end{equation}%
Putting into (\ref{2}), and using,%
\begin{equation}
\begin{tabular}{lllllll}
$Tr\left( J_{a}T_{cd}\right) $ & $=$ & $Tr\left( J_{a}J_{b}T_{cd}\right) $ &
$=$ & $Tr\left( T_{ab}T_{cd}T_{ef}\right) $ & $=$ & $0$ \\
$Tr\left( T_{ab}T_{cd}\right) $ & $=$ & \multicolumn{5}{l}{$\eta _{abcd}=%
\frac{1}{2}\left( \eta _{a(c}\eta _{d)b}-\frac{2}{3}\eta _{ab}\eta
_{cd}\right) $}%
\end{tabular}%
\end{equation}%
we re-discover the field action of the AdS$_{3}$ gravity in terms of the
Dreibein and spin connection, namely
\begin{equation}
\begin{tabular}{lll}
$\mathcal{S}_{grav}\left[ e,\omega \right] $ & $=$ & $\frac{1}{8\pi G}%
\dint\nolimits_{\mathcal{M}_{3D}}e^{a}\left( d\omega _{a}+\frac{1}{2}%
\epsilon _{abc}\omega ^{b}\omega ^{c}+2\epsilon _{abc}\omega ^{bd}\omega
_{d}^{c}\right) +$ \\
&  & $\frac{1}{4\pi G}\dint\nolimits_{\mathcal{M}_{3D}}e^{ab}\left( d\omega
_{ab}+2\epsilon _{dea}\omega ^{d}\omega _{b}^{e}\right) +$ \\
&  & $\frac{1}{48\pi Gl_{{\small AdS}}^{2}}\dint\nolimits_{\mathcal{M}%
_{3D}}\epsilon _{abc}\left( e^{a}e^{b}e^{c}+12e^{a}e^{bd}e_{d}^{c}\right) $%
\end{tabular}
\label{we}
\end{equation}%
Notice here that the higher spins contributions can also be manifested on
the space metric $g_{\mu \nu }$ which gets deformed like
\begin{equation}
\tilde{g}_{\mu \nu }=e_{\mu }^{a}\eta _{ab}e_{\nu }^{a}+\delta g_{\mu \nu }
\end{equation}%
where the variation is given by \cite{44A}%
\begin{equation}
\delta g_{\mu \nu }=\sum_{j}e_{\mu }^{a_{1}...a_{j}}\eta _{\left(
a_{1}...a_{j}\right) \left( b_{1}...b_{j}\right) }e_{\nu }^{b_{1}...b_{j}}
\end{equation}%
with some tensors $\eta _{\left( a_{1}...a_{j}\right) \left(
b_{1}...b_{j}\right) }$ that can be explicitly specified by fixing the rank
of the gauge symmetry.

\subsection{Higher spin AdS$_{3}$ gravity in Chevalley basis}

In this subsection, we draw the lines of a new approach to deal with higher
spins in pure AdS$_{3}$ gravity; this setup has the ability of $\left(
i\right) $ describing gauge symmetries for real forms of the $\boldsymbol{A}%
_{\mathcal{N}}$ family other than $SL(\mathcal{N},\mathbb{R})$, $\left(
ii\right) $ extending to other symmetry families like the orthogonal $%
\boldsymbol{B}_{\mathcal{N}}$ and $\boldsymbol{D}_{\mathcal{N}}$ Lie
algebras developed explicitly in this paper; and $\left( iii\right) $
benefiting from the efficiency of the graphical Tits-Satake description of
Lie algebra of gauge symmetries. This approach is first implemented in the
case of AdS$_{3}$ gravity with $SO(1,2)\simeq SL\left( 2,\mathbb{R}\right) $
symmetry. Then, the construction is extended to $SL\left( 3,\mathbb{R}%
\right) $ by working out the link with the $\boldsymbol{WSL}_{\left( 3,%
\mathbb{R}\right) }$- symmetry of the corresponding CFT$_{2}$.

\ \ \ \

$\bullet $ \emph{SL}$\left( 2,\mathbb{R}\right) $\emph{\ theory } \newline
Our formulation of AdS$_{3}$ gravity using Chern-Simons fields (\ref{1}-\ref%
{2}) is based on eqs. (\ref{19}) describing the projective $SL\left( 2,%
\mathbb{R}\right) $ subalgebra of the Virasoro symmetry
\begin{equation}
\left[ L_{n},L_{m}\right] =\left( n-m\right) L_{n+m}+\frac{c}{12}\left(
n^{3}-n\right) \delta _{n+m}  \label{RC}
\end{equation}%
with labels restricted to $n=0,\pm 1$ for which the central term vanishes;
thus leading to the non anomalous $\left[ L_{0},L_{\pm }\right] =\mp L_{\pm
} $ and $\left[ L_{+},L_{-}\right] =2L_{0}$. For convenience, we set $L_{\mp
}=iE_{\pm \alpha }$ and $L_{0}=H_{\alpha };$ then we write%
\begin{equation}
\left[ H_{\alpha },E_{\pm \alpha }\right] =\pm E_{\pm \alpha },\qquad \left[
E_{+\alpha },E_{-\alpha }\right] =2H_{\alpha }  \label{CR}
\end{equation}%
\textrm{\ }Here,\textrm{\ }$\alpha $ is the positive root of the complex Lie
algebra $\boldsymbol{A}_{1}$ with the property $\alpha ^{2}=2.$ Using the
new generator basis $E_{n\alpha },$ the CS gauge field potential $A_{\mu
}=A_{\mu }^{a}J_{a}$ of $sl(2,\mathbb{R})$ as well as the AdS$_{3}$ gravity
fields expand like $A_{\mu }^{n}E_{n\alpha }$, $\omega _{\mu }^{n}E_{n\alpha
}$ and $e_{\mu }^{n}E_{n\alpha }$ reading explicitly like%
\begin{equation}
\begin{tabular}{lllll}
$A_{\mu }$ & $=$ & $A_{\mu }^{a}J_{a}$ & $=$ & $A_{\mu }^{-}E_{-\alpha
}+A_{\mu }^{0}H_{\alpha }+A_{\mu }^{+}E_{+\alpha }$ \\
$\omega _{\mu }$ & $=$ & $\omega _{\mu }^{a}J_{a}$ & $=$ & $\omega _{\mu
}^{-}E_{-\alpha }+\omega _{\mu }^{0}H_{\alpha }+\omega _{\mu }^{+}E_{+\alpha
}$ \\
$e_{\mu }$ & $=$ & $\omega _{\mu }^{a}J_{a}$ & $=$ & $e_{\mu }^{-}E_{-\alpha
}+e_{\mu }^{0}H_{\alpha }+e_{\mu }^{+}E_{+\alpha }$%
\end{tabular}
\label{A}
\end{equation}%
The distinction between the two real forms of $sl(2,\mathbb{C})$ in eq(\ref%
{A}), that is between $sl(2,\mathbb{R})$ and $su(2),$ arises from formulating%
\textrm{\ their three generators} $J_{a}$ \textrm{in terms of the Chevalley
operators} $\left( H_{\alpha },E_{\pm \alpha }\right) $ as%
\begin{equation}
\begin{tabular}{c||c|c|c||c}
generators & $J_{1}$ & $J_{2}$ & $J_{3}$ & $J_{a}^{\boldsymbol{A}_{1}}$ \\
\hline\hline
$sl(2,\mathbb{R})$ & $E_{+\alpha }+E_{-\alpha }$ & $E_{+\alpha }-E_{-\alpha
} $ & $H_{\alpha }$ & $J_{a}^{sl_{2}}=\mathcal{U}_{a}^{n\alpha }E_{n\alpha }$
\\ \hline
$su\left( 2\right) $ & $i\left( E_{+\alpha }+E_{-\alpha }\right) $ & $%
E_{+\alpha }-E_{-\alpha }$ & $iH_{\alpha }$ & $J_{a}^{su_{2}}=\mathcal{V}%
_{a}^{n\alpha }E_{n\alpha }$ \\ \hline
\end{tabular}
\label{tj}
\end{equation}%
with
\begin{equation}
\mathcal{U}=\frac{1}{2}\left(
\begin{array}{ccc}
1 & 1 & 0 \\
1 & -1 & 0 \\
0 & 0 & 1%
\end{array}%
\right) \qquad ,\qquad \mathcal{V}=\frac{1}{2}\left(
\begin{array}{ccc}
i & i & 0 \\
1 & -1 & 0 \\
0 & 0 & i%
\end{array}%
\right)
\end{equation}%
with $\det \mathcal{U}=-1/4$ and $\det \mathcal{V}=1/4.$ In the last column
of (\ref{tj}), the $\mathcal{U}_{a}^{n\alpha }$ and $\mathcal{V}%
_{a}^{n\alpha }$ are invertible bridge matrices between the Cartesian $%
\left\{ J_{a}\right\} $ and the Chevalley $\left\{ E_{n\alpha }\right\} $
generators. These two changes play an important role in our construction
either for the $\boldsymbol{A}_{\mathcal{N}}$ family; or for the orthogonal
series considered later. Thanks to these transformations, we can work with
the Chevalley generators of $\boldsymbol{A}_{\mathcal{N}}$ using
\begin{equation}
E_{n\alpha }=\left( \mathcal{U}^{-1}\right) _{n\alpha
}^{a}J_{a}^{sl_{2}},\qquad E_{n\alpha }=\left( \mathcal{V}^{-1}\right)
_{n\alpha }^{a}J_{a}^{su_{2}}
\end{equation}%
to move to the Cartesian generators $J_{a}^{sl_{2}}$ and $J_{a}^{su_{2}}$.
These quantities play a key role in differentiating between the $sl(2,%
\mathbb{R})$ and the $su(2)$ theories of 3D gravity; for example, the
Killing forms $Tr(J_{a}^{sl_{2}}J_{b}^{sl_{2}})=q_{ab}^{sl_{2}}$ and Tr$%
(J_{a}^{su_{2}}J_{b}^{su_{2}})=q_{ab}^{su_{2}}$ are respectively related to $%
Tr(E_{n\alpha }E_{m\beta })=\kappa _{nm}$ by the tensors $\mathcal{U}%
_{a}^{n\alpha }\mathcal{U}_{b}^{m\beta }$ and $\mathcal{V}_{a}^{n\alpha }%
\mathcal{V}_{b}^{m\beta }$. In fact, this discrimination can be also
exhibited by the Cartan involution $\vartheta $ acting differently on the
Chevalley generators associated to the simple root $\alpha $ of $\boldsymbol{%
A}_{\mathcal{N}}$ ($\mathcal{N}=1$) as%
\begin{equation}
\begin{tabular}{c|cc||c|c|c}
\multicolumn{2}{c}{Cartan involution $\vartheta $} &  & $\alpha $ & $%
H_{\alpha }$ & $E_{+\alpha }$ \\ \hline\hline
compact $su(2)$ & $\vartheta =+1$ &  & $+\alpha $ & $H_{\alpha }$ & $%
E_{-\alpha }$ \\ \hline
real split $sl(2\emph{,}\mathbb{R}\emph{)}$ & $\vartheta =-1$ &  & $-\alpha $
& $-H_{\alpha }$ & $-E_{-\alpha }$ \\ \hline
\end{tabular}%
\end{equation}%
\textrm{By replacing} $E_{\pm \alpha }=(J_{1}^{sl_{2}}\pm J_{2}^{sl_{2}})/2$
and $H_{\alpha }=J_{3}^{sl_{2}}$ in eq(\ref{A}), we obtain the component
gauge field of the $sl(2,\mathbb{R})$ theory. Instead, by using $E_{\pm
\alpha }=(\mp iJ_{1}^{su_{2}}+J_{2}^{su_{2}})/2$ and $H_{\alpha
}=-iJ_{3}^{su_{2}}$ in eq(\ref{A}), one gets the component gauge field of
the $su(2)$ theory. Substituting into the Chern-Simons 3-form, we obtain
\begin{equation}
Tr\Omega \left[ A\right] =\kappa _{nm}A^{n}dA^{m}+\frac{2}{3}\kappa
_{nml}A^{n}A^{m}A^{l}
\end{equation}%
where the coupling tensor $\kappa _{nm}$ is equal to $Tr\left( E_{n\alpha
}E_{m\alpha }\right) $ and $\kappa _{nml}=Tr\left( E_{n\alpha }E_{m\alpha
}E_{l\alpha }\right) $. For instance, the intersections for the $sl(2,%
\mathbb{R})$ theory become $\mathcal{U}_{a}^{n\alpha }\mathcal{U}%
_{b}^{m\beta }\kappa _{nm}=q_{ab}^{sl_{2}}$ and $\mathcal{U}_{a}^{n\alpha }%
\mathcal{U}_{b}^{m\beta }\mathcal{U}_{c}^{l\gamma }\kappa
_{nml}=q_{abc}^{sl_{2}}$; or equivalently%
\begin{equation}
\kappa _{nm}=(\mathcal{U}^{-1})_{n\alpha }^{a}(\mathcal{U}^{-1})_{m\beta
}^{b}q_{ab}^{sl_{2}},\qquad \kappa _{nml}=(\mathcal{U}^{-1})_{n\alpha }^{a}(%
\mathcal{U}^{-1})_{m\beta }^{b}(\mathcal{U}^{-1})_{l\gamma
}^{c}q_{abc}^{sl_{2}}
\end{equation}

\ \ \

\begin{figure}[h]
\begin{center}
\includegraphics[width=12cm]{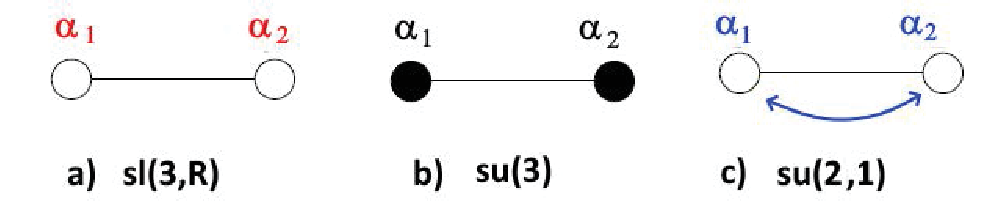}
\end{center}
\par
\vspace{-0.5cm}
\caption{Tits-Satake diagrams of the three real forms of $sl\left( 3,\mathbb{%
C}\right) .$ Useful properties of these graphs are reported in appendix B.}
\label{STA}
\end{figure}

$\bullet $ \emph{Higher spins in} $SL\left( 3,\mathbb{R}\right) $\emph{\
theory}\newline
The complex Lie algebra $\boldsymbol{A}_{2}$ has three real forms described
by the Tits-Satake diagrams in Figure \textbf{\ref{STA},} where $sl\left( 3,%
\mathbb{R}\right) $ is the real split of the complex Lie algebra $sl(3,%
\mathbb{C})$
The expression of the eight generators of $sl\left( 3,\mathbb{R}\right) $ in
terms of the Chevalley operators of the complex Lie algebra $\boldsymbol{A}%
_{2}$ is given in appendix A; see eqs(\ref{75}-\ref{79}). The bridge between
the cartesian generators $\mathfrak{J}_{\text{\textsc{a}}}^{sl_{3}}$ of $%
sl\left( 3,\mathbb{R}\right) $ and the Chevalley $H_{\alpha },$ $E_{\pm
\alpha }$ of $sl\left( 3,\mathbb{C}\right) $ can be formally defined like%
\begin{equation}
\begin{tabular}{lllllll}
$\mathfrak{J}_{\text{\textsc{a}}}^{sl_{3}}$ & $=$ & $\mathcal{U}_{\text{%
\textsc{a}}}^{n_{1},n_{2}}E_{n_{1}\alpha _{1},n_{2}\alpha _{2}}$ & $,$ & $q_{%
\text{\textsc{ab}}}^{sl_{3}}$ & $=$ & $\mathcal{U}_{\text{\textsc{a}}%
}^{n_{1},n_{2}}\mathcal{U}_{\text{\textsc{b}}}^{m_{1},m_{2}}\kappa
_{n_{1}n_{2}m_{1}m_{2}}$ \\
$E_{n_{1}\alpha _{1},n_{2}\alpha _{2}}$ & $=$ & $(\mathcal{U}^{-1}\mathcal{)}%
_{n_{1},n_{2}}^{\text{\textsc{a}}}\mathfrak{J}_{\text{\textsc{a}}}^{sl_{3}}$
& $,$ & $\kappa _{n_{1}n_{2}m_{1}m_{2}}$ & $=$ & $(\mathcal{U}^{-1}\mathcal{)%
}_{n_{1},n_{2}}^{\text{\textsc{a}}}(\mathcal{U}^{-1})_{m_{1},m_{2}}^{\text{%
\textsc{b}}}q_{\text{\textsc{ab}}}^{sl_{3}}$%
\end{tabular}%
\end{equation}%
where $\mathcal{U}_{\text{\textsc{a}}}^{n_{1},n_{2}}$ is an invertible $%
8\times 8$ matrix given by (\ref{u8}) with $\det \mathcal{U}_{\text{\textsc{a%
}}}^{n_{1},n_{2}}=(-2)^{3}$; it is the homologue of the $3\times 3$ matrix $%
\mathcal{U}_{a}^{n\alpha }$ in the $sl\left( 2,\mathbb{R}\right) $ theory.
Notice that similar relations can be written for the two other real forms of
$sl\left( 3,\mathbb{C}\right) ,$ namely the $su(3)$ and $su(2,1)$
represented in Figures \textbf{\ref{STA}}-b, c). We have
\begin{equation}
q_{\text{\textsc{ab}}}^{su_{3}}=\mathcal{V}_{\text{\textsc{a}}}^{n_{1},n_{2}}%
\mathcal{V}_{\text{\textsc{b}}}^{m_{1},m_{2}}\kappa
_{n_{1}n_{2}m_{1}m_{2}},\qquad q_{\text{\textsc{ab}}}^{su_{2,1}}=\mathcal{W}%
_{\text{\textsc{a}}}^{n_{1},n_{2}}\mathcal{W}_{\text{\textsc{b}}%
}^{m_{1},m_{2}}\kappa _{n_{1}n_{2}m_{1}m_{2}}
\end{equation}%
where the $\mathcal{V}_{\text{\textsc{a}}}^{n_{1},n_{2}}$ and $\mathcal{W}_{%
\text{\textsc{a}}}^{n_{1},n_{2}}$ matrices are given by eq(\ref{v8}) and eq(%
\ref{W21}); and where the Killing forms are as follows%
\begin{equation}
\begin{tabular}{lll}
$Tr(\mathfrak{J}_{\text{\textsc{a}}}^{sl_{3}}\mathfrak{J}_{\text{\textsc{b}}%
}^{sl_{3}})$ & $=$ & $q_{\text{\textsc{ab}}}^{sl_{3}}$ \\
$Tr(\mathfrak{J}_{\text{\textsc{a}}}^{su_{3}}\mathfrak{J}_{\text{\textsc{b}}%
}^{su_{3}})$ & $=$ & $q_{\text{\textsc{ab}}}^{su_{3}}$ \\
$Tr(\mathfrak{J}_{\text{\textsc{a}}}^{su_{2,1}}\mathfrak{J}_{\text{\textsc{b}%
}}^{su_{2,1}})$ & $=$ & $q_{\text{\textsc{ab}}}^{su_{2,1}}$ \\
$Tr(E_{n_{1}\alpha _{1},n_{2}\alpha _{2}}E_{m_{1}\alpha _{1},m_{2}\alpha
_{2}})$ & $=$ & $\kappa _{n_{1}n_{2}m_{1}m_{2}}$%
\end{tabular}%
\end{equation}

Next, we use the Chevalley basis $E_{n\alpha }\equiv \left( H_{\alpha
},E_{\pm \alpha }\right) $ of $SL\left( 2\right) $ as the 3D Lorentz-spin
with generators $J_{a}$ to deal with the higher spin gauge fields (\ref{4})
and their interacting field action. To that purpose, we proceed as follows:
\newline
$\left( \mathbf{1}\right) $ We think about the above three operators $%
J_{a}^{sl_{2}}=\mathcal{U}_{a}^{n\alpha }E_{n\alpha }$ in terms of the first
triplet $\left( H_{\alpha _{1}},E_{\pm \alpha _{1}}\right) $ generating the
subalgebra $sl\left( 2\right) $ within $sl\left( 3\right) .$ These Chevalley
generators obey the commutation relations,%
\begin{equation}
\left[ H_{\alpha _{{\small 1}}},E_{\pm \alpha _{{\small 1}}}\right] =\pm
E_{\pm \alpha _{{\small 1}}},\qquad \left[ E_{-\alpha _{{\small 1}%
}},E_{\alpha _{{\small 1}}}\right] =\alpha _{{\small 1}}^{2}H_{\alpha _{%
{\small 1}}}
\end{equation}%
where $\alpha _{1}^{2}=2$ and $\alpha _{{\small 1}}=\epsilon _{1}-\epsilon
_{2}$ with $\epsilon _{i}.\epsilon _{j}=\delta _{ij}$. \newline
$\left( \mathbf{2}\right) $ The real Lie algebra $sl\left( 3,\mathbb{R}%
\right) $ has eight generators $\mathfrak{J}_{\text{\textsc{a}}}^{sl_{3}}=%
\mathfrak{J}_{\text{\textsc{1}}}^{sl_{3}},...,\mathfrak{J}_{\text{\textsc{8}}%
}^{sl_{3}}$ with Killing form $q_{\text{\textsc{ab}}}^{sl_{3}}$; the first
three of these $\mathfrak{J}_{\text{\textsc{a}}}^{sl_{3}}$'s are given by
the Lorentz $J_{a}^{sl_{2}}=\mathcal{U}_{a}^{n\alpha _{1}}E_{n\alpha _{1}}.$
The remaining five $\mathfrak{J}_{\text{\textsc{4}}}^{sl_{3}},$ $\mathfrak{J}%
_{\text{\textsc{5}}}^{sl_{3}},$ $\mathfrak{J}_{\text{\textsc{6}}}^{sl_{3}},$
$\mathfrak{J}_{\text{\textsc{7}}}^{sl_{3}},$ $\mathfrak{J}_{\text{\textsc{8}}%
}^{sl_{3}}$ are obtained by using the two other Chevalley triplets $%
E_{n\alpha _{2}}=\left( H_{\alpha _{2}},E_{\pm \alpha _{2}}\right) $, $%
E_{n\alpha _{3}}=\left( H_{\alpha _{3}},E_{\pm \alpha _{3}}\right) $ with
the constraints $\alpha _{3}=\alpha _{1}+\alpha _{2}$ and $H_{\alpha
_{1}+\alpha _{2}}=H_{\alpha _{1}}+H_{\alpha _{2}}$ and $E_{\pm \alpha
_{3}}=E_{\pm \left( \alpha _{1}+\alpha _{2}\right) }.$ The six elements in
the root system $\Phi _{\boldsymbol{A}_{2}}$ of the Lie algebra $\boldsymbol{%
A}_{2}$ can be splitted as follows%
\begin{equation}
\begin{tabular}{|c|c|c|}
\hline
{\small roots} $\Phi _{\boldsymbol{A}_{2}}$ & $\pm \alpha _{{\small 1}}$ & $%
\left.
\begin{array}{ccc}
\pm \alpha _{{\small 2}} & , & \pm \left( \alpha _{{\small 1}}+\alpha _{%
{\small 2}}\right)%
\end{array}%
\right. $ \\ \hline
{\small L-spin} $\mathfrak{j}\left( \Phi _{\boldsymbol{A}_{2}}\right) $ & $1$
& $2$ \\ \hline
{\small CFT-spin s}$\left( \Phi _{A_{2}}\right) $ & $2$ & $3$ \\ \hline
\end{tabular}%
\end{equation}%
where $\alpha _{2}=\epsilon _{2}-\epsilon _{3}$ and $\alpha _{1}+\alpha
_{2}=\epsilon _{1}-\epsilon _{3}.$ The Cartan involution on these roots as
well as the Chevalley generators $\left( H_{\alpha ,}E_{\pm \alpha ,}\right)
$ are given by eq(\ref{714}) in appendix B.\newline
$\left( \mathbf{3}\right) $ the eight generators $\mathfrak{J}_{\text{%
\textsc{1}}}^{sl_{3}},...,\mathfrak{J}_{\text{\textsc{8}}}^{sl_{3}}$ of the
real form $sl(3,\mathbb{R})$ are related to the Chevalley basis via \textrm{%
\cite{Eric},}%
\begin{equation}
\begin{tabular}{|c|c|c|c|c|}
\hline
$\mathfrak{J}_{\text{\textsc{1}}}^{sl_{3}}\quad ,\quad \mathfrak{J}_{\text{%
\textsc{2}}}^{sl_{3}}$ & $\mathfrak{J}_{\text{\textsc{3}}}^{sl_{3}}$ & $%
\mathfrak{J_{\text{\textsc{4}}}^{sl_{3}}\quad },\mathfrak{\quad J}_{\text{%
\textsc{5}}}^{sl_{3}}$ & $\mathfrak{J}_{\text{\textsc{6}}}^{sl_{3}}$ & $%
\mathfrak{J}_{\text{\textsc{7}}}^{sl_{3}}\quad ,\quad \mathfrak{J}_{\text{%
\textsc{8}}}^{sl_{3}}$ \\ \hline
$E_{\alpha _{{\small 1}}}\pm E_{-\alpha _{{\small 1}}}$ & $H_{\alpha _{%
{\small 1}}}$ & $E_{\alpha _{{\small 2}}}\pm E_{-\alpha _{{\small 2}}}$ & $%
H_{\alpha _{{\small 2}}}$ & $E_{\alpha _{{\small 3}}}\pm E_{-\alpha _{%
{\small 3}}}$ \\ \hline
\end{tabular}%
\end{equation}%
Notice that the root system $\Phi _{\boldsymbol{A}_{{\small 2}}}$ decomposes
into the union of two subsets, $\Phi _{\boldsymbol{A}_{{\small 1}}}=\pm
\alpha _{1};$ and $\Phi _{\boldsymbol{A}_{2}\backslash \boldsymbol{A}%
_{1}}=\pm \alpha _{2},\pm (\alpha _{1}+\alpha _{2}).$ This leads to the
following partition of the $\boldsymbol{A}_{2}$ generators
\begin{equation}
\begin{tabular}{|c|c|c|c|c|c|}
\hline
$\boldsymbol{A}_{2}$ & {\small positive} {\small roots} &
\multicolumn{2}{|c|}{\small new generators} & {\small old generators} &
{\small number} \\ \hline
$\boldsymbol{A}_{1}$ & $\alpha _{1}$ & $E_{\pm }$ & $E_{0}$ & $J_{a}$ & $3$
\\ \hline
$\boldsymbol{A}_{2}\backslash \boldsymbol{A}_{1}$ & $\left.
\begin{array}{c}
\alpha _{2} \\
\alpha _{1}+\alpha _{2}%
\end{array}%
\right. $ & $\left.
\begin{array}{c}
F_{\pm } \\
F_{\pm 2}%
\end{array}%
\right. $ & $F_{0}$ & $T_{ab}$ & $5$ \\ \hline
\end{tabular}
\label{sp}
\end{equation}%
where we have used the root height property $ht(n\alpha _{{\small 1}%
}+m\alpha _{{\small 2}})=n+m$ to denote the step operators $F_{n\alpha _{%
{\small 1}}+m\alpha _{{\small 2}}}$ like $F_{n+m}.$ We illustrate this
splitting (\ref{sp}) by the pattern below,%
\begin{equation}
\begin{tabular}{l|ll|ll}
\cline{2-3}
&  & $F_{\alpha _{1}+\alpha _{2}}$ &  &  \\ \cline{4-4}
&  & $F_{+\alpha _{2}}$ & $E_{+\alpha _{1}}$ & \multicolumn{1}{|l}{} \\
&  & $\ H_{\alpha _{2}}$ & $\ H_{\alpha _{1}}$ & \multicolumn{1}{|l}{} \\
&  & $F_{-\alpha _{2}}$ & $E_{-\alpha _{1}}$ & \multicolumn{1}{|l}{} \\
\cline{4-4}
&  & $F_{-\alpha _{1}-\alpha _{2}}$ &  &  \\ \cline{2-3}
\end{tabular}%
\qquad \equiv \qquad
\begin{tabular}{l|ll|ll}
\cline{2-3}
&  & $F_{+2}$ &  &  \\ \cline{4-4}
&  & $F_{+}$ & $E_{+}$ & \multicolumn{1}{|l}{} \\
&  & $F_{0}$ & $E_{0}$ & \multicolumn{1}{|l}{} \\
&  & $F_{-}$ & $E_{-}$ & \multicolumn{1}{|l}{} \\ \cline{4-4}
&  & $F_{-2}$ &  &  \\ \cline{2-3}
\end{tabular}
\label{sl3}
\end{equation}%
where the two multiplets with L-spins $\mathfrak{j}=1$ and $2$ are
explicitly shown. Using AdS$_{3}$/CFT$_{2}$ correspondence, it follows that $%
\left( \mathbf{1}\right) $ the root $\pm \alpha _{1}$ of the Lie subalgebra $%
\boldsymbol{A}_{1}$, associated with L-spin $\mathfrak{j}=1$ of eqs(\ref{19}%
), is also associated with the conformal current $T\left( z\right) $ living
on the boundary of AdS$_{3}$ with conformal weight $h=2$ \ and conformal
spin $s=2$. This holomorphic current generates the usual Virosoro algebra
satisfying the well known OPE,%
\begin{equation}
T\left( z\right) T\left( w\right) =\frac{c/2}{\left( z-w\right) ^{4}}+\frac{2%
}{\left( z-w\right) ^{2}}T\left( w\right) +\frac{1}{\left( z-w\right) }%
\partial _{w}T\left( w\right) +...  \label{t2}
\end{equation}%
with Laurent modes as%
\begin{equation}
L_{n}=\doint\nolimits_{\gamma _{0}}\frac{dz}{2i\pi }z^{n+1}T\left( z\right)
,\qquad n\in \mathbb{Z}
\end{equation}%
A similar relation is valid for the antiholomorphic current $\bar{T}\left(
\bar{z}\right) $ with conformal weight $\bar{h}=2$ \ and conformal spin $s=h-%
\bar{h}=-2$.\newline
$\left( \mathbf{2}\right) $ the roots $\pm \alpha _{2},$ $\pm (\alpha
_{1}+\alpha _{2})$ sitting in $\boldsymbol{A}_{2}\backslash \boldsymbol{A}%
_{1}$\ are associated with L-spin $\mathfrak{j}=2$ and the conformal spin $%
s=3$ current $W^{\left( 3\right) }\left( z\right) $ obeying amongst others,
\begin{equation}
T\left( z\right) W^{\left( 3\right) }\left( w\right) =\frac{3}{\left(
z-w\right) ^{2}}W^{\left( 3\right) }\left( w\right) +\frac{1}{\left(
z-w\right) }\partial _{w}W^{\left( 3\right) }\left( w\right) +...  \label{t3}
\end{equation}%
with Laurent modes as%
\begin{equation}
W_{n}^{\left( 3\right) }=\doint\nolimits_{\gamma _{0}}\frac{dz}{2i\pi }%
z^{n+2}W\left( z\right) ,\qquad n\in \mathbb{Z}
\end{equation}%
As such, there are two conformal currents $T\left( z\right) $ and $W^{\left(
3\right) }\left( z\right) $ on the boundary of AdS$_{3}$ gravity with $%
\boldsymbol{A}_{2}$ gauge symmetry where we notice that the number of
conformal currents is just the rank of the Lie algebras $\boldsymbol{A}_{2}$%
. Obviously, we also have mirror partners given by the anti-holomorphic
copies often omitted but understood along the presentation. The generators $%
\{E_{n}\}_{n=0,\pm }$ and $\{F_{N}\}_{N=0,\pm 1,\pm 2}$ in (\ref{sl3}) are
related to the Laurent modes $L_{m},$ $W_{m}$ of the conformal $\boldsymbol{%
WA}_{2}$\emph{-} symmetry via%
\begin{equation}
E_{n}=L_{-n}\qquad ,\text{\qquad }F_{N}=W_{-N}
\end{equation}%
As for the standard 3D gravity (\ref{RC}-\ref{CR}) with $\boldsymbol{WA}_{1}$
invariance, these modes describe higher spin AdS$_{3}$ and they correspond
to the vanishing of the conformal anomalies of the $\boldsymbol{WA}_{2}$%
-invariance on the boundary. Recall that the asymptotic $\boldsymbol{WA}_{2}$%
\emph{-} symmetry is generated by the two conformal currents $T\left(
z\right) $ and $W\left( z\right) $ with the Laurent expansions \textrm{\cite%
{23}-\cite{24},}
\begin{equation}
T\left( z\right) =\dsum\limits_{n=-\infty }^{\infty }z^{-n-2}L_{n}\qquad
,\qquad W\left( z\right) =\dsum\limits_{N=-\infty }^{\infty }z^{-N-3}W_{N}
\label{t4}
\end{equation}%
Their Laurent modes satisfy the following commutation relations that close
non linearly as,
\begin{equation}
\begin{tabular}{lll}
$\left[ L_{n},L_{m}\right] $ & $=$ & $\left( n-m\right) L_{n+m}+\frac{c}{12}%
\left( n^{3}-n\right) \delta _{n+m}$ \\
$\left[ L_{n},W_{N}\right] $ & $=$ & $\left( 2n-N\right) W_{n+N}$ \\
$\left[ W_{N},W_{M}\right] $ & $=$ & $\frac{1}{30}\left( N-M\right) \left[
2\left( N+M\right) ^{2}-5NM-8\right] L_{N+M}+$ \\
&  & $\frac{16\left( N-M\right) }{22+5c}\Lambda _{N+M}+\frac{c}{360}\left(
N^{3}-N\right) \left( N^{2}-4\right) \delta _{N+M}$%
\end{tabular}
\label{w3}
\end{equation}%
where the $\Lambda _{M}$'s are non linear in the Virasoro modes \textrm{\cite%
{23,BEN}}%
\begin{equation}
\Lambda _{M}=\sum_{n}:L_{M-N}L_{N}:-\frac{3}{10}\left( M+3\right) \left(
M+2\right) L_{M}
\end{equation}%
The central extensions in these infinite dimensional algebra (\ref{w3})
describe conformal anomalies; they vanish for $n=0,\pm 1$ and $N=0,\pm 1,\pm
2$. By restricting the commutations to these values, the central extensions
disappear and one is left with%
\begin{equation}
\begin{tabular}{lll}
$\left[ E_{i},E_{j}\right] $ & $=$ & $\left( j-i\right) E_{i+j}$ \\
$\left[ E_{i},F_{m}\right] $ & $=$ & $\left( m-2i\right) F_{i+m}$ \\
$\left[ F_{m},F_{n}\right] $ & $=$ & $\frac{\sigma }{3}\left( n-m\right)
\left( 2m^{2}+2n^{2}-mn-8\right) E_{m+n}$%
\end{tabular}%
\end{equation}%
which yield the commutation relations of $sl(3,\mathbb{R})$ for $\sigma =-1$
and $su(2,1)$ for $\sigma =1$. As a consequence of the splitting (\ref{sp}),
the expansion of the Chern-Simons gauge connection decomposes as,%
\begin{equation}
A_{\mu }=\dsum\limits_{n=-1}^{+1}A_{\mu }^{n}E_{n}+\dsum\limits_{N=-2}^{+2}%
\mathcal{B}_{\mu }^{N}F_{N}  \label{AB}
\end{equation}%
where $A_{\mu }^{n}E_{n}$ is valued in $\boldsymbol{A}_{1}$; and $\mathcal{B}%
_{\mu }^{N}F_{N}$ sits into $\boldsymbol{A}_{2}\backslash $ $\boldsymbol{A}%
_{1}$. In this basis, the Dreibein and spin connections of $\boldsymbol{A}%
_{2}$ expand as%
\begin{equation}
\begin{tabular}{lll}
$\omega _{\mu }$ & $=$ & $\dsum\limits_{n=-1}^{+1}\mathcal{\omega }_{\mu
}^{n}E_{n}+\dsum\limits_{N=-2}^{+2}\Omega _{\mu }^{N}F_{N}$ \\
$e_{\mu }$ & $=$ & $\dsum\limits_{n=-1}^{+1}e_{\mu
}^{n}E_{n}+\dsum\limits_{N=-2}^{+2}\mathcal{E}_{\mu }^{N}F_{N}$%
\end{tabular}%
\end{equation}%
The Chern-Simons gauge fields are realised by%
\begin{equation}
\begin{tabular}{lll}
$\left( A_{\mu }^{n}\right) _{L}$ & $=$ & $\omega _{\mu }^{n}+\frac{1}{l_{%
{\small AdS}}}e_{\mu }^{n}$ \\
$\left( A_{\mu }^{n}\right) _{R}$ & $=$ & $\omega _{\mu }^{n}-\frac{1}{l_{%
{\small AdS}}}e_{\mu }^{n}$%
\end{tabular}%
\qquad ,\qquad
\begin{tabular}{lll}
$(\mathcal{A}_{\mu }^{N})_{L}$ & $=$ & $\Omega _{\mu }^{N}+\frac{1}{l_{%
{\small AdS}}}\mathcal{E}_{\mu }^{N}$ \\
$(\mathcal{A}_{\mu }^{N})_{R}$ & $=$ & $\Omega _{\mu }^{N}-\frac{1}{l_{%
{\small AdS}}}\mathcal{E}_{\mu }^{N}$%
\end{tabular}%
\end{equation}%
Substituting (\ref{AB}) into the Chern-Simons 3-form $Tr\left( AdA\right) +%
\frac{2}{3}Tr\left( A^{3}\right) $, we get for the quadratic term $Tr\left(
AdA\right) $,%
\begin{equation}
Tr\left( AdA\right) =\kappa _{pq}A^{p}dA_{\mu }^{q}+\hat{\kappa}_{NM}%
\mathcal{B}^{N}d\mathcal{B}^{M}
\end{equation}%
with%
\begin{equation}
\kappa _{pq}=Tr\left( E_{p}E_{q}\right) ,\qquad \tilde{\kappa}_{pN}=Tr\left(
E_{p}F_{N}\right) =0,\qquad \hat{\kappa}_{NM}=Tr\left( F_{N}F_{M}\right)
\end{equation}%
For the cubic term $Tr\left( A^{3}\right) $, we have%
\begin{equation}
Tr\left( A^{3}\right) =\kappa _{pqr}\left( A^{p}A^{q}A^{r}\right) +3%
\mathring{\kappa}_{rNM}\left( A^{r}\mathcal{B}^{N}\mathcal{B}^{M}\right)
\end{equation}%
with%
\begin{equation}
\begin{tabular}{lll}
$\kappa _{pqr}$ & $=$ & $Tr\left( E_{p}E_{q}E_{r}\right) $ \\
$\kappa _{pNM}$ & $=$ & $Tr\left( E_{p}F_{N}F_{M}\right) $%
\end{tabular}%
\qquad ,\qquad
\begin{tabular}{lll}
$\kappa _{NML}$ & $=$ & $Tr\left( F_{N}F_{M}F_{L}\right) =0$ \\
$\kappa _{pqN}$ & $=$ & $Tr\left( E_{p}E_{q}F_{N}\right) =0$%
\end{tabular}%
\end{equation}%
The Chern-Simons field action of the higher spin gravity has three blocks
like%
\begin{equation}
\mathcal{S}_{\boldsymbol{A}_{2}}^{{\small CS}}=\mathcal{S}_{\boldsymbol{A}%
_{1}}+\mathcal{S}_{\boldsymbol{A}_{2}\backslash \boldsymbol{A}_{1}}+\mathcal{%
S}_{{\small int}}
\end{equation}%
with%
\begin{equation}
\begin{tabular}{lll}
$\mathcal{S}_{\boldsymbol{A}_{1}}$ & $=$ & $\dint\nolimits_{\mathcal{M}%
_{3D}}\kappa _{pq}A^{p}dA^{q}+\frac{2}{3}\kappa _{pqr}A^{p}A^{q}A^{r}$ \\
$\mathcal{S}_{\boldsymbol{A}_{2}\backslash \boldsymbol{A}_{1}}$ & $=$ & $%
\dint\nolimits_{\mathcal{M}_{3D}}\hat{\kappa}_{NM}\mathcal{B}^{N}d\mathcal{B}%
^{M}$%
\end{tabular}%
\end{equation}%
and interacting higher spins%
\begin{equation}
\mathcal{S}_{{\small int}}=2\dint\nolimits_{\mathcal{M}_{3D}}\mathring{\kappa%
}_{rNM}A^{r}\mathcal{B}^{N}\mathcal{B}^{M}
\end{equation}

\begin{figure}[tbph]
\begin{center}
\includegraphics[width=6.5cm]{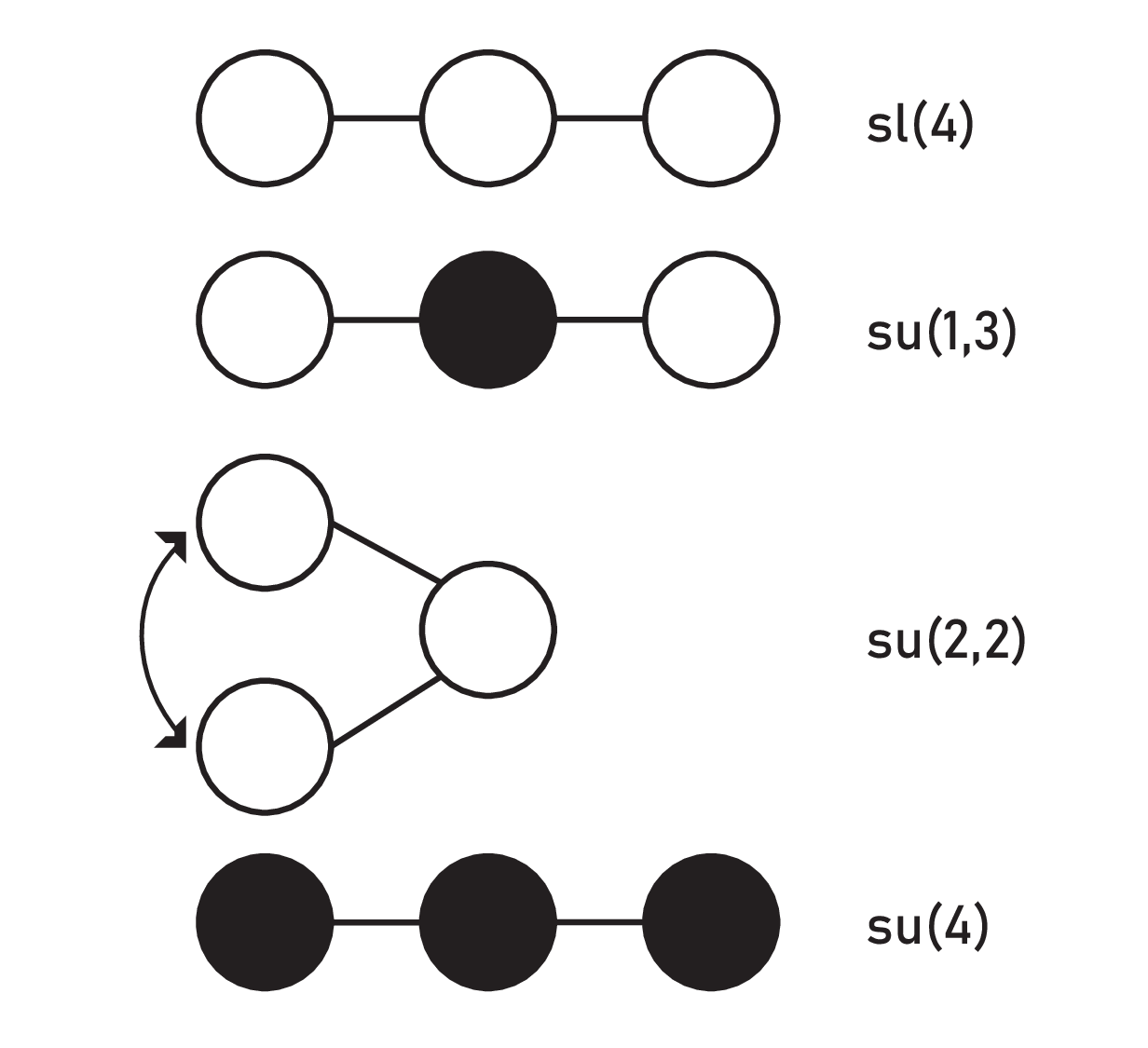}
\end{center}
\par
\vspace{-0.5cm}
\caption{The Tits-Satake diagrams of the three real forms of $sl\left( 4,%
\mathbb{C}\right) .$}
\label{A4}
\end{figure}

$\bullet $ \emph{Higher spins in }$\boldsymbol{A}_{\mathcal{N}-1}$ \emph{%
family}\newline
The extension of the above $\boldsymbol{A}_{1}$ and $\boldsymbol{A}_{2}$
constructions to higher spins within the $\boldsymbol{A}_{\mathcal{N}-1}$
family of gauge symmetries goes straightforwardly. This is done by using the
generators of the $\boldsymbol{A}_{\mathcal{N}-1}$ Lie algebra and the
Tits-Satake diagrams of its real forms. In this basis with Chevalley
generators ($H_{\alpha _{i}},E_{\pm \alpha _{i}}$), the commutation
relations read in terms of the Cartan matrix $A_{ij}$ as
\begin{equation}
\begin{tabular}{lll}
$\left[ H_{i},E_{\pm \alpha _{j}}\right] $ & $=$ & $\pm A_{ji}E_{\pm \alpha
_{j}}$ \\
$\left[ E_{-\alpha _{i}},E_{+\alpha _{i}}\right] $ & $=$ & $H_{i}$%
\end{tabular}%
\end{equation}%
with others obeying the Serre relations. The Lie algebra $\boldsymbol{A}_{%
\mathcal{N}-1}$ has dimension $\mathcal{N}^{2}-1,$ a rank $\mathcal{N}-1$
and $\mathcal{N}^{2}-\mathcal{N}$ roots. The homologue of the graphic (\ref%
{sl3}) has $\mathcal{N}-1$ sectors; this feature follows from the
decomposition of the $\mathcal{N}^{2}-1$ generators of $\boldsymbol{A}_{%
\mathcal{N}-1}$ in terms of the $\boldsymbol{A}_{1}$ multiplets by using the
following identity
\begin{equation}
\mathcal{N}^{2}-1=\sum_{\mathfrak{j}=1}^{\mathcal{N}-1}\left( 2\mathfrak{j}%
+1\right)
\end{equation}%
In this expansion, the highest weight state $\left\vert \mathfrak{j}%
\right\rangle $ of the Lorentz spin $\mathfrak{j}$ of $\boldsymbol{A}_{1}$
is given by the step operator%
\begin{equation}
\left\vert \mathfrak{j}\right\rangle \equiv E_{\alpha _{1}+\ldots +\alpha
_{j}}\qquad ,\qquad 1\leq \mathfrak{j}\leq \mathcal{N}-1
\end{equation}%
For the leading values of the integer $\mathcal{N}$, we have the following
highest weight states (HWS) of the Lorentz spin $\mathfrak{j}$ multiplets
\begin{equation}
\begin{tabular}{|c|c|c|c|}
\hline
$\boldsymbol{A}_{\mathcal{N}-1}$ & $\boldsymbol{A}_{1}$ & $\boldsymbol{A}%
_{2} $ & $\boldsymbol{A}_{3}$ \\ \hline
$\dim $ & $3$ & $8=3+5$ & $15=3+5+7$ \\ \hline
{\small HWS} & $E_{\alpha _{1}}$ & $E_{\alpha _{1}}\oplus E_{\alpha
_{1}+\alpha _{2}}$ & $E_{\alpha _{1}}\oplus E_{\alpha _{1}+\alpha
_{2}}\oplus E_{\alpha _{1}+\alpha _{2}+\alpha _{3}}$ \\ \hline
\end{tabular}%
\end{equation}%
The Tits-Satake diagrams of $\boldsymbol{A}_{1}$ are given by the table
\textbf{\ref{SA1N}}, those of $\boldsymbol{A}_{2}$ are given by Figure
\textbf{\ref{STA}-}(a,b,c); and the four classical ones of $\boldsymbol{A}%
_{3}$ by the pictures of the Figure \textbf{\ref{A4}}.
Below, we give the generators graph using the root system $\Phi _{%
\boldsymbol{A}_{3}}^{\pm }$ of the Lie algebra $\boldsymbol{A}_{3}$ having 6
positive roots that can be splited as follows%
\begin{equation}
\begin{tabular}{|c|c|c|c|}
\hline
$\Phi _{\boldsymbol{A}_{3}}^{+}$ & $\alpha _{{\small 1}}$ & $%
\begin{array}{ccc}
\alpha _{{\small 2}} & , & \alpha _{{\small 1}}+\alpha _{{\small 2}}%
\end{array}%
$ & $%
\begin{array}{ccccc}
\alpha _{{\small 3}} & , & \alpha _{{\small 2}}+\alpha _{{\small 3}} & , &
\alpha _{{\small 1}}+\alpha _{{\small 2}}+\alpha _{{\small 3}}%
\end{array}%
$ \\ \hline
{\small L-spin }$\mathfrak{j}\left( \Phi _{\boldsymbol{A}_{3}}^{+}\right) $
& $1$ & $2$ & $3$ \\ \hline
{\small CFT-spin}$\left( \Phi _{\boldsymbol{A}_{3}}^{+}\right) $ & $2$ & $3$
& $4$ \\ \hline
\end{tabular}
\label{sl4}
\end{equation}%
The three sectors in this higher spin 3D gravity are in one to one with the
three simple roots of $\boldsymbol{A}_{3}$; the generators of these sectors
are put into three $\boldsymbol{A}_{1}$ multiplets with L-spins $\mathfrak{j}%
=1,2,3$ as follows%
\begin{equation}
\begin{tabular}{|lllllll}
\cline{1-2}
& \multicolumn{1}{l|}{$G_{\mathbf{\alpha }_{1}+\mathbf{\alpha }_{2}+\mathbf{%
\alpha }_{3}}$} &  &  &  &  &  \\ \cline{4-5}
& \multicolumn{1}{l|}{$G_{\mathbf{\alpha }_{2}+\mathbf{\alpha }_{3}}$} &  &
\multicolumn{1}{|l}{} & \multicolumn{1}{l|}{$F_{\mathbf{\alpha }_{1}+\mathbf{%
\alpha }_{2}}$} &  &  \\ \cline{7-7}
& \multicolumn{1}{l|}{$G_{\mathbf{\alpha }_{3}}$} &  & \multicolumn{1}{|l}{}
& \multicolumn{1}{l|}{$F_{+\mathbf{\alpha }_{2}}$} &  & \multicolumn{1}{|l|}{%
$E_{+\mathbf{\alpha }_{1}}$} \\
& \multicolumn{1}{l|}{$\ H_{\mathbf{\alpha }_{3}}$} &  & \multicolumn{1}{|l}{%
$\ $} & \multicolumn{1}{l|}{$H_{\mathbf{\alpha }_{2}}$} &  &
\multicolumn{1}{|l|}{$H_{\mathbf{\alpha }_{1}}$} \\
& \multicolumn{1}{l|}{$G_{-\mathbf{\alpha }_{3}}$} &  & \multicolumn{1}{|l}{}
& \multicolumn{1}{l|}{$F_{-\mathbf{\alpha }_{2}}$} &  & \multicolumn{1}{|l|}{%
$E_{-\mathbf{\alpha }_{1}}$} \\ \cline{7-7}
& \multicolumn{1}{l|}{$G_{-\mathbf{\alpha }_{2}-\mathbf{\alpha }_{3}}$} &  &
\multicolumn{1}{|l}{} & \multicolumn{1}{l|}{$F_{-\mathbf{\alpha }_{1}-%
\mathbf{\alpha }_{2}}$} &  &  \\ \cline{4-5}
& \multicolumn{1}{l|}{$G_{-\mathbf{\alpha }_{1}-\mathbf{\alpha }_{2}-\mathbf{%
\alpha }_{3}}$} &  &  &  &  &  \\ \cline{1-2}
\end{tabular}
\label{su4}
\end{equation}%
with Lorentz spin $\mathfrak{j}\left( G_{\mathbf{\alpha }_{1}+\mathbf{\alpha
}_{2}+\mathbf{\alpha }_{3}}\right) =3.$ The generators belonging to the
L-spin $\mathfrak{j}=3$ are given by $G_{N}$ with $-3\leq N\leq 3;$ this
finite set of integer values correspond to the vanishing of the conformal
anomaly in $\boldsymbol{WA}_{3}$ algebra, namely
\begin{equation}
c_{N,M}=\frac{c}{5\times 7!}N\left( N^{2}-1\right) \left( N^{2}-4\right)
\left( N^{2}-9\right) \delta _{N+M}
\end{equation}

\section{$sl\left( \mathcal{N},\mathbb{R}\right) $-HS partition function in
AdS$_{3}$ gravity}

In this section, we work out the factorisation of the higher spin partition
function in AdS$_{3}$ gravity with $\boldsymbol{A}_{\mathcal{N}-1}$ gauge
symmetry by using the AdS$_{3}$/CFT$_{2}$ correspondence while focusing on
the real split form $sl\left( \mathcal{N},\mathbb{R}\right) .$ This real
form has a Tits-Satake diagram with all white nodes as in Figures \textbf{%
\ref{STA}-a} and \textbf{\ref{A4}-a}. We show that the partition function $%
\mathcal{Z}_{A_{\mathcal{N}-1}}$ splits like
\begin{equation}
\mathcal{Z}_{A_{\mathcal{N}-1}}=\left( \mathcal{Z}_{A_{\mathcal{N}-1}/A_{%
\mathcal{N}-2}}\right) \times \ldots \times \left( \mathcal{Z}%
_{A_{2}/A_{1}}\right) \times \mathcal{Z}_{\boldsymbol{A}_{1}}  \label{41}
\end{equation}%
and exploit this decomposition to derive the expressions of $\mathcal{Z}_{%
\boldsymbol{A}_{s-1}/\boldsymbol{A}_{s-2}}$ giving the contributions of each
conformal higher spin $s$ at the AdS$_{3}$ boundary in the full $\mathcal{Z}%
_{A_{\mathcal{N}-1}}$. This HS- partition function can also be interpreted
in the framework of higher spin BTZ black hole \textrm{\cite{BTZ, 6A}}.

\subsection{Computing the partition function $\mathcal{Z}_{\boldsymbol{A}%
_{2}}$}

For convenience, we begin by introducing our approach regarding the
computation of the partition function $\mathcal{Z}_{\boldsymbol{A}_{1}}$
concerning the asymptotic 3D gravity with $sl(2,\mathbb{R})$ gauge symmetry.
Then, we follow the root splitting -based rationale to build the partition
function $\mathcal{Z}_{\boldsymbol{A}_{2}}$ for the $\left\{ \mathfrak{j}%
=1,2\right\} ,$ or equivalently the CFT $\left\{ s=2,3\right\} ,$ higher
spin system with $sl(3,\mathbb{R})$ symmetry. This allows to deduce the
contribution of the spin $\mathfrak{j}=2$ (CFT $s=3$) within $\mathcal{Z}_{%
\boldsymbol{A}_{2}}$, denoted by $\mathcal{Z}_{A_{2}/A_{1}}$.

\ \

$\bullet $ \emph{3D Gravity with }$SL\left( 2,\mathbb{R}\right) $\emph{\
symmetry}\newline
Following Brown and Henneaux \textrm{\cite{1A}}, the symmetry group of the
asymptotically AdS$_{3}$ boundary is given by two copies of the Virasoro
algebra Vir$_{c}\times $Vir$_{\bar{c}}$ as%
\begin{equation}
\begin{tabular}{lllll}
Vir$_{c}$ & : & $\left[ L_{n},L_{m}\right] $ & $=$ & $\left( n-m\right)
L_{n+m}+\frac{c}{12}\left( n^{3}-n\right) \delta _{n+m}$ \\
Vir$_{\bar{c}}$ & : & $\left[ \bar{L}_{n},\bar{L}_{m}\right] $ & $=$ & $%
\left( n-m\right) \bar{L}_{n+m}+\frac{\bar{c}}{12}\left( n^{3}-n\right)
\delta _{n+m}$%
\end{tabular}
\label{v1}
\end{equation}%
These infinite symmetries contain the anomaly free subalgebras sl$\left( 2,%
\mathbb{R}\right) _{L}$ and sl$\left( 2,\mathbb{R}\right) _{R}$\ given by%
\begin{equation}
\begin{tabular}{lllll}
$sl\left( 2,\mathbb{R}\right) _{L}$ & : & $\left[ L_{n},L_{m}\right] $ & $=$
& $\left( n-m\right) L_{n+m}$ \\
$sl\left( 2,\mathbb{R}\right) _{R}$ & : & $\left[ \bar{L}_{n},\bar{L}_{m}%
\right] $ & $=$ & $\left( n-m\right) \bar{L}_{n+m}$%
\end{tabular}
\label{v2}
\end{equation}%
with sub- labels $n,m=0,\pm .$ As such, the partition function of the
boundary CFT$_{2}$ is given by the character of some representation $\phi
_{h,\bar{h}}$ of the Virasoro algebra as follows%
\begin{equation}
\begin{tabular}{lll}
$\mathcal{Z}_{\boldsymbol{A}_{1}}$ & $=$ & $Tr\left[ q^{L_{0}-\frac{c}{24}}%
\bar{q}^{\bar{L}_{0}-\frac{\bar{c}}{24}}\right] $ \\
& $=$ & $\left\vert q\right\vert ^{-\frac{c}{12}}\left( Tr\left[ q^{L_{0}}%
\bar{q}^{\bar{L}_{0}}\right] \right) $%
\end{tabular}
\label{zbt}
\end{equation}%
with $q=e^{2i\pi \tau }$ and $\tau $ being\ the complex parameters of the
boundary 2-torus \textrm{\cite{32}}. Notice that the $sl\left( 2,\mathbb{R}%
\right) $ of the AdS$_{3}$ gravity is given by the diagonal of $sl\left( 2,%
\mathbb{R}\right) _{L}\oplus sl\left( 2,\mathbb{R}\right) _{R}$ (\ref{v2}).
The corresponding conformal spin $\pm 2$ currents describing the CFT$_{2}$
on the boundary of AdS$_{3}$ verify the relations (\ref{t2},\ref{t4}). In
the saddle point approximation where the partition function $\mathcal{Z}_{%
{\small saddle}}$ is factorised like the product of a classical term $e^{-k%
\mathcal{S}^{\left( 0\right) }}$ times quantum contributions $\exp [-\sum
\frac{1}{k^{n}}\mathcal{S}^{\left( n\right) }]$ coming from loop corrections
\textrm{\cite{33}}, the classical term in (\ref{zbt}) is given by%
\begin{equation}
e^{-k\mathcal{S}^{\left( 0\right) }}=\left\vert q\right\vert ^{-\frac{c}{12}}
\end{equation}%
and is interpreted as $\left( i\right) $ the anomaly $L_{0}\left\vert
0,c\right\rangle =\bar{L}_{0}\left\vert 0,c\right\rangle =-k\left\vert
0,c\right\rangle $ (corresponding to $\left\langle 0,c|H|0,c\right\rangle <0$%
), and $\left( ii\right) $ the ground state contribution to the partition
function, namely $\left\langle 0,c|q^{L_{0}}\bar{q}^{\bar{L}%
_{0}}|0,c\right\rangle .$ This leads to the relation $\left\vert
q\right\vert ^{-\frac{c}{12}}=\left\vert q\right\vert ^{-2k}$ requiring $%
c=24k$. The one loop contribution in the saddle point approximation is given
by $\mathcal{Z}_{\boldsymbol{A}_{1}}=\left\vert \mathbf{\chi }_{1}^{%
\boldsymbol{A}_{1}}\right\vert ^{2}$ with character as%
\begin{equation}
\mathbf{\chi }_{1}^{\boldsymbol{A}_{1}}=q^{-\frac{c}{24}}\dprod%
\limits_{n=2}^{\infty }\frac{1}{1-q^{n}}  \label{sl2}
\end{equation}%
Using the Dedekind eta function \textrm{\cite{33A}-\cite{33C}},
\begin{equation}
\mathbf{\eta }\left( q\right) =q^{\frac{1}{24}}\dprod\limits_{n=1}^{\infty
}\left( 1-q^{n}\right)
\end{equation}%
we can put the\ vacuum character into the following form%
\begin{equation}
\mathbf{\chi }_{1}^{\boldsymbol{A}_{1}}=\frac{1}{\mathbf{\eta }\left(
q\right) }q^{-\frac{c-1}{24}}\left( 1-q\right) =q^{-\frac{c}{24}}\frac{q^{+%
\frac{1}{24}}\left( 1-q\right) }{\mathbf{\eta }\left( q\right) }  \label{A1}
\end{equation}%
where $\left( 1-q\right) q^{+1/24}/\mathbf{\eta }\left( q\right) $ encodes
information on the Lie algebra $\boldsymbol{A}_{1}.$ As we will see below,
the factor $\left( 1-q\right) $ can be put in correspondence with the simple
root $\alpha _{1}.$

\ \

$\bullet $ \emph{Higher spin gravity with }$SL\left( 3,\mathbb{R}\right) $%
\emph{\ symmetry}\newline
For higher spin gravity theories, particularly the $SL\left( 3,\mathbb{R}%
\right) $ AdS$_{3}$ with conformal spins $\left\{ s=2,3\right\} $ at the AdS$%
_{3}$ boundary, the partition function $\mathcal{Z}_{\boldsymbol{A}_{2}}$ is
calculated using the boundary conformal $\boldsymbol{WA}_{2}$ invariance ($%
\boldsymbol{WSL}_{3}$ symmetry). Following \textrm{\cite{35,36}} using
thermal AdS$_{3}$ formulation, the one-loop contribution to the partition
function $\mathcal{Z}_{\boldsymbol{A}_{2}}$ for the boundary conformal spins
$2$ and $3$ can be expressed as follows
\begin{equation}
\mathcal{Z}_{\boldsymbol{A}_{2}}=\left\vert \mathbf{\chi }_{1}^{\boldsymbol{A%
}_{2}}\right\vert ^{2}  \label{Z}
\end{equation}%
where $\mathbf{\chi }_{1}^{\boldsymbol{A}_{2}}$ is the vacuum character of
the $\boldsymbol{WA}_{2}$- algebra (\ref{w3}) given by%
\begin{equation}
\mathbf{\chi }_{1}^{\boldsymbol{A}_{2}}=q^{-\frac{c}{24}}\dprod%
\limits_{s=2}^{3}\left( \dprod\limits_{n=s}^{\infty }\frac{1}{1-q^{n}}\right)
\end{equation}%
and expressed in terms of the Dedekind eta function $\mathbf{\eta }\left(
q\right) $ like%
\begin{equation}
\mathbf{\chi }_{1}^{\boldsymbol{A}_{2}}=\frac{1}{\mathbf{\eta }\left(
q\right) ^{2}}q^{-\frac{c-2}{24}}\left( 1-q\right) ^{2}\left( 1-q^{2}\right)
\label{ZZ}
\end{equation}%
This involves three factors, two times of $\left( 1-q\right) $ in one to one
correspondence with the two simple roots $\alpha _{1},$ $\alpha _{2}$; and
one factor $\left( 1-q^{2}\right) $ in relation with the positive root $%
\alpha _{1}+\alpha _{2}.$ In terms of the splitting $(\boldsymbol{A}%
_{2}\backslash \boldsymbol{A}_{1})+\boldsymbol{A}_{1},$ the vacuum character
$\mathbf{\chi }_{1}^{\boldsymbol{A}_{2}}$ can be factorised as%
\begin{equation}
\mathbf{\chi }_{1}^{\boldsymbol{A}_{2}}=\mathbf{\chi }_{1}^{(\boldsymbol{A}%
_{2}\backslash \boldsymbol{A}_{1})}\bullet \mathbf{\chi }_{1}^{\boldsymbol{A}%
_{1}}\qquad ,\qquad \mathbf{\chi }_{1}^{(\boldsymbol{A}_{2}\backslash
\boldsymbol{A}_{1})}=\frac{\mathbf{\chi }_{1}^{\boldsymbol{A}_{2}}}{\mathbf{%
\chi }_{1}^{\boldsymbol{A}_{1}}}
\end{equation}%
Using the expression $\mathbf{\chi }_{1}^{\boldsymbol{A}_{1}}$ given by (\ref%
{A1}), we can calculate the contribution of the boundary conformal spin $3$
to the character $\mathbf{\chi }_{1}^{\boldsymbol{A}_{2}},$ we find%
\begin{equation}
\mathbf{\chi }_{1}^{(\boldsymbol{A}_{2}\backslash \boldsymbol{A}_{1})}\left(
\tau \right) =\frac{1}{\mathbf{\eta }\left( q\right) }q^{\frac{1}{24}}\left(
1-q\right) \left( 1-q^{2}\right)  \label{A21}
\end{equation}%
Compared to eq(\ref{A1}), we learn that the factors $\left( 1-q\right)
\left( 1-q^{2}\right) $ may be also put in correspondence with the positive
roots of $\boldsymbol{A}_{2}\backslash \boldsymbol{A}_{1}$: the factor $%
\left( 1-q\right) $ is associated with the simple root $\alpha _{2}$ and $%
\left( 1-q^{2}\right) $ with the root $\alpha _{1}+\alpha _{2}.$

\subsection{Results for the partition function $\mathcal{Z}_{\boldsymbol{A}%
_{N-1}}$}

The generalisation of eqs(\ref{Z}-\ref{ZZ}) to the special linear $SL\left(
\mathcal{N},\mathbb{R}\right) $ family with $\mathcal{N}\geq 3$ gives the
partition function $\mathcal{Z}_{\boldsymbol{A}_{\mathcal{N}-1}}=|\mathbf{%
\chi }_{1}^{\boldsymbol{A}_{\mathcal{N}-1}}\mathbf{|}^{2}$ with vacuum
character $\mathbf{\chi }_{1}^{\boldsymbol{A}_{\mathcal{N}-1}}$ as follows%
\begin{equation}
\mathbf{\chi }_{1}^{\boldsymbol{A}_{\mathcal{N}-1}}=q^{-\frac{c}{24}%
}\dprod\limits_{s=1}^{\mathcal{N}-1}\left( \dprod\limits_{n=s}^{\infty }%
\frac{1}{1-q^{n}}\right)  \label{Ann}
\end{equation}%
It also reads in terms of the Dedekind eta function like%
\begin{equation}
\mathbf{\chi }_{1}^{\boldsymbol{A}_{\mathcal{N}-1}}=\frac{1}{\left[ \mathbf{%
\eta }\left( q\right) \right] ^{\mathcal{N}-1}}q^{-\frac{c+1-\mathcal{N}}{24}%
}\dprod\limits_{j=1}^{\mathcal{N}}\left( 1-q^{j}\right) ^{\mathcal{N}-j}
\label{An}
\end{equation}%
For the case $\mathcal{N}=2$, we get the $\mathbf{\chi }_{1}^{\boldsymbol{A}%
_{1}}$ of eq(\ref{sl2}); and for $\mathcal{N}=3$ we obtain $\mathbf{\chi }%
_{1}^{\boldsymbol{A}_{2}}.$ For generic $N\geq 3,$ the factors in the
product $\dprod\nolimits_{j=1}^{N-1}\left( 1-q^{j}\right) ^{N-j}\equiv
\mathbf{\psi }_{N}$ can be put in correspondence with the $N\left(
N-1\right) /2$ positive roots of the Lie algebra $\boldsymbol{A}_{N-1}$ like%
\begin{equation}
\begin{tabular}{l||l|l|l|l|l|}
factor & $\left( 1-q\right) ^{N-1}$ & $\left( 1-q^{2}\right) ^{N-2}$ & $%
\left( 1-q^{3}\right) ^{N-3}$ & $\ldots $ & $\left( 1-q^{N-1}\right) $ \\
\hline
root & $\alpha _{1},...,\alpha _{N-1}$ & $\alpha _{i}+\alpha _{i+1}$ & $%
\alpha _{i}+\alpha _{i+1}+\alpha _{i+2}$ & $\ldots $ & $\alpha
_{1}+...+\alpha _{N-1}$%
\end{tabular}%
\end{equation}%
\begin{equation*}
\end{equation*}%
Consequently, the AdS$_{3}$ gravity theory has $\left( N-1\right) $
conformal currents $W^{\left( s\right) }\left( z\right) $ with conformal
spins $s=2,3,\ldots ,N.$ In this family, the one loop contribution to the
partition function%
\begin{equation}
\mathcal{Z}_{\boldsymbol{A}_{N-1}}=\left\vert \mathbf{\chi }_{1}^{%
\boldsymbol{A}_{N-1}}\right\vert ^{2}
\end{equation}%
is given by the vacuum character $\mathbf{\chi }_{1}^{\boldsymbol{A}_{N-1}}$
of the $\boldsymbol{WA}_{N-1}$ algebra. Using (\ref{An}), we can calculate
the contribution of each higher conformal spin current with $s=2,...,N$ to
the vacuum character. For the CFT$_{2}$- spin $s=N,$ the contribution is
given by $\mathbf{\chi }_{1}^{(\boldsymbol{A}_{N-1}\backslash \boldsymbol{A}%
_{N-2})}$ reading like%
\begin{equation}
\begin{tabular}{lll}
$\mathbf{\chi }_{1}^{\boldsymbol{A}_{N-1}}$ & $=$ & $\mathbf{\chi }_{1}^{(%
\boldsymbol{A}_{N-1}\backslash \boldsymbol{A}_{N-2})}\bullet \mathbf{\chi }%
_{1}^{\boldsymbol{A}_{N-2}}$ \\
$\mathbf{\chi }_{1}^{(\boldsymbol{A}_{N-1}\backslash \boldsymbol{A}_{N-2})}$
& $=$ & $\frac{\mathbf{\chi }_{1}^{\boldsymbol{A}_{N-1}}}{\mathbf{\chi }%
_{1}^{\boldsymbol{A}_{N-2}}}$%
\end{tabular}%
\end{equation}%
while for the spin $s=N-1,$ the corresponding character is
\begin{equation}
\mathbf{\chi }_{1}^{(\boldsymbol{A}_{N-2}\backslash \boldsymbol{A}_{N-3})}=%
\frac{\mathbf{\chi }_{1}^{\boldsymbol{A}_{N-2}}}{\mathbf{\chi }_{1}^{%
\boldsymbol{A}_{N-3}}}
\end{equation}%
Using (\ref{An}), we obtain%
\begin{equation}
\mathbf{\chi }_{1}^{(\boldsymbol{A}_{N-1}\backslash \boldsymbol{A}_{N-2})}=%
\frac{1}{\left[ \mathbf{\eta }\left( q\right) \right] }q^{\frac{1}{24}%
}\dprod\limits_{j=1}^{N-1}\left( 1-q^{j}\right)
\end{equation}%
Here, the factors in the product $\dprod\nolimits_{j=1}^{N-1}\left(
1-q^{j}\right) $ correspond to $N-1$ positive roots of $A_{N-1}\backslash
A_{N-2}.$\ These roots share the simple $\alpha _{N-1}$ as shown by the
following table%
\begin{equation}
\begin{tabular}{c||c|c|c|c|c|}
factor & $1-q$ & $1-q^{2}$ & $1-q^{3}$ & $\ldots $ & $1-q^{N-1}$ \\ \hline
root & $\alpha _{N-1}$ & $\alpha _{N-2}+\alpha _{N-1}$ & $\alpha
_{N-3}+\alpha _{N-2}+\alpha _{N-1}$ & $\ldots $ & $\alpha _{1}+...+\alpha
_{N-1}$%
\end{tabular}%
\end{equation}%
\begin{equation*}
\end{equation*}

$\bullet $ \emph{Higher spin gravity with }$SL\left( 4,\mathbb{R}\right) $%
\emph{\ symmetry}\newline
As an application, we consider the case of AdS$_{3}$\ gravity with $SL\left(
4,\mathbb{R}\right) $\ symmetry which will be of utility when we will
generalise the construction to orthogonal symmetries, thanks to its
isomorphism with the $SO\left( 6,\mathbb{R}\right) $\ group. Putting $N=4$
into (\ref{An}), we obtain the following vacuum character for the $SL\left(
4,\mathbb{R}\right) $ theory
\begin{equation}
\mathbf{\chi }_{1}^{\boldsymbol{A}_{3}}=\frac{1}{\left[ \mathbf{\eta }\left(
q\right) \right] ^{3}}q^{-\frac{c-3}{24}}\left( 1-q\right) ^{3}\left(
1-q^{2}\right) ^{2}\left( 1-q^{3}\right)  \label{A3}
\end{equation}%
involving six remarkable factors; three times of $\left( 1-q\right) $ in one
to one correspondence with the three simple roots $\alpha _{1},$ $\alpha
_{2},$ $\alpha _{3}$; two factors $\left( 1-q^{2}\right) $ in relation with
the positive roots $\alpha _{1}+\alpha _{2},$ $\alpha _{2}+\alpha _{3}$; and
one $\left( 1-q^{3}\right) $ for $\alpha _{1}+\alpha _{2}+\alpha _{3}$. From
this relation, we can determine the contributions of the conformal spins $%
s=2,$ $3,$ $4$ at the boundary of AdS$_{3}$. For $s=4$, the contribution $%
\mathbf{\chi }_{1}^{(\boldsymbol{A}_{3}\backslash \boldsymbol{A}_{2})}$ is
obtained by splitting the roots systems of $\boldsymbol{A}_{3}$ like $(%
\boldsymbol{A}_{3}\backslash \boldsymbol{A}_{2})+\boldsymbol{A}_{2}$. This
way, the vacuum character is factorized like%
\begin{equation}
\begin{tabular}{lll}
$\mathbf{\chi }_{1}^{\boldsymbol{A}_{3}}$ & $=$ & $\mathbf{\chi }_{1}^{(%
\boldsymbol{A}_{3}\backslash \boldsymbol{A}_{2})}\bullet \mathbf{\chi }_{1}^{%
\boldsymbol{A}_{2}}$ \\
$\mathbf{\chi }_{1}^{(\boldsymbol{A}_{3}\backslash \boldsymbol{A}_{2})}$ & $%
= $ & $\frac{\mathbf{\chi }_{1}^{\boldsymbol{A}_{3}}}{\mathbf{\chi }_{1}^{%
\boldsymbol{A}_{2}}}$%
\end{tabular}%
\end{equation}%
where%
\begin{equation}
\mathbf{\chi }_{1}^{(\boldsymbol{A}_{3}\backslash \boldsymbol{A}_{2})}=\frac{%
1}{\left[ \mathbf{\eta }\left( q\right) \right] }q^{\frac{1}{24}}\left(
1-q\right) \left( 1-q^{2}\right) \left( 1-q^{3}\right)  \label{SU4}
\end{equation}%
Here as well, the three factors $\left( 1-q\right) \left( 1-q^{2}\right)
\left( 1-q^{3}\right) $ in the above equation may be associated with
positive roots of $\boldsymbol{A}_{3}$. The factor $\left( 1-q\right) $
corresponds to $\alpha _{3},$\ the factor $\left( 1-q^{2}\right) $ is
associated with $\alpha _{2}+\alpha _{3}$ and the factor $\left(
1-q^{3}\right) $ with $\alpha _{1}+\alpha _{2}+\alpha _{3}.$\newline
In what follows, we use these results to study the HS-AdS$_{3}$ gravity with
orthogonal $B_{\mathcal{N}}$ and $D_{\mathcal{N}}$ gauge symmetries.

\section{Higher spins with $B_{\mathcal{N}}$ symmetry}

In this section, we generalize the above construction of higher spin AdS$%
_{3} $ gravity with the $\boldsymbol{A}_{\mathcal{N}}$ family to the
orthogonal $\boldsymbol{B}_{\mathcal{N}}$ series with $\mathcal{N}\geq 2.$%
The theory for $\boldsymbol{B}_{1}$ ($\mathcal{N}=1$) coincides precisely
with the $\boldsymbol{A}_{1}$ spin 2 AdS$_{3}$ gravity; thanks to the
homomorphism $SO\left( 1,2\right) \simeq SL\left( 2,\mathbb{R}\right) $ and $%
SO\left( 3\right) \simeq SU\left( 2\right) $. \newline
In subsection 4.1, we give general aspects on real forms of the complex Lie
algebra $\boldsymbol{B}_{\mathcal{N}}$ while focussing on the $SO\left(
\mathcal{N},1+\mathcal{N}\right) $ family. We derive the higher spin gauge
fields of the AdS$_{3}$ gravity with orthogonal gauge symmetry, and the
conserved conformal currents at asymptotic AdS$_{3}$ using the extremal node
decompositions (LEND and REND). \newline
In subsection 4.2, we focus on the leading $SO\left( 2,3\right) $ and $%
SO\left( 3,4\right) $ gravity models described by Tits-Satake diagrams with
all white nodes; then we give application to the calculation of the HS-
partition function in $SO\left( \mathcal{N},1+\mathcal{N}\right) $ theory.

\subsection{AdS$_{3}$ gravity with SO$\left( \mathcal{N},1+\mathcal{N}%
\right) $ symmetry}

\begin{figure}[tbph]
\begin{center}
\includegraphics[width=7cm]{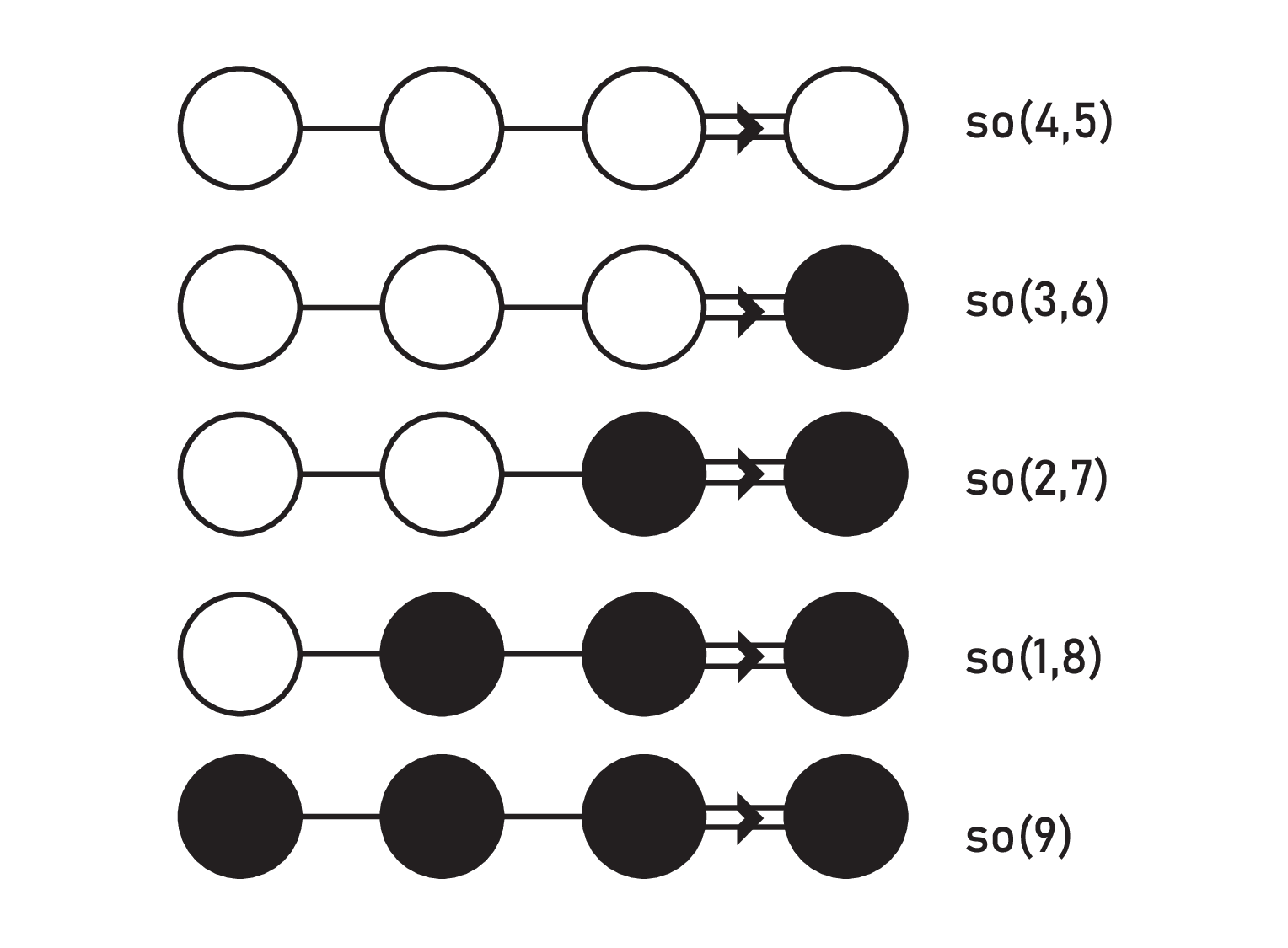}
\end{center}
\par
\vspace{-0.5cm}
\caption{The Tits-Satake diagrams for standard real forms of $\boldsymbol{B}%
_{4}.$}
\label{B3}
\end{figure}

We start by recalling that the simple Lie algebra $\boldsymbol{B}_{\mathcal{N%
}}$ has rank $\mathcal{N}$ and $\mathcal{N}\left( 2\mathcal{N}+1\right) $
dimensions. By following \textrm{\cite{SPIN,BEH,BHE}}, it also has $\mathcal{%
N}+1$ standard real forms including the real compact $SO(1+2\mathcal{N})$
and the real split form $SO(\mathcal{N},1+\mathcal{N})$ as well as $SO(p,q)$
with $p+q=1+2\mathcal{N}$ and $p<q.$ For illustration, we give in Figure
\textbf{\ref{B3}} the Tits-Satake diagrams of $\boldsymbol{B}_{4}.$
From now on, we focus on the real split form $so(\mathcal{N},1+\mathcal{N})$%
. This is motivated from $\left( i\right) $ the analogy with previous
sections concerning the 3D gravity with $SL(\mathcal{N},\mathbb{R})$
symmetry which is the real split from of $\boldsymbol{A}_{\mathcal{N}}$, and
$\left( ii\right) $ the appearance of the Lorentz group $SO(1,2)$ as a
leading member of the $SO(\mathcal{N},1+\mathcal{N})$ family.

\begin{figure}[tbph]
\begin{center}
\includegraphics[width=8cm]{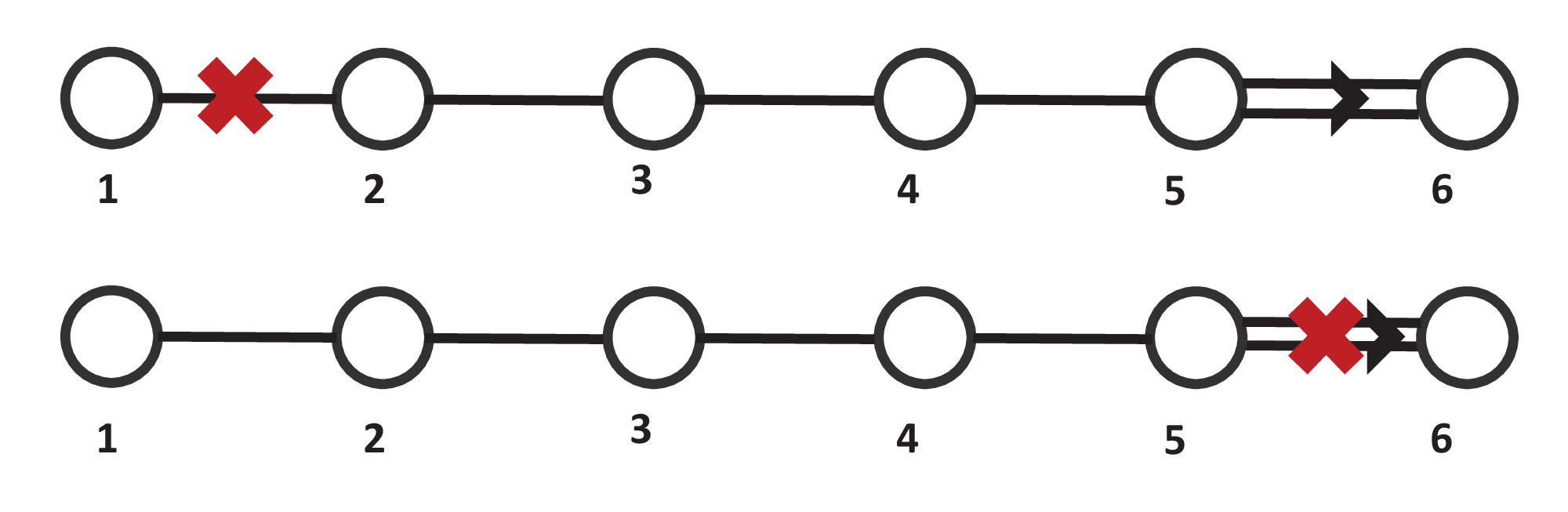}
\end{center}
\par
\vspace{-0.5cm}
\caption{Tits-Satake diagram of $so(6,7)$. (a) The first node corresponds to
$so(1,2)$ whose cutting leaves $so(5,6)$. (b) The cutting of the last node
gives $sl(6)$.}
\label{B6}
\end{figure}

We show below that 3D gravity with $SO(\mathcal{N},1+\mathcal{N})$ gauge
symmetry has $\mathcal{N}$ multiplets $\mathfrak{M}_{\mathfrak{j}}$ of $%
SO(1,2)$ whose contents follow from splitting the $\mathcal{N}\left( 2%
\mathcal{N}+1\right) $ dimensions of $\boldsymbol{B}_{\mathcal{N}}$. Recall
that the Tits-Satake diagram of $so(\mathcal{N},1+\mathcal{N})$ has $%
\mathcal{N}$ white nodes (no black node), and therefore looks like the
Dynkin diagram of the Lie algebra $\boldsymbol{B}_{\mathcal{N}}.$ In the
Left (resp. Right) Extremal Node Decomposition LEND (resp. REND) shown by
Figure \textbf{\ref{B6}, }the subgroup $SO\left( 1,2\right) \simeq SL(2,%
\mathbb{R})$ within $SO(\mathcal{N},1+\mathcal{N})$ corresponds to the first
(resp.last) node and is associated with $\pm \alpha _{1}$ (resp. $\pm \alpha
_{\mathcal{N}}$) in the root system $\Phi _{\boldsymbol{B}_{\mathcal{N}}}$.
This set has $2\mathcal{N}^{2}$ roots generated by $\mathcal{N}$ simple
roots $\alpha _{1},...,\alpha _{\mathcal{N}}$ realised as%
\begin{equation}
\alpha _{1}=\epsilon _{1}-\epsilon _{2},\qquad \ldots ,\qquad \alpha _{%
\mathcal{N}-1}=\epsilon _{\mathcal{N}-1}-\epsilon _{\mathcal{N}},\qquad
\alpha _{\mathcal{N}}=\epsilon _{\mathcal{N}}
\end{equation}%
The first $\mathcal{N}-1$ ones have length $\alpha _{i}^{2}=2$ and the N-th
has length $\alpha _{\mathcal{N}}^{2}=1.$ The full set of roots is given by $%
\pm \left( \epsilon _{i}-\epsilon _{j}\right) $, $\pm \left( \epsilon
_{i}+\epsilon _{j}\right) $ and $\pm \epsilon _{i}$ with $1\leq i<j\leq
\mathcal{N}$; their explicit content will be given when considering
particular models.

\subsubsection{Higher spin content in $SO(\mathcal{N},1+\mathcal{N})$ theory}

To describe the higher spins in the $SO(\mathcal{N},1+\mathcal{N})$ 3D
gravity theory, we split its $\mathcal{N}\left( 2\mathcal{N}+1\right) $
generators $\{\mathfrak{T}_{\text{\texttt{A}}}\}$ in terms of multiplets $%
\mathfrak{M}_{\mathfrak{j}}$\ of $SO\left( 1,2\right) $; in a similar way to
the treatment of the $\boldsymbol{A}_{\mathcal{N}}$ theory. In this regard,
recall the two basis generators of $SL(2,\mathbb{R})$ used before and which
will also be used here to deal with the orthogonal symmetry:

$\left( \mathbf{1}\right) $ The cartesian generators $J_{a}$ obeying the
commutation relations (\ref{K}) and which in terms of, we realise the $%
\mathcal{N}\left( 2\mathcal{N}+1\right) $ generators of $SO\left( \mathcal{N}%
,1+\mathcal{N}\right) $ as polynomials of $J_{a}$. In fact, using the LEND,
we show that the set $\{\mathfrak{T}_{\text{\texttt{A}}}\}$ can be realised
like
\begin{equation}
\begin{tabular}{lll}
$\{\mathfrak{T}_{\text{\texttt{A}}}\}$ & $=$ & $\oplus _{n=1}^{\mathcal{N}%
}T_{(a_{1}...a_{2n-1})}$ \\
& $=$ & $T_{(a_{1})}\oplus T_{(a_{1}a_{2}a_{3})}\oplus ...\oplus
T_{(a_{1}...a_{2\mathcal{N}-1})}$%
\end{tabular}%
\end{equation}%
with $T_{(a_{1}...a_{2n-1})}$ being completely symmetric and traceless
polynomials
\begin{equation}
T_{\left( a_{1}\ldots a_{2n-1}\right) }=\mathcal{P}_{2n-1}\left[ J_{a}\right]
,\qquad n=1,\ldots ,\mathcal{N}  \label{tr}
\end{equation}%
Typical monomials of\textrm{\ }$J_{a}$\textrm{\ }realising the $T_{\left(
a_{1}\ldots a_{2n-1}\right) }$'s are given by
\begin{equation}
J_{(a_{1}}J_{a_{2}}\ldots J_{a_{2n-1})},\qquad \eta _{(a_{1}a_{2}}\mathrm{T}%
_{a_{3}\ldots a_{2n-1})}
\end{equation}%
where $\mathrm{T}_{(a_{3}\ldots a_{2n-1})}$ is a completely symmetric tensor
of rank $2n-3.$ For the example of the rank $3$ tensor, we have
\begin{equation}
\mathrm{T}_{\left( abc\right) }=J_{(a}J_{b}J_{c)}-\frac{3}{5}\boldsymbol{J}%
^{2}\eta _{(ab}J_{c)}
\end{equation}%
where $\boldsymbol{J}^{2}$ is the Casimir $J_{a}\eta ^{ab}J_{b}$. \newline
If instead we use the REND\textrm{, }the the set $\{\mathfrak{T}_{\text{%
\texttt{A}}}\}$ can be realised like
\begin{equation}
\begin{tabular}{lll}
$\{\mathfrak{T}_{\text{\texttt{A}}}\}$ & $=$ & $\oplus _{n=1}^{\mathcal{N}%
-1}T_{(a_{1}...a_{n})}\oplus T_{\left[ \mathcal{N}\left( \mathcal{N}%
+1\right) /2\right] }$ \\
& $=$ & $T_{(a_{1})}\oplus T_{(a_{1}a_{2})}\oplus ...\oplus T_{(a_{1}...a_{%
\mathcal{N}-1})}\oplus T_{\left[ \mathcal{N}\left( \mathcal{N}+1\right) /2%
\right] }$%
\end{tabular}%
\end{equation}%
where $T_{\left[ \mathcal{N}\left( \mathcal{N}+1\right) /2\right] }$ refers
to an isolated spin multiplet with $\mathfrak{j}_{\mathcal{N}}=\mathcal{N}%
\left( \mathcal{N}+1\right) /2.$ Illustrating examples will be given later.

$\left( \mathbf{2}\right) $ The Chevalley generators given by the usual $%
\mathcal{N}$ triplets $H_{\alpha _{i}},E_{\pm \alpha _{i}}$ associated with
the simple roots, together with the Serre relations and the root system $%
\Phi _{\boldsymbol{B}_{\mathcal{N}}}=\Phi _{\boldsymbol{B}_{\mathcal{N}%
}}^{+}\cup \Phi _{\boldsymbol{B}_{\mathcal{N}}}^{-}$ of the Lie algebra of $%
SO\left( \mathcal{N},1+\mathcal{N}\right) .$ As for the $\boldsymbol{A}_{%
\mathcal{N}}$ Lie algebras, the set of positive $\Phi _{\boldsymbol{B}_{%
\mathcal{N}}}^{+}$ (resp. negative $\Phi _{\boldsymbol{B}_{\mathcal{N}}}^{-}$%
) roots splits into $\mathcal{N}$ subsets as for the example of $SO\left(
2,3\right) $ having positive $\Phi _{\boldsymbol{B}_{2}}^{+}$ decomposing
like $\Phi _{\boldsymbol{B}_{1}}^{+}+(\Phi _{\boldsymbol{B}%
_{2}}^{+}\backslash \Phi _{\boldsymbol{B}_{1}}^{+})$ where the R-END gives%
\begin{equation}
(\Phi _{B_{2}}^{+})_{R}\quad :\quad \alpha _{1}\quad ,\quad \left.
\begin{array}{l}
\alpha _{2} \\
\alpha _{2}+\alpha _{1} \\
\alpha _{2}+\alpha _{2}+\alpha _{1}%
\end{array}%
\right.  \label{al1}
\end{equation}%
and the LEND is,\textrm{\ }%
\begin{equation}
(\Phi _{B_{2}}^{+})_{L}\quad :\quad \alpha _{2}\quad ,\quad \left.
\begin{array}{l}
\alpha _{1} \\
\alpha _{1}+\alpha _{2} \\
\alpha _{1}+2\alpha _{2}%
\end{array}%
\right.
\end{equation}%
The commutation relations of the SO$\left( \mathcal{N},1+\mathcal{N}\right) $
Chevalley generators $\left( h_{\alpha _{i}},e_{\pm \alpha _{i}}\right) $
read in terms of the $\mathcal{N}$ simple roots $\alpha _{i}$ as
\begin{equation}
\left[ H_{\alpha _{i}},E_{\pm \alpha _{i}}\right] =\pm B_{ji}E_{\pm \alpha
_{j}},\qquad \left[ E_{+\alpha _{i}},E_{-\alpha _{i}}\right] =H_{\alpha _{i}}
\end{equation}%
where $B_{ij}$ is the Cartan matrix of $\boldsymbol{B}_{\mathcal{N}}$.%
\textrm{\ }Recall also that for the Chevalley generators\textrm{\ }$%
H_{\alpha _{i}},E_{\pm \alpha _{i}},$\textrm{\ }the Serre relations read as
follows%
\begin{equation}
ad\left( E_{+\alpha _{i}}\right) ^{1-B_{ij}}\left( E_{+\alpha _{j}}\right)
=0,\qquad ad\left( E_{-\alpha _{i}}\right) ^{1-B_{ij}}\left( E_{-\alpha
_{j}}\right) =0
\end{equation}%
indicating that\textrm{\ }$ad\left( E_{\pm \alpha _{i}}\right) $\textrm{\ }%
are nilpotent operators.\textrm{\ }With these ingredients at hand, we turn
now to study the higher spins in 3D gravity and the associated conserved $%
\boldsymbol{WB}_{\mathcal{N}}$-currents of the boundary CFT$_{2}$.\newline
The higher spins in $SO(\mathcal{N},1+\mathcal{N})$ gauge theory are
obtained by decomposing its $\mathcal{N}\left( 2\mathcal{N}+1\right) $
dimensions with respect to the spins of $SL\left( 2,\mathbb{R}\right) .$
Unlike the $\boldsymbol{A}_{\mathcal{N}}$ family, we find here two different
series as described below:\textrm{\ }

\paragraph{\textbf{A) Vector series:}\newline
}

This series correspond to the LEND\textrm{\ }portrayed in Figure \textbf{\ref%
{B6}-a)} leading to the following expansion%
\begin{equation}
\mathcal{N}\left( 2\mathcal{N}+1\right) =\sum_{l=1}^{\mathcal{N}}\left(
4l-1\right)  \label{dec}
\end{equation}%
It involves $\left( 4l-1\right) $- dimensional multiplets of SO(1,2). By
setting $\mathfrak{j}=2l-1$ with $\mathfrak{j}\geq 1,$ we can rewrite the
above relation like
\begin{equation}
\mathcal{N}\left( 2\mathcal{N}+1\right) =\sum_{\mathfrak{j}=odd}^{2\mathcal{N%
}-1}\left( 2\mathfrak{j}+1\right)  \label{ec}
\end{equation}%
indicating that higher spins in 3D gravity with $SO(\mathcal{N},1+\mathcal{N}%
)$ gauge symmetry involve only odd integer $SL\left( 2,\mathbb{R}\right) $
spins $\mathfrak{j}=2l-1$ like
\begin{equation}
\begin{tabular}{|c|c|c|c|c|c|c|}
\hline
$l$ & $=$ & $1$ & $2$ & $3$ & $\ldots $ & $\mathcal{N}$ \\ \hline
$\mathfrak{j}$ & $=$ & $1$ & $3$ & $5$ & $\ldots $ & $2\mathcal{N}-1$ \\
\hline
$2\mathfrak{j}+1$ & $=$ & $3$ & $7$ & $11$ & $\ldots $ & $4\mathcal{N}-1$ \\
\hline
\end{tabular}%
\end{equation}%
For $\mathcal{N}=1,$ we recover the $SO\left( 1,2\right) $ gauge symmetry of
the standard AdS$_{3}$ gravity with L-spin $\mathfrak{j}=1$. For the case of
$\mathcal{N}=2,$ the 10 dimensions $\{\mathfrak{T}_{\text{\texttt{A}}%
}\}_{1\leq \text{\texttt{A}}\leq 10}$ of the gauge symmetry $SO(2,3)$ split
in terms of polynomials of the $SO\left( 1,2\right) $ generators $J_{a}$ as
follows%
\begin{equation}
\begin{tabular}{lllll}
$\mathfrak{T}_{\text{\texttt{A}}}$ & $=$ & $J_{a}$ & $\oplus $ & $T_{\left(
abc\right) }$ \\
$10$ & $=$ & $3$ & $+$ & $7$%
\end{tabular}%
\end{equation}%
where the tensor $T_{\left( abc\right) }$ is traceless and completely
symmetric. Therefore, the $SO(2,3)$ involves two $SO\left( 1,2\right) $
multiplets with Lorentz spins $\mathfrak{j}=1$ (for $J_{a}$) and $\mathfrak{j%
}=3$ (for $T_{\left( abc\right) }$). In terms of the Cartan-Weyl generators (%
$E_{\pm \alpha },H_{\alpha }$) , the $J_{a}$ and the $T_{\left( abc\right) }$
associated to (\ref{al1}) are related to the following multiplets
\begin{equation}
J_{a}\sim \left(
\begin{array}{c}
E_{+\alpha _{1}} \\
H_{\alpha _{1}} \\
E_{-\alpha _{1}}%
\end{array}%
\right) ,\qquad T_{\left( abc\right) }\sim \left. \left(
\begin{array}{l}
E_{+\left( \alpha _{1}+2\alpha _{2}\right) } \\
E_{+\left( \alpha _{1}+\alpha _{2}\right) } \\
E_{+\alpha _{2}} \\
H_{\alpha _{2}} \\
E_{-\alpha _{2}} \\
E_{-\left( \alpha _{1}+\alpha _{2}\right) } \\
E_{-\left( \alpha _{1}+2\alpha _{2}\right) }%
\end{array}%
\right) \right.
\end{equation}%
At the asymptotic AdS$_{3},$ there are two conserved currents generating the
symmetries of the boundary CFT$_{2}$. They are given by $\left( \mathbf{i}%
\right) $ the usual energy momentum $T\left( z\right) $ with conformal spin $%
s=2$ which includes the three $J_{a}$'s as the non anomalous $L_{0,\pm }$
Laurent modes%
\begin{equation}
L_{n}=\doint\nolimits_{\gamma _{0}}\frac{dz}{2i\pi }z^{n+1}T\left( z\right)
\end{equation}%
$\left( \mathbf{ii}\right) $ a holomorphic current $W_{4}\left( z\right) $
with conformal weight $h=4$ satisfying amongst others the following OPE,%
\begin{equation}
T\left( z\right) W^{\left( 4\right) }\left( w\right) =\frac{4}{\left(
z-w\right) ^{2}}W^{\left( 4\right) }\left( w\right) +\frac{1}{\left(
z-w\right) }\partial _{w}W^{\left( 4\right) }\left( w\right) +...
\end{equation}%
with Laurrent modes as%
\begin{equation}
W_{n}^{\left( 4\right) }=\doint\nolimits_{\gamma _{0}}\frac{dz}{2i\pi }%
z^{n+3}W^{\left( 4\right) }\left( z\right)
\end{equation}%
For a generic rank of $\boldsymbol{B}_{\mathcal{N}},$ the higher spin
gravity with $SO(\mathcal{N},1+\mathcal{N})$ gauge symmetry has $\mathcal{N}$
\ multiplets with $\mathfrak{j}=2l-1$ in terms of which, the $\mathcal{N}%
\left( 2\mathcal{N}+1\right) $ generators $\{\mathfrak{T}_{\text{\texttt{A}}%
}\}$ decompose like in (\ref{ec}). For a given $\mathfrak{j}=2m-1$, the
generators $T_{\left( a_{1}\ldots a_{2m-1}\right) }$ of the gauge symmetry
are realised by completely symmetric and traceless polynomials $J_{a_{i}}$
as in eq(\ref{tr}); it has $4m-1$ degrees. In fact, the completely symmetric
$J_{(a_{1}}J_{a_{2}}\ldots J_{a_{2m-1})}$ carries $m\left( 2m+1\right) $
degrees of freedom; the extra undesired degrees are killed by demanding the
traceless condition, thus reducing $J_{(a_{1}}J_{a_{2}}\ldots J_{a_{2m-1})}$
down to the $2m-3$ rank tensor $J_{(b_{1}}J_{a_{2}}\ldots J_{b_{2m-3})}$
having $\left( m-1\right) \left( 2m-1\right) $ degrees. By substracting, we
obtain the desired number of degrees, namely%
\begin{equation}
m\left( 2m+1\right) -\left( m-1\right) \left( 2m-1\right) =4m-1
\end{equation}%
The boundary CFT$_{2}$ at the asymptotic AdS$_{3}$ has $\mathcal{N}$
conserved holomorphic currents $W^{(s)}\left( z\right) $ with conformal
weight $s=2l$ and integer $l=1,2,...,\mathcal{N}.$ For $s>2,$ we have the
following OPE,%
\begin{equation}
T\left( z\right) W^{(s)}\left( w\right) =\frac{s}{\left( z-w\right) ^{2}}%
W^{(s)}\left( w\right) +\frac{1}{\left( z-w\right) }\partial
_{w}W^{(s)}\left( w\right) +...  \label{sn}
\end{equation}%
and
\begin{equation}
W_{n}^{\left( 2l\right) }=\doint\nolimits_{\gamma _{0}}\frac{dz}{2i\pi }%
z^{n+2l-1}W^{\left( 2l\right) }\left( z\right)  \label{mw}
\end{equation}%
as well as the vacuum expectation value%
\begin{equation}
\left\langle W^{\left( 2l\right) }\left( z\right) W^{(2l)}\left( w\right)
\right\rangle \sim \frac{c/\text{\texttt{a}}_{\mathcal{N}}}{\left(
z-w\right) ^{4l}}  \tag{c}
\end{equation}%
giving the anomalies of the conformal W-algebra at the AdS$_{3}$ boundary.
Notice that here there are $\mathcal{N}$ conserved currents $%
\{W^{(2l)}\}_{1\leq l\leq \mathcal{N}}$ in the boundary CFT$_{2}$, they
generate the $\boldsymbol{WB}_{\mathcal{N}}$ invariance at the asymptotic
higher spin AdS$_{3}$ gravity. In terms of the $W_{n}^{(2l)}$ Laurent modes,
we have%
\begin{equation}
W^{(2l)}\left( z\right) =\dsum\limits_{n=-\infty }^{\infty
}z^{-n-2l}W_{n}^{(2l)}  \label{wl}
\end{equation}%
The vanishing conditions of the central extensions (\ref{c}) of the $%
\boldsymbol{WB}_{\mathcal{N}}$ algebra expressed as,
\begin{equation}
c_{n,m}^{\left( 2l\right) }\sim \frac{c}{\text{\texttt{a}}_{2l}}n\left[
n^{2}-1\right] \left[ n^{2}-3^{2}\right] ...\left[ n^{2}-\left( 2l-1\right)
^{2}\right] \delta _{n+m}
\end{equation}%
are solved by the Laurent modes $W_{n}^{(2l)}$ with subscripts $n=0,$ $\pm
1, $ $\pm 3,\ldots ,$ $\pm \left( 2l-1\right) .$ These restrictions lead to
a finite set of $W_{n}^{(2l)}$ generators%
\begin{equation}
\left(
\begin{array}{c}
W_{2l-1}^{(2l)} \\
\vdots \\
W_{1}^{(2l)} \\
W_{0}^{(2l)} \\
W_{-1}^{(2l)} \\
\vdots \\
W_{1-2l}^{(2l)}%
\end{array}%
\right) \qquad ,\qquad l=1,...,\mathcal{N}
\end{equation}%
giving precisely the generators of $SO(\mathcal{N},1+\mathcal{N})$.

\paragraph{\textbf{B) Spinorial series:}\newline
}

The spinorial series is obtained by the REND in Figure \textbf{\ref{B6}-b};
here the extremal node decomposition of $SO(\mathcal{N},1+\mathcal{N})$
generates $SL(\mathcal{N},\mathbb{R})$; as such the $\mathcal{N}\left( 2%
\mathcal{N}+1\right) $ orthogonal dimensions are splitted as
\begin{equation}
\mathcal{N}\left( 2\mathcal{N}+1\right) =\left( \mathcal{N}^{2}-1\right) +1+%
\frac{\mathcal{N}\left( \mathcal{N}+1\right) }{2}+\frac{\mathcal{N}\left(
\mathcal{N}+1\right) }{2}
\end{equation}%
where the $\mathcal{N}^{2}-1$ dimensions organise like
\begin{equation}
\mathcal{N}^{2}-1=\sum_{\mathfrak{j}=1}^{\mathcal{N}-1}\left( 2\mathfrak{j}%
+1\right)
\end{equation}%
and the extra $1+\mathcal{N}\left( \mathcal{N}+1\right) $ can be imagined as
\begin{equation}
1+\mathcal{N}\left( \mathcal{N}+1\right) =2\mathfrak{\tilde{j}}^{spinor}+1
\end{equation}%
where $\mathfrak{\tilde{j}}^{spinor}$ is an isolated Lorentz spin (ILS)
given by $\mathfrak{\tilde{j}}^{spinor}=\frac{\mathcal{N}\left( \mathcal{N}%
+1\right) }{2}.$ Therefore, the two higher spin families for $SO\left(
\mathcal{N},\mathcal{N}+1\right) $ AdS$_{3}$ gravity are as collected in the
following table%
\begin{equation}
\begin{tabular}{c|c|c}
{\small series} & {\small Lorentz- spin} & $\text{boundary CFT}_{2}\text{%
-spin}$ \\ \hline\hline
{\small vector } & $\left.
\begin{array}{ccccc}
\mathfrak{j}_{m} & = & 2{\small m-1}\text{\ } & \text{ }; & 1{\small \leq
m\leq }\mathcal{N}%
\end{array}%
\right. $ \ \  & $\left.
\begin{array}{ccc}
s_{m} & = & 2m%
\end{array}%
\right. $ \\ \hline\hline
{\small spinor} & $\left.
\begin{array}{ccccc}
\mathfrak{j}_{n} & = & n & \text{ \ }; & {\small 1\leq n\leq }\mathcal{N}-1
\\
\mathfrak{\tilde{j}}_{\mathcal{N}} & = & \frac{\mathcal{N}\left( \mathcal{N}%
+1\right) }{2} & \text{ \ }; & n=\mathcal{N}%
\end{array}%
\right. $ & $\left.
\begin{array}{ccc}
s_{n} & = & n+1 \\
\tilde{s}_{\mathcal{N}} & = & \frac{\mathcal{N}\left( \mathcal{N}+1\right) }{%
2}+1%
\end{array}%
\right. $ \\ \hline\hline
\end{tabular}%
\end{equation}

\subsubsection{Higher spin $SO(2,3)$ and $SO(3,4)$ models}

The three leading gauge group members within the orthogonal $SO(\mathcal{N}%
,1+\mathcal{N})$ family are given by%
\begin{equation}
SO(1,2),\qquad SO(2,3),\qquad SO(3,4)
\end{equation}%
and should be thought of in terms of embedding as follows%
\begin{equation}
SO(1,2)\subset SO(2,3)\subset SO(3,4)  \label{em}
\end{equation}%
The root systems and the Chevalley-Serre generators of the $SO(2,3)$ and $%
SO(3,4)$ are described below as they draw the path for the generalisation to
generic rank $\mathcal{N}$.

\ \ \ \

$\bullet $ $SO(2,3)$ \emph{symmetry: }\newline
Because of the embedding (\ref{em}), the two roots $\pm \alpha _{1}$ of $%
SO(1,2)$ are part of the root systems of $SO(2,3)$ and $SO(3,4)$. For the
case of $SO(2,3)$ having 8 roots, the positive ones read in terms of the two
simple as follows%
\begin{equation}
\alpha _{1},\quad \alpha _{2},\quad \alpha _{1}+\alpha _{2},\quad \alpha
_{1}+2\alpha _{2}
\end{equation}%
Notice that for this model, both decompositions LEND and REND coincide; and
therefore we only have one higher spin conformal spectrum. In fact, $SO(2,3)$
has ten generators given by the two Cartans $H_{\alpha _{1}},H_{\alpha _{2}}$
and eight step operators $E_{\pm \alpha _{1}},$ $E_{\pm \alpha _{2}},$ $%
E_{\pm \left( \alpha _{1}+\alpha _{2}\right) },$ $E_{\pm \left( \alpha
_{1}+2\alpha _{2}\right) }$; they are realised in our notations by \newline
$\left( \mathbf{i}\right) $ three $H_{\alpha _{1}},$ $E_{\pm \alpha _{1}}$
associated with $\alpha _{1}$ which are precisely the $E_{0},E_{\pm }$ used
before; these correspond in the asymptotic limit of AdS$_{3}$ to the non
anomalous three Virasoro mode generators
\begin{equation}
L_{0}=W_{0}^{(2)},\quad L_{+}=W_{+1}^{(2)},\quad L_{-}=W_{-1}^{(2)}
\end{equation}%
$\left( \mathbf{ii}\right) $ seven $H_{\alpha _{2}},E_{\pm \alpha _{2}},$ $%
E_{\pm \left( \alpha _{1}+\alpha _{2}\right) },$ $E_{\pm \left( \alpha
_{1}+2\alpha _{2}\right) }$ associated with the roots $\alpha _{2},$ $\alpha
_{1}+\alpha _{2}$ and $\alpha _{1}+2\alpha _{2};$ they are given by $F_{0},$
$F_{\pm },$ $F_{\pm 2},$ $F_{\pm 3}$ and are realised in terms of the non
anomalous holomorphic current modes (\ref{wl})%
\begin{equation}
W_{0}^{(4)},\quad W_{\pm 1}^{(4)},\quad W_{\pm 2}^{(4)},\quad W_{\pm 3}^{(4)}
\end{equation}

$\bullet $ $SO(3,4)$ \emph{symmetry}\newline
The $SO(3,4)$ model is the first leading model with two different higher
spin series. The vector series is given by the LEND reading in Lie algebra
language in terms of the following branching pattern%
\begin{equation}
so(3,4)\quad \rightarrow \quad so(2,3)\oplus so\left( 1,1\right) \oplus
2\times 5
\end{equation}%
with $5\equiv \left( 2,3\right) $; it corresponds to cutting the left node $%
\alpha _{1}$ in the Dynkin diagram of $\boldsymbol{B}_{3}$ as depicted in
Figure \textbf{\ref{B6}-a}. The simple root $\alpha _{1}$ should be put in
correspondence with $so\left( 1,1\right) ;$ and the $2\times 5$ with roots $%
\beta $ in $\Phi _{\boldsymbol{B}_{3}}$ depending on $\alpha _{1}$, that is
\begin{equation}
\frac{\partial \beta }{\partial \alpha _{1}}\neq 0
\end{equation}%
In other words, the 8 roots of $so(2,3)$ are generated by $\alpha
_{2},\alpha _{3}$; as such they correspond to $\frac{\partial \beta }{%
\partial \alpha _{1}}=0$. Using eqs(\ref{dec}-\ref{ec}), the 21 dimensions
of $so(3,4)$ split as follows%
\begin{equation}
21=10+11=\left( 3+7\right) +11
\end{equation}%
with $10$ referring to the dimension of $so(2,3)$ and $11=1+\left( 2\times
5\right) $. By using LEND\ (vectorial series),\textrm{\ }the 9 positive
roots of $\Phi _{B_{3}}^{+}$ splits like%
\begin{equation}
(\Phi _{B_{3}}^{+})_{{\small LEND}}\quad :\quad \left. \alpha _{3}\right.
\emph{\quad ,\qquad }\left.
\begin{array}{l}
\alpha _{2} \\
\alpha _{2}+\alpha _{3} \\
\alpha _{2}+2\alpha _{3}%
\end{array}%
\right. \emph{\quad ,\qquad }\left.
\begin{array}{l}
\alpha _{1} \\
\alpha _{1}+\alpha _{2} \\
\alpha _{1}+\alpha _{2}+\alpha _{3} \\
\alpha _{1}+\alpha _{2}+2\alpha _{3} \\
\alpha _{1}+2\alpha _{2}+2\alpha _{3}%
\end{array}%
\right.  \label{321}
\end{equation}%
They describe three SO(1,2) multiplets of spin $j=1,3,5.$ On the other hand,
by using REND (spinorial series),\textrm{\ }the 9 positive roots of $\Phi
_{B_{3}}^{+}$ decomposes as\textrm{\ }%
\begin{equation}
(\Phi _{B_{3}}^{+})_{{\small REND}}\quad :\quad \left. \alpha _{1}\right.
\quad ,\qquad \left.
\begin{array}{l}
\alpha _{2} \\
\alpha _{1}+\alpha _{2}%
\end{array}%
\right. \quad ,\qquad \left.
\begin{array}{l}
\alpha _{3} \\
\alpha _{2}+\alpha _{3} \\
\alpha _{1}+\alpha _{2}+\alpha _{3} \\
\alpha _{2}+2\alpha _{3} \\
\alpha _{1}+\alpha _{2}+2\alpha _{3} \\
\alpha _{1}+2\alpha _{2}+2\alpha _{3}%
\end{array}%
\right.  \label{spp}
\end{equation}%
They describe three SO(1,2) multiplets of spin $j=1,2,6.$ \newline
In sum, the 21 generators of the $SO(3,4)$ gauge symmetry of the higher spin
gravity model are given by the three Cartans $H_{\alpha _{1}},H_{\alpha
_{2}},H_{\alpha _{3}}$ and 18 step operators as follows:

\textbf{A) Vectorial series} for which the 18 Cartan-Weyl operators $E_{\pm
\alpha }$ are realised in terms of the Laurent modes $W_{n}^{\left(
2l\right) }$ of eq(\ref{mw}) as follows:

\begin{itemize}
\item The three $H_{\alpha _{1}},$ $E_{\pm \alpha _{1}}$ associated with the
$\alpha _{1}$ are precisely given by the $E_{0},E_{\pm }$ we used before to
generate $so(1,2)$. They correspond to
\begin{equation}
L_{0}=W_{0}^{(2)},\quad L_{+}=W_{+1}^{(2)},\quad L_{-}=W_{-1}^{(2)}
\label{a}
\end{equation}

\item The seven Cartan-Weyl operators $H_{\alpha _{2}},$ $E_{\pm \alpha
_{2}},$ $E_{\pm \left( \alpha _{2}+\alpha _{1}\right) },$ $E_{\pm \left(
\alpha _{2}+2\alpha _{1}\right) }$ generating the coset space $%
SO(2,3)\backslash SO(1,2);$ they are associated to the spin 4 current modes $%
W^{(4)}\left( z\right) $ given by eq(\ref{mw}) for $l=2.$:%
\begin{equation}
W_{0}^{(4)},\quad W_{\pm 1}^{(4)},\quad W_{\pm 2}^{(4)},\quad W_{\pm 3}^{(4)}
\label{b}
\end{equation}

\item the remaining 11 operators%
\begin{equation}
\begin{tabular}{lllll}
$H_{\alpha _{3}}$ & $,\quad $ & $E_{\pm \left( \alpha _{3}+\alpha
_{2}\right) }$ & $,\quad $ & $E_{\pm \left( \alpha _{3}+\alpha _{2}+2\alpha
_{1}\right) }$ \\
$E_{\pm \alpha _{3}}$ & $,\quad $ & $E_{\pm \left( \alpha _{1}+\alpha
_{2}+\alpha _{3}\right) }$ & $,\quad $ & $E_{\pm \left( \alpha _{3}+2\alpha
_{2}+2\alpha _{1}\right) }$%
\end{tabular}%
\end{equation}%
generate the space $SO(3,4)\backslash SO(2,3),$ and are given by%
\begin{equation}
W_{0}^{(6)},\quad W_{\pm 1}^{(6)},\quad W_{\pm 2}^{(6)},\quad W_{\pm
3}^{(6)},\quad W_{\pm 4}^{(6)},\quad W_{\pm 5}^{(6)}  \label{c}
\end{equation}
\end{itemize}

\textbf{B) Spinorial series} given by the REND corresponding to cutting $%
\alpha _{3}$\ in the Dynkin diagram of $\boldsymbol{B}_{3}$ depicted in
Figure \textbf{\ref{B6}-b. }This decomposition yields%
\begin{equation}
so(3,4)\quad \rightarrow \quad sl(3)\oplus so\left( 1,1\right) \oplus
2\times 6
\end{equation}%
where the 21 dimensions of $so(3,4)$ decompose as%
\begin{equation}
21=8+1+12=\left( 3+5\right) +13
\end{equation}%
meaning we have a spectrum of Lorentz spins $\mathfrak{j}=1,2$ and an
isolated $\mathfrak{\tilde{j}}_{3}=6$ as given by (\ref{spp}). Therefore we
have three CFT$_{2}$ currents $T\left( z\right) ,$ $W^{(3)}\left( z\right) ,$
$W^{(7)}\left( z\right) $ living at the boundary of the AdS$_{3}$; the non
anomalous Laurent modes of these W- currents give the 18 Cartan-Weyl
operators; they are given by $\left( i\right) $ the three $L_{0,\pm },$ $%
\left( ii\right) $\ the five $W_{0,\pm 1,\pm 2}^{(3)};$ and $\left(
iii\right) $ the thirteen $W_{0,\pm 1,\pm 2,\pm 3,\pm 4,\pm 5,\pm 6}^{(7)}$.

\subsection{$SO(\mathcal{N},1+\mathcal{N})$ theory and HS- partition function%
}

In this subsection, we give the field content of the $SO(\mathcal{N},1+%
\mathcal{N})$ theory while focussing on the leading $SO(2,3)$ and $SO(3,4)$
members of the family. We also give application regarding the explicit
computation of the partition function of higher spins in AdS3 gravity with $%
\boldsymbol{B}_{\mathcal{N}}$ family for $\mathcal{N}=2,3$.

\subsubsection{$SO(2,3)$ theory}

This is a particular AdS$_{3}$ gauge theory in the sense that it is also the
leading member in the symplectic $\boldsymbol{C}_{\mathcal{N}}$ family.
Since the two split real forms of both algebras $\boldsymbol{B}_{\mathcal{N}%
} $ and $\boldsymbol{C}_{\mathcal{N}}$ are identical, the following study
for the orthogonal leading models is easily replicated for the symplectic
case.

\paragraph{$SO(2,3)$ gauge fields :\newline
}

The 10 Chern-Simons gauge potentials of the $SO(2,3)$ theory couple to the
generators $H_{\alpha _{1}},$ $E_{\pm \alpha _{1}},$ and $H_{\alpha _{2}},$ $%
E_{\pm \alpha _{2}},$ $E_{\pm \left( \alpha _{1}+\alpha _{2}\right) },$ $%
E_{\pm \left( \alpha _{1}+2\alpha _{2}\right) }$ as described in section 2;
these fields are organized in two $SO(1,2)$ multiplets as follows
\begin{equation}
\begin{tabular}{|c|c|c|}
\hline
multiplet & gauge fields & number \\ \hline
$\mathfrak{j}=1$ & $\mathcal{A}_{\mu }^{0},\text{ }\mathcal{A}_{\mu }^{\pm }$
& 3 \\ \hline
$\mathfrak{j}=3$ & $\mathcal{B}_{\mu }^{0},\text{ }\mathcal{B}_{\mu }^{\pm },%
\text{ }\mathcal{B}_{\mu }^{\pm 2},\text{ }\mathcal{B}_{\mu }^{\pm 3}$ & 7
\\ \hline
\end{tabular}%
\end{equation}%
From the view of the boundary CFT$_{2}$, they couple to the non anomalous
generators $L_{n}$ and $W_{N}^{\left( 4\right) }$ as follows%
\begin{equation}
A_{\mu }=\dsum\limits_{n=-1}^{+1}\mathcal{A}_{\mu
}^{n}L_{n}+\dsum\limits_{N=-3}^{+3}\mathcal{B}_{\mu }^{N}W_{N}^{\left(
4\right) }
\end{equation}%
The expansions of the Dreibein and the spin connections of $SO(2,3)$ expand
in a similar way as follows%
\begin{equation}
\begin{tabular}{lll}
$\omega _{\mu }$ & $=$ & $\dsum\limits_{n=-1}^{+1}\mathcal{\omega }_{\mu
}^{n}L_{n}+\dsum\limits_{N=-3}^{+3}\Omega _{\mu }^{N}W_{N}^{\left( 4\right)
} $ \\
$e_{\mu }$ & $=$ & $\dsum\limits_{n=-1}^{+1}e_{\mu
}^{n}L_{n}+\dsum\limits_{N=-2}^{+2}\mathcal{E}_{\mu }^{N}W_{N}^{\left(
4\right) }$%
\end{tabular}%
\end{equation}%
The relations with the Chern-Simons gauge fields are given by%
\begin{equation}
\begin{tabular}{lllllll}
$\left( A_{\mu }^{n}\right) _{L}$ & $=$ & $\omega _{\mu }^{n}+\frac{1}{l_{%
{\small AdS}}}e_{\mu }^{n}$ & $\qquad ,\qquad $ & $(\mathcal{A}_{\mu
}^{N})_{L}$ & $=$ & $\Omega _{\mu }^{N}+\frac{1}{l_{{\small AdS}}}\mathcal{E}%
_{\mu }^{N}$ \\
$\left( A_{\mu }^{n}\right) _{R}$ & $=$ & $\omega _{\mu }^{n}-\frac{1}{l_{%
{\small AdS}}}e_{\mu }^{n}$ & $\qquad ,\qquad $ & $(\mathcal{A}_{\mu
}^{N})_{R}$ & $=$ & $\Omega _{\mu }^{N}-\frac{1}{l_{{\small AdS}}}\mathcal{E}%
_{\mu }^{N}$%
\end{tabular}%
\end{equation}%
By substituting (\ref{AB}) into the Chern-Simons 3-form $Tr\left( AdA\right)
+\frac{2}{3}Tr\left( A^{3}\right) $, we get the following Lagrangian density%
\begin{equation}
\begin{tabular}{lll}
$\mathcal{L}_{{\small tot}}^{{\small CS}}$ & $=$ & $\kappa _{pq}A^{p}dA_{\mu
}^{q}+\frac{2}{3}\kappa _{pqr}A^{p}A^{q}A^{r}+\tilde{\kappa}_{pN}A^{p}d%
\mathcal{B}^{N}+\mathcal{B}^{N}dA^{p}+$ \\
&  & $2\tilde{\kappa}_{qrN}A^{q}A^{r}\mathcal{B}^{N}+2\mathring{\kappa}%
_{rNM}A^{r}\mathcal{B}^{N}\mathcal{B}^{M}+$ \\
&  & $\hat{\kappa}_{NM}\mathcal{B}^{N}d\mathcal{B}^{M}+\frac{2}{3}\hat{\kappa%
}_{NML}\mathcal{B}^{N}\mathcal{B}^{M}\mathcal{B}^{L}$%
\end{tabular}%
\end{equation}%
with $\kappa _{pq}=Tr\left( E_{p}E_{q}\right) $ and%
\begin{equation}
\begin{tabular}{llllllll}
$\tilde{\kappa}_{pN}$ & $=$ & $Tr\left( E_{p}F_{N}\right) $ & $=0$ & $%
,\qquad $ & $\kappa _{pqr}$ & $=$ & $Tr\left( E_{p}E_{q}E_{r}\right) $ \\
$\kappa _{pqN}$ & $=$ & $Tr\left( E_{p}E_{q}F_{N}\right) $ & $=0$ & $,\qquad
$ & $\kappa _{pNM}$ & $=$ & $Tr\left( E_{p}F_{N}F_{M}\right) $ \\
$\kappa _{NML}$ & $=$ & $Tr\left( F_{N}F_{M}F_{L}\right) $ & $=0$ & $,\qquad
$ & $\hat{\kappa}_{NM}$ & $=$ & $Tr\left( F_{N}F_{M}\right) $%
\end{tabular}%
\end{equation}%
with small labels $p,q=0,\pm $ and capital labels $N,M=0,\pm ,\pm 2,\pm 3.$

\paragraph{Partition function $\mathcal{Z}_{\boldsymbol{B}_{2}}$ :\newline
}

The partition function of the $SO(2,3)$ higher spin theory is given by $%
\mathcal{Z}_{\boldsymbol{B}_{2}}=\left\vert \mathbf{\chi }_{1}^{\boldsymbol{B%
}_{2}}\left( q\right) \right\vert ^{2}$; it is determined by using results
from the $\boldsymbol{A}_{\mathcal{N}}$ family. Because $\boldsymbol{B}%
_{1}\simeq \boldsymbol{A}_{1}$, we have%
\begin{equation}
\mathbf{\chi }_{1}^{\boldsymbol{B}_{1}}\left( q\right) =q^{-\frac{c}{24}%
}\dprod\limits_{n=2}^{\infty }\frac{1}{1-q^{n}}
\end{equation}%
To obtain $\mathbf{\chi }_{1}^{\boldsymbol{B}_{2}},$ we use the factorisation%
\begin{equation}
\mathbf{\chi }_{1}^{\boldsymbol{B}_{2}}=\mathbf{\chi }_{1}^{\boldsymbol{B}%
_{2}\backslash \boldsymbol{B}_{1}}\bullet \mathbf{\chi }_{1}^{\boldsymbol{B}%
_{1}}
\end{equation}%
with%
\begin{equation}
\mathbf{\chi }_{1}^{\boldsymbol{B}_{1}}=q^{-\frac{c}{24}}\frac{q^{\frac{1}{24%
}}\left( 1-q\right) }{\mathbf{\eta }\left( q\right) }
\end{equation}%
The factor $\mathbf{\chi }_{1}^{\boldsymbol{B}_{2}\backslash \boldsymbol{B}%
_{1}}$ is given by the contribution of the $so(1,2)$ multiplet with $%
\mathfrak{j}=3$ made of the positive roots \{$\alpha _{2},\alpha _{1}+\alpha
_{2},\alpha _{1}+2\alpha _{2}$\}. It is given by
\begin{equation}
\begin{tabular}{lll}
$\mathbf{\chi }_{1}^{\boldsymbol{B}_{2}\backslash \boldsymbol{B}_{1}}$ & $=$
& $\frac{1}{\mathbf{\eta }\left( q\right) }q^{\frac{1}{24}}\left( 1-q\right)
\left( 1-q^{2}\right) \left( 1-q^{3}\right) $ \\
$\mathbf{\chi }_{1}^{\boldsymbol{B}_{2}}$ & $=$ & $\frac{q^{-\frac{c}{24}}}{[%
\mathbf{\eta }\left( q\right) ]^{2}}q^{\frac{1}{12}}\left( 1-q\right)
^{2}\left( 1-q^{2}\right) \left( 1-q^{3}\right) $%
\end{tabular}%
\end{equation}%
Actually, these calculations of the HS- partition functions can be extended
to $SO(\mathcal{N},1+\mathcal{N})$ with generic $\mathcal{N}.$ The HS
partition function for the vectorial models is given by $\mathcal{Z}_{%
\boldsymbol{B}_{\mathcal{N}}}^{vect}=|\mathbf{\chi }_{1}^{\boldsymbol{B}_{%
\mathcal{N}}^{vect}}\left( q\right) |^{2}$ with vacuum character factorizing
as%
\begin{equation}
\mathbf{\chi }_{1}^{\boldsymbol{B}_{\mathcal{N}}^{vect}}=\mathbf{\chi }_{1}^{%
\boldsymbol{B}_{\mathcal{N}}^{vect}\backslash \boldsymbol{B}_{\mathcal{N}%
-1}^{vect}}\bullet \mathbf{\chi }_{1}^{\boldsymbol{B}_{\mathcal{N}-1}^{vect}}
\end{equation}%
For example, the calculation of $\mathbf{\chi }_{1}^{\boldsymbol{B}%
_{3}^{vect}}$ and $\mathbf{\chi }_{1}^{\boldsymbol{B}_{4}^{vect}}$ follow
from the factorisation $\mathbf{\chi }_{1}^{\boldsymbol{B}%
_{3}^{vect}\backslash \boldsymbol{B}_{2}^{vect}}\bullet \mathbf{\chi }_{1}^{%
\boldsymbol{B}_{2}^{vect}}$ and $\mathbf{\chi }_{1}^{\boldsymbol{B}%
_{4}^{vect}\backslash \boldsymbol{B}_{3}^{vect}}\bullet \mathbf{\chi }_{1}^{%
\boldsymbol{B}_{3}^{vect}}$. For the first example see below.\newline
Regarding the spinorial models higher spin partition function $\mathcal{Z}_{%
\boldsymbol{B}_{\mathcal{N}}}^{spin}$, it is given by $|\mathbf{\chi }_{1}^{%
\boldsymbol{B}_{\mathcal{N}}^{spin}}\left( q\right) |^{2}$ with vacuum
character factorizing as%
\begin{equation}
\mathbf{\chi }_{1}^{\boldsymbol{B}_{\mathcal{N}}^{spin}}=\mathbf{\chi }_{1}^{%
\boldsymbol{B}_{\mathcal{N}}^{spin}\backslash \boldsymbol{A}_{\mathcal{N}%
-1}^{spin}}\bullet \mathbf{\chi }_{1}^{\boldsymbol{A}_{\mathcal{N}-1}^{spin}}
\end{equation}%
See below for further explicit details illustrated on the $SO(3,4)$ three
dimensional gravity.

\subsubsection{$SO(3,4)$ theory}

In this theory, the LEND and the REND are different; for that we describe
the resulting models separately.

\paragraph{Left end node decomposition :\newline
}

This construction gives the first way to embed $SO(1,2)$ into $SO(3,4)$; it
is termed as the vectorial model for which the positive root system is as in
(\ref{321}); and the 21 CS gauge fields sitting into three multiplets with
cardinals 3+7+11 as follows%
\begin{equation}
\begin{tabular}{|c|c|c|}
\hline
multiplet & gauge fields & number \\ \hline
$\mathfrak{j}=1$ & $\mathcal{A}_{\mu }^{0},\text{ }\mathcal{A}_{\mu }^{\pm }$
& 3 \\ \hline
$\mathfrak{j}=3$ & $\mathcal{B}_{\mu }^{0},\text{ }\mathcal{B}_{\mu }^{\pm },%
\text{ }\mathcal{B}_{\mu }^{\pm 2},\text{ }\mathcal{B}_{\mu }^{\pm 3}$ & 7
\\ \hline
$\mathfrak{j}=5$ & $\mathcal{C}_{\mu }^{0},\text{ }\mathcal{C}_{\mu }^{\pm },%
\text{ }\mathcal{C}_{\mu }^{\pm 2},\text{ }\mathcal{C}_{\mu }^{\pm 3},\text{
}\mathcal{C}_{\mu }^{\pm 4},\text{ }\mathcal{C}_{\mu }^{\pm 5}$ & 11 \\
\hline
\end{tabular}%
\end{equation}%
These gauge fields couple to the 21 generators of $SO(3,4)$ realised in
terms of the non anomalous generators of the boundary W-algebra like%
\begin{equation}
A_{\mu }=\dsum\limits_{n=-1}^{+1}\mathcal{A}_{\mu
}^{n}L_{n}+\dsum\limits_{n=-3}^{+3}\mathcal{B}_{\mu }^{n}W_{n}^{\left(
4\right) }+\dsum\limits_{n=-5}^{+5}\mathcal{C}_{\mu }^{n}W_{n}^{\left(
6\right) }
\end{equation}%
where $L_{n},$ $W_{n}^{\left( 4\right) }$ and $W_{n}^{\left( 6\right) }$ are
as in eqs(\ref{a},\ref{b}\ref{c}). The expansions of the Dreibein and the
spin connections of $SO(3,4)$ decompose as follows%
\begin{equation}
\begin{tabular}{lll}
$\omega _{\mu }$ & $=$ & $\dsum\limits_{n=-1}^{+1}\mathcal{\omega }_{\mu
}^{n}L_{n}+\dsum\limits_{N=-3}^{+3}\Omega _{\mu }^{N}W_{n}^{\left( 4\right)
}+\dsum\limits_{n=-5}^{+5}\Theta _{\mu }^{N}W_{n}^{\left( 6\right) }$ \\
$e_{\mu }$ & $=$ & $\dsum\limits_{n=-1}^{+1}e_{\mu
}^{n}L_{n}+\dsum\limits_{N=-3}^{+3}\mathcal{E}_{\mu }^{N}W_{n}^{\left(
4\right) }+\dsum\limits_{n=-5}^{+5}\digamma _{\mu }^{n}W_{n}^{\left(
6\right) }$%
\end{tabular}%
\end{equation}%
The partition function of the vector series of $SO(3,4)$ higher spin theory
is given by $\mathcal{Z}_{\boldsymbol{B}_{3}}=\left\vert \mathbf{\chi }_{1}^{%
\boldsymbol{B}_{3}}\left( q\right) \right\vert ^{2}$; it is determined by
using $\mathcal{Z}_{\boldsymbol{B}_{2}}$ and the factorisation%
\begin{equation}
\mathbf{\chi }_{1}^{\boldsymbol{B}_{3}^{vect}}=\mathbf{\chi }_{1}^{%
\boldsymbol{B}_{3}^{vect}\backslash \boldsymbol{B}_{2}^{vect}}\bullet
\mathbf{\chi }_{1}^{\boldsymbol{B}_{2}^{vect}}
\end{equation}%
Notice that by using LEND, the root system of $SO(3,4)$ given by eq(\ref{321}%
) decomposes like%
\begin{equation}
\left. \pm \alpha _{3}\right. \emph{\quad ,\qquad }\left.
\begin{array}{l}
\pm \alpha _{2} \\
\pm \left( \alpha _{2}+\alpha _{3}\right) \\
\pm \left( \alpha _{2}+2\alpha _{3}\right)%
\end{array}%
\right.
\end{equation}%
giving the roots system of $SO(2,3)$ and an extra multiplet involving the
following positive roots
\begin{equation}
\left.
\begin{array}{l}
\alpha _{1} \\
\alpha _{1}+\alpha _{2} \\
\alpha _{1}+\alpha _{2}+\alpha _{3} \\
\alpha _{1}+\alpha _{2}+2\alpha _{3} \\
\alpha _{1}+2\alpha _{2}+2\alpha _{3}%
\end{array}%
\right.
\end{equation}%
and their opposite interpreted in terms of an $SO(1,2)$ multiplet with $%
\mathfrak{j}=5.$ So, the contribution of the factor $\mathbf{\chi }_{1}^{%
\boldsymbol{B}_{3}^{vect}\backslash \boldsymbol{B}_{2}^{vect}}$ is given by%
\begin{equation}
\mathbf{\chi }_{1}^{\boldsymbol{B}_{3}^{vect}\backslash \boldsymbol{B}%
_{2}^{vect}}=\frac{1}{\mathbf{\eta }\left( q\right) }q^{\frac{1}{24}%
}\dprod\limits_{m=1}^{5}\left( 1-q^{m}\right)
\end{equation}%
So, we have for $\mathbf{\chi }_{1}^{\boldsymbol{B}_{3}^{vect}\backslash
\boldsymbol{A}_{1}^{vect}}$ the following%
\begin{equation}
\begin{tabular}{lll}
$\mathbf{\chi }_{1}^{\boldsymbol{B}_{3}^{vect}\backslash \boldsymbol{A}%
_{1}^{vect}}$ & $=$ & $\frac{1}{[\mathbf{\eta }\left( q\right) ]^{2}}q^{%
\frac{1}{12}}\left( 1-q\right) ^{2}\left( 1-q^{2}\right) ^{2}\left(
1-q^{3}\right) ^{^{2}}\left( 1-q^{4}\right) \left( 1-q^{5}\right) $ \\
$\mathbf{\chi }_{1}^{\boldsymbol{B}_{3}^{vect}}$ & $=$ & $\frac{q^{-\frac{c}{%
24}}}{[\mathbf{\eta }\left( q\right) ]^{3}}q^{\frac{1}{8}}\left( 1-q\right)
^{3}\left( 1-q^{2}\right) ^{2}\left( 1-q^{3}\right) ^{^{2}}\left(
1-q^{4}\right) \left( 1-q^{5}\right) $%
\end{tabular}%
\end{equation}

\paragraph{Right end node decomposition :\newline
}

This decomposition gives another way to embed $SO(1,2)$ within $SO(3,4)$; it
is termed as the spinorial model for which the positive root system is as in
(\ref{spp}); and the 21 CS gauge fields sitting into three multiplets with
cardinals 3+5+13 as follows%
\begin{equation}
\begin{tabular}{|c|c|c|}
\hline
multiplet & gauge fields & number \\ \hline
$\mathfrak{j}=1$ & $\mathcal{A}_{\mu }^{0},\text{ }\mathcal{A}_{\mu }^{\pm }$
& 3 \\ \hline
$\mathfrak{j}=2$ & $\mathcal{B}_{\mu }^{0},\text{ }\mathcal{B}_{\mu }^{\pm },%
\text{ }\mathcal{B}_{\mu }^{\pm 2},\text{ }$ & 5 \\ \hline
$\mathfrak{j}=6$ & $\mathcal{C}_{\mu }^{0},\text{ }\mathcal{C}_{\mu }^{\pm },%
\text{ }\mathcal{C}_{\mu }^{\pm 2},\text{ }\mathcal{C}_{\mu }^{\pm 3},\text{
}\mathcal{C}_{\mu }^{\pm 4},\text{ }\mathcal{C}_{\mu }^{\pm 5},\mathcal{C}%
_{\mu }^{\pm 6}$ & 13 \\ \hline
\end{tabular}%
\end{equation}%
Their coupling to the non anomalous generators of the boundary W-symmetry
follows a similar scheme as above. \newline
The partition function of the spinorial model of the $SO(3,4)$ higher spin
theory is given by $\mathcal{Z}_{\boldsymbol{B}_{3}^{spin}}=\left\vert
\mathbf{\chi }_{1}^{\boldsymbol{B}_{3}}\left( q\right) \right\vert ^{2}$; it
is determined by using $\mathcal{Z}_{\boldsymbol{B}_{2}^{spin}}$ and the
factorisation%
\begin{equation}
\mathbf{\chi }_{1}^{\boldsymbol{B}_{3}^{spin}}=\mathbf{\chi }_{1}^{%
\boldsymbol{B}_{3}^{spin}\backslash \boldsymbol{A}_{2}}\bullet \mathbf{\chi }%
_{1}^{\boldsymbol{A}_{2}}
\end{equation}%
Notice that by using the REND, the root system of $SO(3,4)$ given by eq(\ref%
{321}) is decomposed like%
\begin{equation}
\left. \pm \alpha _{1}\right. \quad ,\qquad \left.
\begin{array}{l}
\pm \alpha _{2} \\
\pm \left( \alpha _{1}+\alpha _{2}\right)%
\end{array}%
\right.
\end{equation}%
giving the roots system of $SL(3)$ and an extra multiplet involving the
following positive roots
\begin{equation}
\left.
\begin{array}{l}
\alpha _{3} \\
\alpha _{2}+\alpha _{3} \\
\alpha _{1}+\alpha _{2}+\alpha _{3} \\
\alpha _{2}+2\alpha _{3} \\
\alpha _{1}+\alpha _{2}+2\alpha _{3} \\
\alpha _{1}+2\alpha _{2}+2\alpha _{3}%
\end{array}%
\right.
\end{equation}%
and their opposite, interpreted in term of an $SO(1,2)$ multiplet with $%
\mathfrak{j}=6.$ So, the contribution of the factor $\mathbf{\chi }_{1}^{%
\boldsymbol{B}_{3}^{spin}\backslash \boldsymbol{A}_{2}}$ is given by
\begin{equation}
\mathbf{\chi }_{1}^{\boldsymbol{B}_{3}^{spin}\backslash \boldsymbol{A}_{2}}=%
\frac{1}{\mathbf{\eta }\left( q\right) }q^{\frac{1}{24}}\left( 1-q\right)
\left( 1-q^{2}\right) \left( 1-q^{3}\right) ^{2}\left( 1-q^{4}\right) \left(
1-q^{5}\right)
\end{equation}%
Combining the above relations, we have
\begin{equation}
\mathbf{\chi }_{1}^{\boldsymbol{B}_{3}^{spin}\backslash \boldsymbol{A}_{1}}=%
\frac{1}{[\mathbf{\eta }\left( q\right) ]^{2}}q^{\frac{1}{12}}\left(
1-q\right) ^{2}\left( 1-q^{2}\right) ^{2}\left( 1-q^{3}\right) ^{2}\left(
1-q^{4}\right) \left( 1-q^{5}\right)
\end{equation}%
and%
\begin{equation}
\mathbf{\chi }_{1}^{\boldsymbol{B}_{3}^{spin}}=\frac{q^{-\frac{c}{24}}}{[%
\mathbf{\eta }\left( q\right) ]^{3}}q^{\frac{1}{8}}\left( 1-q\right)
^{3}\left( 1-q^{2}\right) ^{2}\left( 1-q^{3}\right) ^{^{2}}\left(
1-q^{4}\right) \left( 1-q^{5}\right)
\end{equation}

\section{Higher spins with $\boldsymbol{D}_{\mathcal{N}}$ symmetry}

This section is dedicated to the study of higher spin AdS$_{3}$ gravity with
orthogonal $SO(\mathcal{N},\mathcal{N})$ symmetry. This represents an
application of the graphic description detailed above and a further
illustration of its efficiency regarding the $\boldsymbol{D}_{\mathcal{N}}$
symmetries. This completes the study initiated in the previous section
regarding orthogonal higher spin gravity. We also calculate the HS partition
function $\mathcal{Z}_{\boldsymbol{D}_{\mathcal{N}}}$.

\subsection{AdS$_{3}$ gravity with $SO\left( \mathcal{N},\mathcal{N}\right) $
symmetry}

\begin{figure}[tbph]
\begin{center}
\includegraphics[width=7cm]{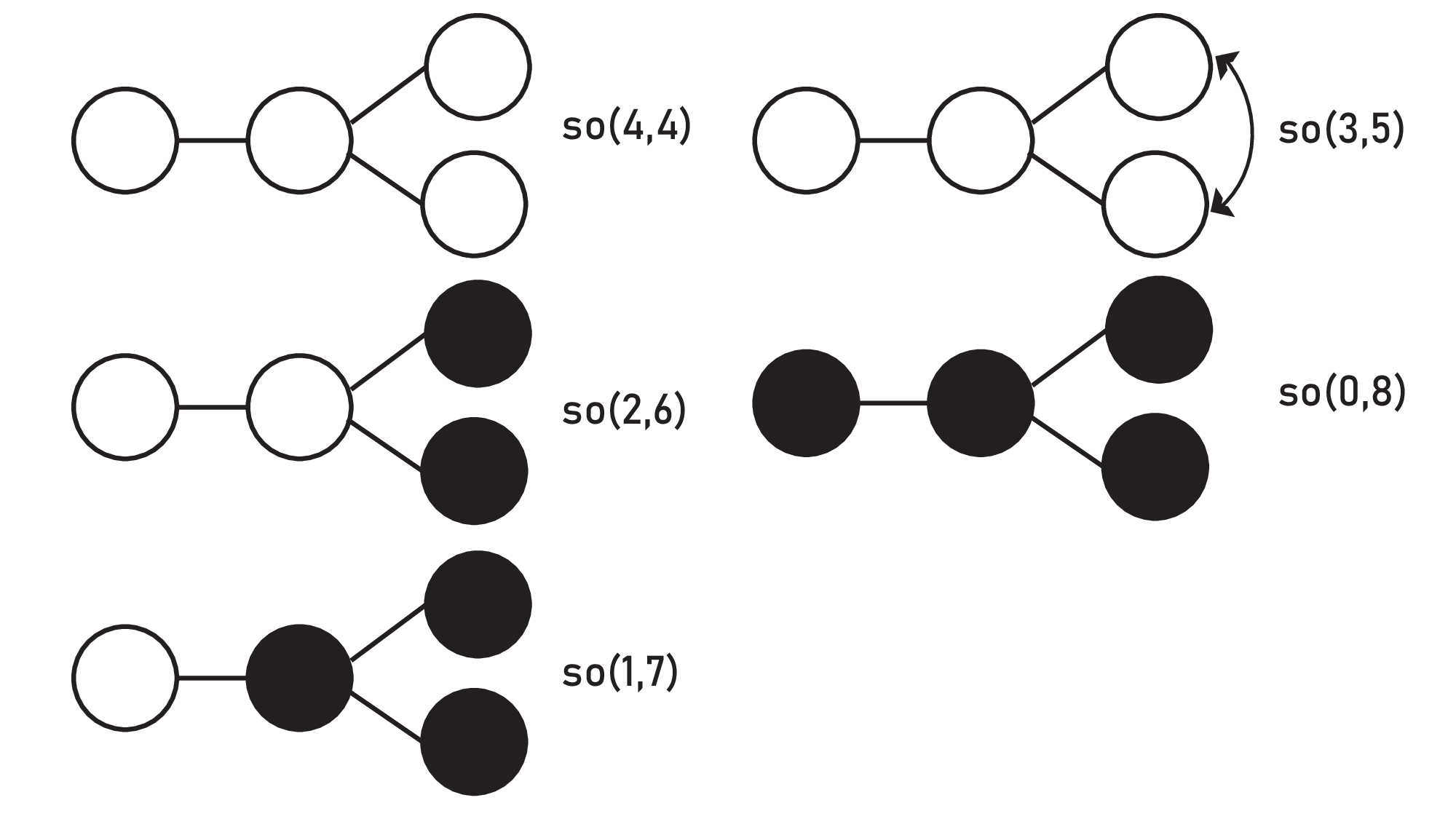}
\end{center}
\par
\vspace{-0.5cm}
\caption{The Tits-Satake diagrams associated to standard real forms of $%
\boldsymbol{D}_{4}.$}
\label{TSD}
\end{figure}

Lie algebra $\boldsymbol{D}_{\mathcal{N}}$ has rank $\mathcal{N}$ and $%
\mathcal{N}\left( 2\mathcal{N}-1\right) $ dimensions. It has $\mathcal{N}+1$
standard real forms including the real compact $SO(2\mathcal{N})$, the real
split form $SO(\mathcal{N},\mathcal{N})$ and $SO(p,q)$ where $p+q=2\mathcal{N%
}$ with $p<q.$ For an illustration, we give in Figure \textbf{\ref{TSD}} the
Tits-Satake diagrams of $\boldsymbol{D}_{4}.$
In the present investigation, we focus on the real split $SO(\mathcal{N},%
\mathcal{N})$ containing the $SL(\mathcal{N},\mathbb{R})$ as a gauge
subsymmetry and having $SO\left( 2,2\right) $ as the leading member
describing the AdS$_{3}$ isometry. For $\mathcal{N}=3$, we have the $%
SO\left( 3,3\right) $ symmetry group which is homomorphic to $SL\left( 4,%
\mathbb{R}\right) $ sitting in the $\boldsymbol{A}_{\mathcal{N}}$- series in
the Cartan classification.

\subsubsection{Higher spin content in $SO\left( \mathcal{N},\mathcal{N}%
\right) $ theory}

\begin{figure}[tbph]
\begin{center}
\includegraphics[width=8cm]{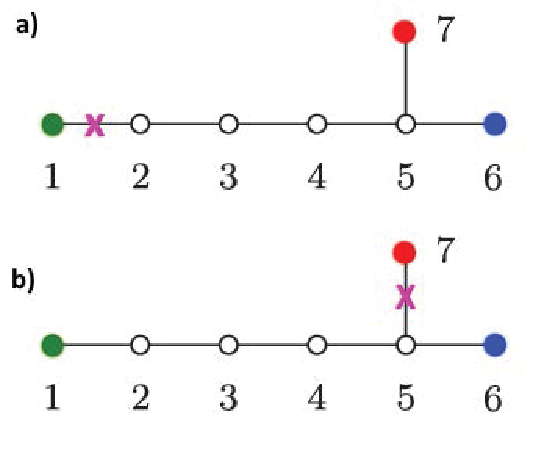}
\end{center}
\par
\vspace{-0.5cm}
\caption{The extremal node decompositions of the Dynkin diagram of $%
\boldsymbol{D}_{7}$. a) the vector series corresponding to the LEND. b) the
spinorial series corresponding to the REND (cutting the red node or
equivalently the blue one).}
\label{D70}
\end{figure}

As for the Lie algebras $\boldsymbol{A}_{\mathcal{N}}$ and $\boldsymbol{B}_{%
\mathcal{N}}$ series,\ 3D gravity at asymptotic limit of AdS$_{3}$ with $%
SO\left( \mathcal{N},\mathcal{N}\right) $ gauge symmetry has $\mathcal{N}$
conformal currents $W^{\left( s\right) }\left( z\right) $ living on the
frontier of AdS$_{3}$. These boundary conformal currents generate the $%
\boldsymbol{WD}_{\mathcal{N}}$-invariance \textrm{\cite{WDN}}. We show that
in the AdS$_{3}$ gravity with $SO\left( \mathcal{N},\mathcal{N}\right) $
Chern Simons description, one distinguishes two families of higher spins
termed as the vectorial series and the spinorial series. These have
different higher spin contents and are interestingly interpreted in terms of
the END of the Tits-Satake (Dynkin) diagram of the $SO\left( \mathcal{N},%
\mathcal{N}\right) $ as illustrated in Figure \textbf{\ref{D70} }for $%
\boldsymbol{D}_{\mathcal{7}}.$

\paragraph{\textbf{A) Spinorial series:}\newline
}

As portrayed in Figure \textbf{\ref{D70}-b)}, the spinorial series
correspond to omitting the red (or equivalently the blue) node in the Dynkin
diagram of the $\boldsymbol{D}_{\mathcal{N}}$ Lie algebra\textbf{.} This
REND leads to splitting the $\mathcal{N}\left( 2\mathcal{N}-1\right) $
dimensions of the $so\left( \mathcal{N},\mathcal{N}\right) $ Lie algebra as
follows
\begin{equation}
\begin{tabular}{lll}
$so\left( \mathcal{N},\mathcal{N}\right) $ & $:$ & $sl\left( \mathcal{N},%
\mathbb{R}\right) \oplus so\left( 1,1\right) \oplus 2\left[ \frac{\mathcal{N}%
\left( \mathcal{N}-1\right) }{2}\right] $ \\
$\mathcal{N}\left( 2\mathcal{N}-1\right) $ & $:$ & $\left( \mathcal{N}%
^{2}-1\right) +1+\mathcal{N}\left( \mathcal{N}-1\right) $%
\end{tabular}%
\end{equation}%
Here, the $\mathcal{N}^{2}-1$\ dimensions of $sl\left( \mathcal{N},\mathbb{R}%
\right) $ decompose in terms of spins $\mathfrak{j}$ like
\begin{equation}
\mathcal{N}^{2}-1=\sum_{\mathfrak{j}=1}^{\mathcal{N}-1}\left( 2\mathfrak{j}%
+1\right)
\end{equation}%
and the extra $1+\mathcal{N}\left( \mathcal{N}-1\right) $ thought of as
\begin{equation}
1+\mathcal{N}\left( \mathcal{N}-1\right) =2\mathfrak{\mathring{j}}_{\mathcal{%
N}}^{spinor}+1
\end{equation}%
This gives $\mathfrak{\mathring{j}}_{\mathcal{N}}^{spinor}=\mathcal{N}\left(
\mathcal{N}-1\right) /2$ which describes an isolated spin (ILS). By setting $%
\mathcal{N}=7$ for example, these features read as follows%
\begin{equation}
\begin{tabular}{ccc}
$so\left( 7,7\right) $ & $:$ & $sl\left( 7,\mathbb{R}\right) \oplus so\left(
1,1\right) \oplus 21_{+}\oplus 21_{-}$ \\
$91$ & $:$ & $48+43$ \ \ \ \ \ \ \ \ \ \ \ \ \ \ \ \ \ \ \ \ \ \ \ \ \ \ \ \
\ \
\end{tabular}%
\end{equation}%
with%
\begin{equation}
\begin{tabular}{lll}
$48$ & $=$ & $3+5+7+9+11+13$ \\
$\mathfrak{\mathring{j}}_{{\small spinor}}$ & $=$ & $\frac{7\times 6}{2}=21$
\\
$43$ & $=$ & $2\mathfrak{\mathring{j}}_{{\small spinor}}+1$%
\end{tabular}%
\end{equation}%
where the $\mathfrak{\mathring{j}}_{{\small spinor}}=21$ is the\textrm{\ }%
isolated spin multiplet. Notice that by setting $\mathcal{N}=4,$ we have%
\begin{equation}
\begin{tabular}{ccc}
$so\left( 4,4\right) $ & $:$ & $sl\left( 4,\mathbb{R}\right) \oplus so\left(
1,1\right) \oplus 6_{+}\oplus 6_{-}$ \\
$28$ & $:$ & $\left( 3+5+7\right) +13$ \ \ \ \ \ \ \ \ \ \ \ \ \ \
\end{tabular}
\label{8}
\end{equation}%
with ILS given by $\mathfrak{\mathring{j}}_{{\small spinor}}=6.$

\paragraph{\textbf{B) Vector series: \newline
}}

The vector family corresponds to the cutting of the green node ($\mathbf{%
\alpha }_{1}$) in (\textbf{\ref{D70}-a).} This LEND leads to breaking the $%
so\left( \mathcal{N},\mathcal{N}\right) $ Lie algebra like%
\begin{equation}
\begin{tabular}{lll}
$so\left( \mathcal{N},\mathcal{N}\right) $ & $:$ & $so\left( \mathcal{N}-1,%
\mathcal{N}-1\right) \oplus so\left( 1,1\right) $ \\
&  & $\oplus 2(\mathcal{N}-1)_{+}\oplus 2(\mathcal{N}-1)_{-}$%
\end{tabular}%
\end{equation}%
For the $\mathcal{N}=4,7$ examples, we have%
\begin{equation}
\begin{tabular}{lll}
$so\left( 4,4\right) $ & $:$ & $so\left( 3,3\right) \oplus so\left(
1,1\right) \oplus 6_{+}\oplus 6_{-}$ \\
$so\left( 7,7\right) $ & $:$ & $so\left( 6,6\right) \oplus so\left(
1,1\right) \oplus 12_{+}\oplus 12_{-}$%
\end{tabular}
\label{11}
\end{equation}%
Regarding the splitting of the $\mathcal{N}\left( 2\mathcal{N}-1\right) $
dimensions of the $so\left( \mathcal{N},\mathcal{N}\right) $, we find that
there are two interesting ways to do it: $\left( \mathbf{i}\right) $ We
either have%
\begin{equation}
\mathcal{N}\left( 2\mathcal{N}-1\right) =\sum_{\mathfrak{j}=0}^{\mathcal{N}%
-1}\left( 4j+1\right) =\sum_{\mathfrak{j}=0}^{\mathcal{N}-1}\left[ 2\left(
2j\right) +1\right]
\end{equation}%
It reads for $\mathcal{N}=7$ as%
\begin{equation}
1+5+9+13+17+21+25
\end{equation}%
lacking the triplet $3$ which is highly demanded as it is associated with
the $SO(1,2)$ symmetry. $\left( \mathbf{ii}\right) $ Or we have the expansion%
\begin{equation}
\begin{tabular}{lll}
$\mathcal{N}\left( 2\mathcal{N}-1\right) $ & $=$ & $3+5+7+\dsum%
\limits_{n=4}^{\mathcal{N}}\left( 4n-3\right) $ \\
& $=$ & $3+5+7+\dsum\limits_{n=4}^{\mathcal{N}}\left[ 2\left( 2n-2\right) +1%
\right] $%
\end{tabular}%
\end{equation}%
which gives for $\mathcal{N}=7$%
\begin{equation}
91=\left( 3+5+7\right) +13+17+21+25
\end{equation}%
To single out the right decomposition to retain, we need a constraint
relation; it is given by the particular case $\mathcal{N}=4$ where the two
ENDs (vector and spinor) should coincide thanks to the triality property of $%
\boldsymbol{D}_{4}$ (Figure \textbf{\ref{D4})}.\newline
As a result of this analysis, the two higher spin families for $SO\left(
\mathcal{N},\mathcal{N}\right) $ AdS$_{3}$ gravity are as collected in the
following table%
\begin{equation}
\begin{tabular}{c|c|c}
{\small series} & {\small spacetime spin} & {\small boundary CFT}$_{2}$%
{\small -spin} \\ \hline\hline
{\small vector } & $\left.
\begin{array}{ccccc}
\mathfrak{\mathring{j}}_{n} & = & n & \text{ }; & n=1,2,3 \\
\mathfrak{j}_{n} & = & {\small 2n-}2\text{ \ } & \text{ }; & {\small 4\leq
n\leq }\mathcal{N}%
\end{array}%
\right. $ \ \ \  & $\left.
\begin{array}{ccc}
\mathring{s}_{n} & = & n+1 \\
s_{n} & = & 2n-1%
\end{array}%
\right. $ \ \ \ \ \ \  \\ \hline\hline
{\small spinor} & $\left.
\begin{array}{ccccc}
\mathfrak{j}_{n} & = & n & \text{ \ }; & {\small 1\leq n\leq }\mathcal{N}-1
\\
\mathfrak{\mathring{j}}_{\mathcal{N}} & = & \frac{\mathcal{N}\left( \mathcal{%
N}-1\right) }{2} & \text{ \ }; & n=\mathcal{N}%
\end{array}%
\right. $ & $\left.
\begin{array}{ccc}
s_{n} & = & n+1 \\
\mathring{s}_{\mathcal{N}} & = & \frac{\mathcal{N}\left( \mathcal{N}%
-1\right) }{2}+1%
\end{array}%
\right. $ \\ \hline\hline
\end{tabular}
\label{tabD}
\end{equation}%
\begin{equation*}
\end{equation*}

\subsubsection{Higher spin $SO\left( 4,4\right) $ model}

\begin{figure}[tbph]
\begin{center}
\includegraphics[width=4cm]{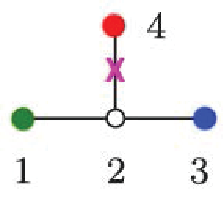}
\end{center}
\par
\vspace{-0.5cm}
\caption{The extremal node decompositions of the Dynkin diagram of D$_{4}$
where the LEND and REND coincide.}
\label{D4}
\end{figure}

The $SO\left( 4,4\right) $ gauge symmetry is the leading member in the
family $SO\left( \mathcal{N},\mathcal{N}\right) $; it has 28 dimensions with
the properties as given by (\ref{8}). The corresponding Tits-Satake diagram
has four nodes with an outer- automorphism symmetry $\mathbb{S}_{3}$
permuting the three external nodes ($\alpha _{1},$ $\alpha _{3},$ $\alpha
_{4}$) while fixing $\alpha _{2}$ as illustrated in Figure \textbf{\ref{D4}}%
.
Each node in this $\boldsymbol{D}_{4}$ diagram is associated with one of the
four simple roots $\alpha _{i}$ ($i=1,2,3,4$) giving the four Chevalley
triplets of $SO\left( 4,4\right) ,$ namely $H_{\alpha _{i}},$ $E_{\pm \alpha
_{i}}.$ The full root system of $SO\left( 4,4\right) $ contains 24 roots;
the 12 positive ones $\Phi _{\left( \boldsymbol{D}_{4}\right) }^{+}$ can be
organised in six different ways given by the permutation of the roots $%
\left( \alpha _{1},\alpha _{3},\alpha _{4}\right) ;$ two of them are as
follows

$\bullet $ \emph{First way}%
\begin{equation}
\alpha _{1},\quad \left.
\begin{array}{l}
\alpha _{2} \\
\alpha _{1}+\alpha _{2}%
\end{array}%
\right. ,\quad \left.
\begin{array}{l}
\alpha _{3} \\
\alpha _{2}+\alpha _{3} \\
\alpha _{1}+\alpha _{2}+\alpha _{3}%
\end{array}%
\right. ,\quad \left.
\begin{array}{l}
\alpha _{4} \\
\alpha _{4}+\alpha _{2} \\
\alpha _{4}+\alpha _{1}+\alpha _{2} \\
\alpha _{4}+\alpha _{2}+\alpha _{3} \\
\alpha _{4}+\alpha _{1}+\alpha _{2}+\alpha _{3} \\
\alpha _{4}+\alpha _{1}+2\alpha _{2}+\alpha _{3}%
\end{array}%
\right.  \label{4b}
\end{equation}

$\bullet $ \emph{Second way}
\begin{equation}
\alpha _{4},\quad \left.
\begin{array}{l}
\alpha _{2} \\
\alpha _{4}+\alpha _{2}%
\end{array}%
\right. ,\quad \left.
\begin{array}{l}
\alpha _{3} \\
\alpha _{2}+\alpha _{3} \\
\alpha _{4}+\alpha _{2}+\alpha _{3}%
\end{array}%
\right. ,\quad \left.
\begin{array}{l}
\alpha _{1} \\
\alpha _{1}+\alpha _{2} \\
\alpha _{1}+\alpha _{2}+\alpha _{3} \\
\alpha _{1}+\alpha _{2}+\alpha _{4} \\
\alpha _{1}+\alpha _{2}+\alpha _{3}+\alpha _{4} \\
\alpha _{1}+2\alpha _{2}+\alpha _{3}+\alpha _{4}%
\end{array}%
\right.
\end{equation}%
The first three blocks in (\ref{4b}) have no dependence into the simple root
$\mathbf{\alpha }_{4},$ they belong to the subset
\begin{equation}
\frac{\partial \Phi _{\boldsymbol{D}_{4}}}{\partial \alpha _{4}}=0
\end{equation}%
which is nothing but the $A_{3}$ root system of eq(\ref{sl4}). The fourth
block involves $\alpha _{4}$\ and contains six positive roots indicating
that it sits in the Lorentz multiplet with spin $\mathfrak{j}_{4}=6.$

\paragraph{Spin content in the asymptotic $SO\left( 4,4\right) $ model:%
\newline
}

The CFT$_{2}$ spin content living at the boundary of the $SO\left(
4,4\right) $ model is determined by splitting the 28 dimensions of $so\left(
4,4\right) $ in terms of $sl_{2}$ multiplets namely%
\begin{equation}
28=3+5+7+13
\end{equation}%
In the Cartan- Weyl basis of the $SO\left( 4,4\right) $ gauge symmetry, the
28 generators are splited like $E_{\left( 1\right) }\oplus F_{\left(
2\right) }\oplus G_{\left( 3\right) }\oplus H_{\left( 6\right) }$; they are
collected in the following table
\begin{equation}
\begin{tabular}{|c|c|c|c|c|c|}
\hline
spin $\mathfrak{j}$ & {\small roots} & \multicolumn{2}{|c|}{\small Chevalley
generators} & {\small old generators} & {\small number} \\ \hline\hline
1 & $\left. \alpha _{1}\right. $ & $E_{\pm }$ & $E_{0}\ $ & $J_{a}$ & $3$ \\
\hline
2 & $\left.
\begin{array}{c}
\alpha _{2} \\
\alpha _{1}+\alpha _{2}%
\end{array}%
\right. $ & $\left.
\begin{array}{c}
F_{\pm } \\
F_{\pm 2}%
\end{array}%
\right. $ & $F_{0}$ & ${\small T}_{\left( 2\right) }$ & $5$ \\ \hline
3 & $\left.
\begin{array}{c}
\alpha _{3} \\
\alpha _{2}+\alpha _{3} \\
\alpha _{1}+\alpha _{2}+\alpha _{3}%
\end{array}%
\right. $ & $\left.
\begin{array}{c}
G_{\pm } \\
G_{\pm 2} \\
G_{\pm 3}%
\end{array}%
\right. $ & $G_{0}$ & ${\small T}_{\left( 3\right) }\oplus {\small T}%
_{\left( 6\right) }$ & $7$ \\ \hline
\end{tabular}
\label{tabC}
\end{equation}%
\begin{equation*}
\end{equation*}%
and%
\begin{equation}
\begin{tabular}{|c|c|c|c|c|c|}
\hline
spin $\mathfrak{j}_{4}$ & {\small roots} & \multicolumn{2}{|c|}{\small %
Chevalley} & {\small old } & {\small number} \\ \hline
6 & \multicolumn{1}{|l|}{$%
\begin{array}{c}
\alpha _{4} \\
\alpha _{2}+\alpha _{4} \\
\alpha _{1}+\alpha _{2}+\alpha _{4} \\
\alpha _{2}+\alpha _{3}+\alpha _{4} \\
\alpha _{1}+\alpha _{2}+\alpha _{3}+\alpha _{4} \\
\alpha _{1}+2\alpha _{2}+\alpha _{3}+\alpha _{4}%
\end{array}%
$} & $\left.
\begin{array}{c}
H_{\pm } \\
H_{\pm 2} \\
H_{\pm 3} \\
H_{\pm 4} \\
H_{\pm 5} \\
H_{\pm 6}%
\end{array}%
\right. $ & $\ \ H_{0}$ \ \  & ${\small T}_{\left( 6\right) }$ & $13$ \\
\hline
\end{tabular}%
\end{equation}%
with%
\begin{equation}
\begin{tabular}{lllllll}
$E_{\pm }$ & $=$ & $E_{\pm \mathbf{\mathbf{\alpha }}_{1}}$ & $\qquad ,\qquad
$ & $2E_{0}$ & $=$ & $\left[ E_{-},E_{+}\right] $ \\
$F_{\pm }$ & $=$ & $E_{\pm \mathbf{\mathbf{\alpha }}_{2}}$ & $\qquad ,\qquad
$ & $2F_{0}$ & $=$ & $\left[ F_{-},F_{+}\right] $ \\
$G_{\pm }$ & $=$ & $E_{\pm \mathbf{\mathbf{\alpha }}_{3}}$ & $\qquad ,\qquad
$ & $2G_{0}$ & $=$ & $\left[ G_{-},G_{+}\right] $ \\
$H_{\pm }$ & $=$ & $E_{\pm \mathbf{\mathbf{\alpha }}_{4}}$ & $\qquad ,\qquad
$ & $2H_{0}$ & $=$ & $\left[ H_{-},H_{+}\right] $%
\end{tabular}%
\end{equation}

\paragraph{Gauge fields in the $SO(4,4)$ gravity model:\newline
}

The gauge fields in the AdS$_{3}$ gravity with gauge symmetry $SO(4,4)$
involves $28_{L}+28_{R}$ Chern-Simons gauge potentials. They describe the
two types of gauge fields, namely the left $A_{\mu }^{L}$ and the right $%
A_{\mu }^{R}$ Chern-Simons potentials. The Lorentz spins of the CS gauge
fields as well as the CFT$_{2}$ currents are described below.

\ \ \

$\bullet $ \emph{Chern-Simons gauge potentials}\newline
The 28 components of the $SO(4,4)$ gauge fields and the conformal W-currents
at the boundary of AdS$_{3}$ are given by%
\begin{equation}
\begin{tabular}{|c|c|}
\hline
gauge fields & L-spin \\ \hline
$\left\{ \mathcal{A}_{\mu }^{n}\right\} _{-1\leq n\leq 1}$ & 1 \\ \hline
$\left\{ \mathcal{B}_{\mu }^{n}\right\} _{-2\leq n\leq 2}$ & 2 \\ \hline
$\left\{ \mathcal{C}_{\mu }^{n}\right\} _{-3\leq n\leq 3}$ & 3 \\ \hline
$\left\{ \mathcal{D}_{\mu }^{n}\right\} _{-6\leq n\leq 6}$ & 6 \\ \hline
\end{tabular}%
\qquad ,\qquad
\begin{tabular}{|c|c|}
\hline
CFT currents & CFT$_{2}$ \\ \hline
$T\left( z\right) $ & 2 \\ \hline
$W_{3}\left( z\right) $ & 3 \\ \hline
$W_{4}\left( z\right) $ & 4 \\ \hline
$W_{7}\left( z\right) $ & 7 \\ \hline
\end{tabular}%
\end{equation}%
These 3D Chern-Simons gauge potentials couple to the 28 generators of $%
SO(4,4)$ in the table (\ref{tabC}) like%
\begin{equation}
A_{\mu }=\dsum\limits_{n=-1}^{+1}\mathcal{A}_{\mu
}^{n}E_{n}+\dsum\limits_{N=-2}^{+2}\mathcal{B}_{\mu
}^{N}F_{N}+\dsum\limits_{\Lambda =-3}^{+3}\mathcal{C}_{\mu }^{\Lambda
}G_{\Lambda }+\dsum\limits_{\Delta =-6}^{+6}\mathcal{D}_{\mu }^{\Delta
}H_{\Delta }
\end{equation}%
The dynamics of these potentials are given by the action%
\begin{equation}
\mathcal{S}=\dint\nolimits_{AdS_{3}}Tr\Omega \left[ A_{L}\right]
-\dint\nolimits_{AdS_{3}}Tr\Omega \left[ A_{R}\right]  \label{sa}
\end{equation}%
with Chern-Simons 3-form $\Omega \left[ A\right] =AdA+\frac{2}{3}A^{3}.$

$\bullet $ \emph{Gravity fields}\newline
These are given by the Dreibeins $e_{\mu }$ and the $SO(4,4)$ spin
connections $\omega _{\mu }$, they expand as%
\begin{equation}
\begin{tabular}{lll}
$\omega _{\mu }$ & $=$ & $\dsum\limits_{n=-1}^{+1}\mathcal{\omega }_{\mu
}^{n}E_{n}+\dsum\limits_{N=-2}^{+2}\Omega _{\mu
}^{N}F_{N}+\dsum\limits_{\Lambda =-3}^{+3}\Theta _{\mu }^{\Lambda
}G_{\Lambda }+\dsum\limits_{\Lambda =-6}^{+6}\Psi _{\mu }^{\Delta }H_{\Delta
}$ \\
$e_{\mu }$ & $=$ & $\dsum\limits_{n=-1}^{+1}e_{\mu
}^{n}E_{n}+\dsum\limits_{N=-2}^{+2}\mathcal{E}_{\mu
}^{N}F_{N}+\dsum\limits_{\Lambda =-3}^{+3}\digamma _{\mu }^{\Lambda
}G_{\Lambda }+\dsum\limits_{\Lambda =-6}^{+6}\Phi _{\mu }^{\Delta }H_{\Delta
}$%
\end{tabular}%
\end{equation}%
The relations between these gauge fields and the Chern-Simons follow eqs(\ref%
{14},\ref{25}).

\subsection{$SO\left( \mathcal{N},\mathcal{N}\right) $ HS- partition function%
}

Here, we build the higher spin partition function for the $SO\left( \mathcal{%
N},\mathcal{N}\right) $\ AdS$_{3}$ gravity by emphasizing on the two leading
models, namely $\mathcal{N}=4,5$. This quantity is given in general by $%
\mathcal{Z}_{\boldsymbol{D}_{\mathcal{N}}}=\left\vert \mathbf{\chi }_{1}^{%
\boldsymbol{D}_{\mathcal{N}}}\left( q\right) \right\vert ^{2},$ where the
factorization of the vacuum character $\mathbf{\chi }_{1}^{\boldsymbol{D}_{%
\mathcal{N}}}$\ follows from the splitting of the $so\left( \mathcal{N},%
\mathcal{N}\right) $\ real form. We give two different computations of the
HS- partition function based on the end node decompositions. To visualize
this, we begin by treating the $SO(4,4)$ theory for which the spinor and the
vector series coincide. Then, we study the $SO(5,5)$ theory where we can
differentiate between these series.

\subsubsection{$SO(4,4)$ theory}

On the boundary of the AdS$_{3}$ gravity with $SO(4,4)_{L}\times SO(4,4)_{R}$
gauge symmetry, there are four conserved conformal currents. The Virasoro
and primary field currents with CFT$_{2}$ spin $s=2,3,4$ are collectively
given by%
\begin{equation}
W^{\left( 2\right) }\left( z\right) ,\quad W^{\left( 3\right) }\left(
z\right) ,\quad W^{\left( 4\right) }\left( z\right)
\end{equation}%
In addition, we have an isolated conformal current $W^{\left( 7\right)
}\left( z\right) $ with CFT$_{2}$-spin 7 and Laurent expansion $%
\sum_{n}z^{-n-7}W_{n}^{\left( 7\right) }$ where the Laurent modes are%
\begin{equation}
W_{n}^{\left( 7\right) }=\doint \frac{dz}{2i\pi }z^{n+6}W^{\left( 7\right)
}\left( z\right)
\end{equation}%
The higher spin partition function $\mathcal{Z}_{\boldsymbol{D}_{4}}=%
\mathcal{Z}_{\boldsymbol{D}_{4}^{vect}}=\mathcal{Z}_{\boldsymbol{D}%
_{4}^{spin}}$ in the AdS$_{3}$ gravity with gauge symmetry $SO(4,4)$ is
given by%
\begin{equation}
\mathcal{Z}_{\boldsymbol{D}_{4}}=\left\vert \mathbf{\chi }_{1}^{\boldsymbol{D%
}_{4}}\right\vert ^{2}  \label{zs}
\end{equation}%
In this relation, the $\mathbf{\chi }_{1}^{\boldsymbol{D}_{4}}$ is the
vacuum character of the $\boldsymbol{WD}_{4}$ algebra living on the boundary
of the $SO(4,4)$ theory of 3D AdS$_{3}$ gravity. Using properties of the
extremal node decomposition of $\boldsymbol{D}_{4}$, namely%
\begin{equation}
so(4,4)\qquad \rightarrow \qquad sl(4,\mathbb{R})\oplus sl\left( 1,\mathbb{R}%
\right) \oplus 2\times \left[ \mathbf{6}\right]
\end{equation}%
we can relate the character $\mathbf{\chi }_{1}^{\boldsymbol{D}_{4}}$ to the
character $\mathbf{\chi }_{1}^{\boldsymbol{A}_{3}}$ such that%
\begin{equation}
\mathbf{\chi }_{1}^{\boldsymbol{D}_{4}}=\left[ \mathbf{\chi }_{1}^{%
\boldsymbol{A}_{3}}\right] \bullet \left[ \mathbf{\chi }_{1}^{\boldsymbol{D}%
_{4}/\boldsymbol{A}_{3}}\right]
\end{equation}%
Putting back into (\ref{zs}), one gets the expression of the partition
function%
\begin{equation}
\mathcal{Z}_{\boldsymbol{D}_{4}}=\mathcal{Z}_{\boldsymbol{A}_{3}}\bullet
\mathcal{Z}_{\boldsymbol{D}_{4}/\boldsymbol{A}_{3}}
\end{equation}%
where
\begin{equation}
\mathcal{Z}_{\boldsymbol{D}_{4}/\boldsymbol{A}_{3}}=\left\vert \mathbf{\chi }%
_{1}^{\boldsymbol{D}_{4}/\boldsymbol{A}_{3}}\right\vert ^{2}
\end{equation}%
To calculate the\textrm{\ }$\mathcal{Z}_{\boldsymbol{D}_{4}},$ we use the
character $\mathbf{\chi }_{1}^{\boldsymbol{A}_{3}}$ computed from \ref{An}%
\begin{equation}
\mathbf{\chi }_{1}^{\boldsymbol{A}_{3}}=\frac{1}{\left[ \mathbf{\eta }\left(
q\right) \right] ^{3}}q^{-\frac{c-3}{24}}\left( 1-q\right) ^{3}\left(
1-q^{2}\right) ^{2}\left( 1-q^{3}\right)
\end{equation}%
The contribution of the $\mathbf{\chi }_{1}^{\boldsymbol{D}_{4}/\boldsymbol{A%
}_{3}}$ factor is given by%
\begin{equation}
\mathbf{\chi }_{1}^{\boldsymbol{D}_{4}/\boldsymbol{A}_{3}}=\frac{1}{\mathbf{%
\eta }\left( q\right) }q^{\frac{1}{24}}\frac{\left( 1-q^{3}\right) }{\left(
1-q^{6}\right) }\dprod\limits_{n=1}^{6}\left( 1-q^{n}\right)
\end{equation}%
leading in turn to%
\begin{equation}
\mathbf{\chi }_{1}^{\boldsymbol{D}_{4}}=\frac{q^{-\frac{c}{24}}}{[\mathbf{%
\eta }\left( q\right) ]^{4}}\frac{q^{\frac{1}{6}}}{\left( 1-q\right) \left(
1-q^{2}\right) \left( 1-q^{4}\right) }\dprod\limits_{n=1}^{6}\left(
1-q^{n}\right) ^{6-n}  \label{zd4}
\end{equation}%
These calculations of the HS- partition functions $\mathcal{Z}_{\boldsymbol{D%
}_{\mathcal{N}}}^{vect}$ and $\mathcal{Z}_{\boldsymbol{D}_{\mathcal{N}%
}}^{spin}$ can be extended to $SO(\mathcal{N},\mathcal{N})$ with generic $%
\mathcal{N}.$ The HS partition function for the vectorial models is given by
$\mathcal{Z}_{\boldsymbol{D}_{\mathcal{N}}}^{vect}=|\mathbf{\chi }_{1}^{%
\boldsymbol{D}_{\mathcal{N}}^{vect}}\left( q\right) |^{2}$ with vacuum
character factorising as%
\begin{equation}
\mathbf{\chi }_{1}^{\boldsymbol{D}_{\mathcal{N}}^{vect}}=\mathbf{\chi }_{1}^{%
\boldsymbol{D}_{\mathcal{N}}^{vect}\backslash \boldsymbol{D}_{\mathcal{N}%
-1}^{vect}}\bullet \mathbf{\chi }_{1}^{\boldsymbol{D}_{\mathcal{N}-1}^{vect}}
\end{equation}%
For example, the calculation of $\mathbf{\chi }_{1}^{\boldsymbol{D}%
_{5}^{vect}}$ follow from the factorisation $\mathbf{\chi }_{1}^{\boldsymbol{%
D}_{5}^{vect}\backslash \boldsymbol{D}_{4}^{vect}}\bullet \mathbf{\chi }%
_{1}^{\boldsymbol{D}_{4}^{vect}}$ as detailed below.

\subsubsection{$SO\left( 5,5\right) $ theory}

For this model, the field content of the vector and the spinorial (or
equivalently co-spinorial) series are different. The calculation of the HS-
partition for these models is given below.

\paragraph{Left node decomposition:\newline
}

This decomposition leads to the vector series described by cutting the left
node in the Tits-Satake diagram of $so\left( 5,5\right) ;$ thus leading to%
\begin{equation}
\begin{tabular}{ccc}
$so\left( 5,5\right) $ & $:$ & $so\left( 4,4\right) \oplus sl\left( 1,%
\mathbb{R}\right) \oplus 2\left[ 8\right] $ \\
$45$ & $:$ & $28+1+16$%
\end{tabular}%
\end{equation}%
The higher spins in the vectorial model are%
\begin{equation}
\begin{tabular}{c|c|c}
{\small series} & {\small Lorentz- spin} & $\text{boundary }${\small CFT}$%
_{2}\text{-spin}$ \\ \hline\hline
{\small vector } & $1,2,3;\quad 6,8$ & $2,3,4;\quad 7,9$ \\ \hline\hline
\end{tabular}%
\end{equation}%
Recall that the root system of $so\left( 5,5\right) $ has 40 roots $\pm
\left( \varepsilon _{i}\pm \varepsilon _{j}\right) _{i<j}$ generated by five
simple roots $\varepsilon _{i}-\varepsilon _{i+1}$ with $i=1,...,4$; and $%
\alpha _{5}=\varepsilon _{4}+\varepsilon _{5}$. \ The positive roots of $%
\Phi _{\boldsymbol{D}_{5}}^{+}$ are dispatched as follows%
\begin{equation}
\Phi _{\boldsymbol{D}_{4}}^{+}:\quad
\begin{tabular}{llll}
$\alpha _{2},\quad $ & $\left.
\begin{array}{c}
\alpha _{3} \\
\alpha _{2}+\alpha _{3}%
\end{array}%
\right. ,\quad $ & $\left.
\begin{array}{c}
\alpha _{4} \\
\alpha _{3}+\alpha _{4} \\
\alpha _{2}+\alpha _{3}+\alpha _{4}%
\end{array}%
\right. ,\quad $ & $\left.
\begin{array}{c}
\alpha _{5} \\
\alpha _{3}+\alpha _{5} \\
\alpha _{2}+\alpha _{3}+\alpha _{5} \\
\alpha _{3}+\alpha _{4}+\alpha _{5} \\
\alpha _{2}+\alpha _{3}+\alpha _{4}+\alpha _{5} \\
\alpha _{2}+2\alpha _{3}+\alpha _{4}+\alpha _{5}%
\end{array}%
\right. $%
\end{tabular}%
\end{equation}%
and
\begin{equation}
\Phi _{\boldsymbol{D}_{5}/\boldsymbol{D}_{4}}^{+}:\quad
\begin{tabular}{ccc}
\multicolumn{3}{c}{$\alpha _{1}$} \\
\multicolumn{3}{c}{$\alpha _{1}+\alpha _{2}$} \\
\multicolumn{3}{c}{$\alpha _{1}+\alpha _{2}+\alpha _{3}$} \\
$\alpha _{1}+\alpha _{2}+\alpha _{3}+\alpha _{5}$ & $,$ & $\alpha
_{1}+\alpha _{2}+\alpha _{3}+\alpha _{4}$ \\
\multicolumn{3}{c}{$\alpha _{1}+\alpha _{2}+\alpha _{3}+\alpha _{4}+\alpha
_{5}$} \\
\multicolumn{3}{c}{$\alpha _{1}+\alpha _{2}+2\alpha _{3}+\alpha _{4}+\alpha
_{5}$} \\
\multicolumn{3}{c}{$\alpha _{1}+2\alpha _{2}+2\alpha _{3}+\alpha _{4}+\alpha
_{5}$}%
\end{tabular}%
\end{equation}%
The partition function of this vectorial model is given by $\mathcal{Z}_{%
\boldsymbol{D}_{5}}^{vect}=|\mathbf{\chi }_{1}^{\boldsymbol{D}%
_{5}^{vect}}|^{2};$ it factorises like%
\begin{equation}
\mathcal{Z}_{\boldsymbol{D}_{5}^{vect}}=\mathcal{Z}_{\boldsymbol{D}%
_{5}^{vect}/\boldsymbol{D}_{4}^{vect}}\bullet \mathcal{Z}_{\boldsymbol{D}%
_{4}^{vect}}
\end{equation}%
where $\mathcal{Z}_{\boldsymbol{D}_{4}^{vect}}$ is given by eq(\ref{zd4});
and where $\mathcal{Z}_{\boldsymbol{D}_{5}^{vect}/\boldsymbol{D}%
_{4}^{vect}}=|\mathbf{\chi }_{1}^{\boldsymbol{D}_{5}^{vect}/\boldsymbol{D}%
_{4}^{vect}}|^{2}$ as follows%
\begin{equation}
\mathbf{\chi }_{1}^{\boldsymbol{D}_{5}^{vect}/\boldsymbol{D}_{4}}=\frac{1}{%
\mathbf{\eta }\left( q\right) }q^{\frac{1}{24}}\frac{\left( 1-q^{4}\right) }{%
\left( 1-q^{8}\right) }\dprod\limits_{n=1}^{8}\left( 1-q^{n}\right)
\end{equation}%
Using (\ref{zd4}), we end up with%
\begin{equation}
\mathbf{\chi }_{1}^{\boldsymbol{D}_{5}}=\frac{q^{-\frac{c}{24}}}{[\mathbf{%
\eta }\left( q\right) ]^{5}}q^{\frac{5}{24}}\frac{\left( 1-q^{7}\right)
\left( 1-q^{8}\right) }{\left( 1-q\right) \left( 1-q^{2}\right) }%
\dprod\limits_{n=1}^{8}\left( 1-q^{n}\right) ^{7-n}
\end{equation}

\paragraph{Right node decomposition:\newline
}

This decomposition describes the spinorial model where the principal
embedding is realised as
\begin{equation}
\begin{tabular}{ccc}
$so\left( 5,5\right) $ & $:$ & $sl\left( 5,\mathbb{R}\right) \oplus sl\left(
1,\mathbb{R}\right) \oplus 2\left[ 10\right] $ \\
$45$ & $:$ & $24+1+20$%
\end{tabular}%
\end{equation}%
Here, the higher spins fields are given by%
\begin{equation}
\begin{tabular}{c|c|c}
{\small series} & {\small Lorentz- spin} & $\text{boundary }${\small CFT}$%
_{2}\text{-spin}$ \\ \hline\hline
{\small spinorial} & $1,2,3;\quad 4,10$ & $2,3,4;\quad 5,11$ \\ \hline\hline
\end{tabular}%
\end{equation}%
and the 20 positive roots of $\Phi _{\boldsymbol{D}_{5}}^{+}$ decompose into
$so(1,2)$ multiplets as follows%
\begin{equation}
\Phi _{\boldsymbol{A}_{4}}^{+}:\quad
\begin{tabular}{llll}
$\alpha _{1},\quad $ & $\left.
\begin{array}{c}
\alpha _{2} \\
\alpha _{1}+\alpha _{2}%
\end{array}%
\right. ,\quad $ & $\left.
\begin{array}{c}
\alpha _{3} \\
\alpha _{2}+\alpha _{3} \\
\alpha _{1}+\alpha _{2}+\alpha _{3}%
\end{array}%
\right. ,\quad $ & $\left.
\begin{array}{c}
\alpha _{4} \\
\alpha _{3}+\alpha _{4} \\
\alpha _{2}+\alpha _{3}+\alpha _{4} \\
\alpha _{1}+\alpha _{2}+\alpha _{3}+\alpha _{4}%
\end{array}%
\right. $%
\end{tabular}%
\end{equation}%
and%
\begin{equation}
\Phi _{\boldsymbol{D}_{5}/\boldsymbol{A}_{4}}^{+}:\quad
\begin{tabular}{ccc}
\multicolumn{3}{c}{$\alpha _{5}$} \\
\multicolumn{3}{c}{$\alpha _{3}+\alpha _{5}$} \\
$\alpha _{2}+\alpha _{3}+\alpha _{5}$ & $,$ & $\alpha _{3}+\alpha
_{4}+\alpha _{5}$ \\
$\alpha _{1}+\alpha _{2}+\alpha _{3}+\alpha _{5}$ & $,$ & $\alpha
_{2}+\alpha _{3}+\alpha _{4}+\alpha _{5}$ \\
$\alpha _{1}+\alpha _{2}+\alpha _{3}+\alpha _{4}+\alpha _{5}$ & $,$ & $%
\alpha _{2}+2\alpha _{3}+\alpha _{4}+\alpha _{5}$ \\
\multicolumn{3}{c}{$\alpha _{1}+\alpha _{2}+2\alpha _{3}+\alpha _{4}+\alpha
_{5}$} \\
\multicolumn{3}{c}{$\alpha _{1}+2\alpha _{2}+2\alpha _{3}+\alpha _{4}+\alpha
_{5}$}%
\end{tabular}%
\end{equation}%
The higher spin partition function is calculated by using the factorisation%
\begin{equation}
\mathcal{Z}_{\boldsymbol{D}_{5}^{spin}}=\mathcal{Z}_{\boldsymbol{A}%
_{4}}\times \mathcal{Z}_{\boldsymbol{D}_{5}^{spin}/\boldsymbol{A}_{4}}
\end{equation}%
where $\mathcal{Z}_{\boldsymbol{D}_{5}^{spin}/\boldsymbol{A}_{4}}=\left\vert
\mathbf{\chi }_{1}^{\boldsymbol{D}_{5}/\boldsymbol{A}_{4}}\right\vert ^{2}$
and%
\begin{equation}
\mathbf{\chi }_{1}^{\boldsymbol{D}_{5}^{spin}/\boldsymbol{A}_{4}}=\frac{1}{%
\mathbf{\eta }\left( q\right) }q^{\frac{1}{24}}\frac{\left( 1-q^{3}\right)
\left( 1-q^{4}\right) \left( 1-q^{5}\right) }{\left( 1-q^{8}\right) \left(
1-q^{9}\right) \left( 1-q^{10}\right) }\dprod\limits_{n=1}^{10}\left(
1-q^{n}\right)
\end{equation}%
Using eq(\ref{An}) namely%
\begin{equation}
\mathbf{\chi }_{1}^{\boldsymbol{A}_{4}}=\frac{q^{-\frac{c}{24}}}{\left[
\mathbf{\eta }\left( q\right) \right] ^{4}}q^{\frac{4}{24}%
}\dprod\limits_{n=1}^{5}\left( 1-q^{n}\right) ^{5-n}
\end{equation}%
we end up with%
\begin{equation}
\mathbf{\chi }_{1}^{\boldsymbol{D}_{5}^{spin}}=\frac{q^{-\frac{c}{24}}}{%
\left[ \mathbf{\eta }\left( q\right) \right] ^{5}}q^{\frac{5}{24}}\frac{%
\left( 1-q^{7}\right) \left( 1-q^{8}\right) }{\left( 1-q\right) \left(
1-q^{2}\right) }\dprod\limits_{n=1}^{8}\left( 1-q^{n}\right) ^{7-n}
\end{equation}

\section{Conclusion and comments}

In the present inquiry, we proposed a novel approach \textrm{based on} the
rich structure of AdS$_{3}$/CFT$_{2}$ correspondence and Tits-Satake
diagrams of gauge symmetry to construct the higher spin gravity theories in
the framework of AdS$_{3}$ space time within the Chern Simons formulation.
This approach was first \textrm{employed} in recasting the\ higher spin
theory with $SL(\mathcal{N},\mathbb{R})$ gauge symmetry which is the real
split form of the complex $\boldsymbol{A}_{\mathcal{N}-1}$ Lie algebras. The
revisiting of this gravity theory allowed us to establish a link between the
higher spin fields and the principal embedding algorithm realised in terms
of the extremal node decomposition (LEND and REND) of Tits-Satake graphs
describing gauge symmetry.

The generalisation of the $\boldsymbol{A}_{\mathcal{N}-1}$ theory in this
construction to higher spin gravity theories with orthogonal $\boldsymbol{B}%
_{\mathcal{N}}$\ and $\boldsymbol{D}_{\mathcal{N}}$\ symmetries was based on
a parallel rationale where we focused on the real split forms $SO(\mathcal{N}%
,\mathcal{N}+1)$ and $SO(\mathcal{N},\mathcal{N})$. We showed that these
orthogonal models have different field contents according to\textrm{\ }sets $%
\mathfrak{M}_{\mathfrak{j}}$ of $SO(1,2)$ spins $\mathfrak{j}$. Recall that
the L-spin $\mathfrak{j}$ is linked to the conformal spin $s$ via
the relation $s=\mathfrak{j}+1.$ While the $\boldsymbol{A}_{\mathcal{N}-1}$
theory is known to include all the integer $SO(1,2)$ spins $\mathfrak{j}$ up
to its rank as,%
\begin{equation}
\boldsymbol{A}_{\mathcal{N}-1}:\quad \{\mathfrak{M}_{\mathfrak{j}}\}\qquad
with\qquad \mathfrak{j}=1,2,3,4,\ldots ,\mathcal{N}-1
\end{equation}%
the $\boldsymbol{B}_{\mathcal{N}}$ theory divulged two series: $\left(
i\right) $ a vectorial containing only $\mathfrak{M}_{\mathfrak{j}}$'s with
odd spins $\mathfrak{j},$ and at the boundary W-currents $W^{(s)}$ with even
spins s, as follows
\begin{equation}
\begin{tabular}{lllll}
$\boldsymbol{B}_{\mathcal{N}}^{vect}$ & $:$ & $\mathfrak{j}$ & $=$ & $%
1,3,5,7,\ldots ,2\mathcal{N}-1$ \\
&  & $s$ & $=$ & $2,4,6,8,\ldots ,2\mathcal{N}$%
\end{tabular}%
\end{equation}%
This resulted from the left extremal node decomposition of the $SO(\mathcal{N%
},\mathcal{N}+1)$ Tits-Satake diagram. $\left( ii\right) $ a spinorial
series containing only spins $\mathfrak{j}$ and conformal spins s as follows
\begin{equation}
\begin{tabular}{lllllll}
$\boldsymbol{B}_{\mathcal{N}}^{spin}$ & $:$ & $\mathfrak{j}$ & $=$ & $%
1,2,3,4,\ldots ,\mathcal{N}-1$ & ; & $\frac{\mathcal{N}\left( \mathcal{N}%
+1\right) }{2}$ \\
&  & $s$ & $=$ & $2,3,4,5,\ldots ,\mathcal{N}$ & ; & $\frac{\mathcal{N}%
\left( \mathcal{N}+1\right) }{2}+1$%
\end{tabular}%
\end{equation}%
which resulted from the right extremal node decomposition of the $SO(%
\mathcal{N},\mathcal{N}+1)$ Tits-Satake diagram.

The investigation of the $\boldsymbol{D}_{\mathcal{N}}$\ theory also lead to
two series for the higher spin fields. $\left( i\right) $ The vectorial
follows from the decomposition of $SO(\mathcal{N},\mathcal{N})$ by cutting
the vector node in the Tits-Satake diagram that looks like the $\boldsymbol{D%
}_{\mathcal{N}}$'s Dynkin diagram. It shares the first three elements with
the linear family spectrum ($\mathfrak{j}=1,2,3$) and then adopts a
2-periodicity for $6\leq \mathfrak{j}\leq 2N-2:$
\begin{equation}
\begin{tabular}{lllllll}
$\boldsymbol{D}_{\mathcal{N}}^{vect}$ & $:$ & $\mathfrak{j}$ & $=$ & $1,2,3$
& $;\quad $ & $6,8,10,\ldots ,2\mathcal{N}-2$ \\
&  & $s$ & $=$ & $2,3,4$ & $;\quad $ & $7,9,11,\ldots ,2\mathcal{N}-1$%
\end{tabular}%
\end{equation}%
$\left( ii\right) $ The spinorial set emerges from decomposing $SO(\mathcal{N%
},\mathcal{N})$ with respect to the spinorial node; it has a remarkable
isolated $\mathfrak{j}=\mathcal{N}\left( \mathcal{N}-1\right) /2$. This is
written as%
\begin{equation}
\begin{tabular}{lllllll}
$\boldsymbol{D}_{\mathcal{N}}^{spin}$ & $:$ & $\mathfrak{j}$ & $=$ & $%
1,2,3,4,\ldots ,\mathcal{N}-1$ & $;$ & $\frac{\mathcal{N}\left( \mathcal{N}%
-1\right) }{2}$ \\
&  & $s$ & $=$ & $2,3,4,5,\ldots ,\mathcal{N}$ & $;$ & $\frac{\mathcal{N}%
\left( \mathcal{N}-1\right) }{2}+1$%
\end{tabular}%
\end{equation}%
These results were moreover implemented into the calculation of the higher
spin partition function using the correspondence with the CFT$_{2}$ where
the symmetry is given by $\boldsymbol{WA}_{\mathcal{N}-1},$ $\boldsymbol{WB}%
_{\mathcal{N}}$- and $\boldsymbol{WD}_{\mathcal{N}}$ algebras. The
interpretation of the characters in terms of root systems allowed to
identify the contribution of higher spin and to write the full HS-partition
function as a factorization of these contributions. The explicit computation
was given for $SL(\mathcal{N},\mathbb{R})$ and for the leading models $%
\mathcal{N}=2,3$ of $SO(\mathcal{N},\mathcal{N}+1)$\ and $\mathcal{N}=4,5$
of $SO(\mathcal{N},\mathcal{N}).$ We found that the higher spin partition
functions are given by $Z^{\mathbf{g}}=|\mathbf{\chi }_{1}^{\mathbf{g}}|^{2}$
with%
\begin{equation}
\begin{tabular}{|c|c|}
\hline
$\mathbf{g}$ & $\mathbf{\chi }_{1}^{\mathbf{g}}$ \\ \hline
$\boldsymbol{so(2,3)}$ & $\frac{q^{-\frac{c}{24}}}{[\mathbf{\eta }\left(
q\right) ]^{2}}q^{\frac{1}{12}}\left( 1-q\right)
\dprod\limits_{n=1}^{3}\left( 1-q^{n}\right) $ \\ \hline
$\boldsymbol{so(3,4)}$ & $\frac{q^{-\frac{c}{24}}}{[\mathbf{\eta }\left(
q\right) ]^{3}}q^{\frac{1}{8}}\left( 1-q\right) ^{2}\left( 1-q^{2}\right)
\left( 1-q^{3}\right) \dprod\limits_{n=1}^{5}\left( 1-q^{n}\right) $ \\
\hline
$\boldsymbol{so(4,4)}$ & $\frac{q^{-\frac{c}{24}}}{[\mathbf{\eta }\left(
q\right) ]^{4}}\frac{q^{\frac{1}{6}}}{\left( 1-q\right) \left(
1-q^{2}\right) \left( 1-q^{4}\right) }\dprod\limits_{n=1}^{6}\left(
1-q^{n}\right) ^{6-n}$ \\ \hline
$\boldsymbol{so(5,5)}$ & $\frac{q^{-\frac{c}{24}}}{[\mathbf{\eta }\left(
q\right) ]^{5}}\frac{q^{\frac{5}{24}}\left( 1-q^{7}\right) \left(
1-q^{8}\right) }{\left( 1-q\right) \left( 1-q^{2}\right) }%
\dprod\limits_{n=1}^{8}\left( 1-q^{n}\right) ^{7-n}$ \\ \hline
\end{tabular}%
\end{equation}%
These higher spin partition functions can be further applied in the
framework of the BTZ black hole particularly in the computation of the
HS-BTZ black hole partition function and the derivation of the gravitational
exclusion principle \textrm{\cite{exclusion}} for the orthogonal symmetries
and therefore validate the supposition made in \textrm{\cite{exclusion2,41}}.

As perspective of this investigation, notice that this analysis is valid for
HS-AdS$_{3}$ gravities based on different real split forms of the complex
Lie algebras. Moreover, the construction can be enlarged to deal with other
higher spin 3D gravities. Higher spins for AdS$_{3}$ gravity with
exceptional gauge symmetries and application to exceptional BTZ black hole
will be reported in a future occasion.

\section{Appendices}

In this section, we give two appendices A and B where we collect useful
tools employed in the core of the paper. In appendix A, we give explicit
realisations for generators of non compact groups as well as the properties
of their Lie algebras with regards to higher spin AdS$_{3}$ gravity. In
appendix B, we describe general aspects of real forms of Lie algebras,
Cartan involution and Tits-Satake diagrams.

\subsection{Appendix A: higher spin 3 algebra of $so\left( 1,2\right) $}

The $so\left( 1,2\right) $ is the Lie algebra of the Lorentz group in 3D
spacetime $\mathbb{R}^{1,2}$. This is a non compact group with one diagonal
generator H \textrm{generating so(1,1)}. In AdS$_{3}$ gravity, the three
generators of $so\left( 1,2\right) $ turn out to be intimately related with
the non anomalous generators $L_{0}$, $L_{\pm }$ of the conformal spin 2
current of the CFT$_{2}$ living on the boundary of AdS$_{3};$ thanks to AdS$%
_{3}$/CFT$_{2}$ correspondence. In addition, the $so\left( 1,2\right) $ is
the building block in the principal embedding of \cite{4A} used in the study
of higher spin 3D gravity. In this higher spin generalisation, the $SO\left(
1,2\right) $ is embedded in \textrm{bigger groups} like for instance the $%
SL\left( 3,\mathbb{R}\right) $ with rank two whose two commuting diagonal
generators H$_{1}$, H$_{2}$ are the non anomalous generators L$_{0}$ and W$%
_{0}$ generators of the W$_{3}$- conformal symmetry living at the boundary
of AdS$_{\mathrm{3}}$. Below, we give some details regarding $so\left(
1,2\right) $ with homomorphisms%
\begin{equation}
so\left( 1,2\right) \simeq su\left( 1,1\right) \simeq sl\left( 2,\mathbb{R}%
\right)
\end{equation}%
\ and its embedding into\textrm{\ }$sl\left( 3,\mathbb{R}\right) $ and
generally in $sl\left( \mathcal{N},\mathbb{R}\right) .$ This embedding has
been extended in this paper to the orthogonal symmetries.

\subsubsection{The so$\left( 1,2\right) $\ Lorentz algebra}

It is generated by three real matrix operators $\boldsymbol{t}^{\left[ ab%
\right] }\sim \varepsilon ^{abc}\boldsymbol{t}_{c}$ with label $a$ lifted by
the metric $\eta _{ab}=(-,++)$ like $t_{a}=\eta _{ab}\boldsymbol{t}^{b}$;
thus constituting a basic difference with so$\left( 3\right) $ of the
Euclidian 3D space with metric $\delta _{ab}=(+,++)$. The commutation
relations are given by%
\begin{equation}
\left[ J_{0},J_{1}\right] =+J_{2},\qquad \left[ J_{0},J_{2}\right]
=-J_{1},\qquad \left[ J_{1},J_{2}\right] =-J_{0}  \label{rea}
\end{equation}%
The vector realisation of this algebra is given by the infinitesimal
rotation $\delta x_{c}=$ $\mathbf{\lambda }_{c}^{d}x_{d}$ with rotation
matrix $\mathbf{\lambda }=\psi _{{\small [ab]}}J^{{\small [ab]}}+O\left(
2\right) $ ($\mathbf{\lambda }^{T}=-\mathbf{\lambda }$) expanding like $\psi
_{{\small [12]}}J_{{\small [12]}}-\psi _{{\small [20]}}J_{{\small [20]}%
}-\psi _{{\small [01]}}J_{{\small [01]}}$ with the $\psi _{{\small [ab]}}$'s
giving the group parameters. This expansion shows that $J_{{\small [12]}}$
is a compact generator while the two others are non compact. By setting $%
\theta _{{\small 3}}=\psi _{{\small [12]}}$ while $\theta _{{\small a}%
}=i\psi _{{\small [0a]}}$ with $a=1,2$ as well as $\boldsymbol{M}_{{\small 3}%
}=J_{{\small [12]}}$ and $\boldsymbol{M}_{{\small a}}=iJ_{{\small [0a]}}$,
one can present the above $\mathbf{\lambda }$ rotation matrix like $\theta
_{1}\boldsymbol{M}_{1}+\theta _{2}\boldsymbol{M}_{2}+\theta _{3}\boldsymbol{M%
}_{3}$ giving another way to think about the difference between so$\left(
1,2\right) $ and so$\left( 3\right) $. By using the homomorphism $so\left(
1,2\right) \simeq sl\left( 2,\mathbb{R}\right) ,$ we can work out a
realisation of (\ref{rea}) in terms of 2$\times $2 matrices as follows \cite%
{Kitaev},%
\begin{equation}
J_{1}=\frac{1}{2}\left(
\begin{array}{cc}
0 & 1 \\
1 & 0%
\end{array}%
\right) ,\qquad J_{2}=\frac{1}{2}\left(
\begin{array}{cc}
0 & i \\
-i & 0%
\end{array}%
\right) ,\qquad J_{0}=\frac{i}{2}\left(
\begin{array}{cc}
1 & 0 \\
0 & -1%
\end{array}%
\right)
\end{equation}%
that is $J_{1}=\sigma ^{x}/2,$ $J_{2}=\sigma ^{y}/2,$ and $J_{0}=i\sigma
^{z}/2$ where $J_{0}$ is the compact generator. They are related to the non
anomalous Virasoro generators as $J_{1}=(L_{-}-L_{+})/2,$ $%
J_{2}=i(L_{-}+L_{+})/2,$ and $J_{0}=iL_{0}$. By \textrm{choosing }$J_{2}$%
\textrm{\ as the compact} generator while following \textrm{\cite{Eric}} we
can express the three $K_{a}$ ($\sim 2J_{a}^{\prime }$) generating $sl\left(
2,\mathbb{R}\right) $ in terms of the usual 2$\times $2 matrices like $%
K_{1}=\sigma ^{1},$ $K_{2}=i\sigma ^{2},$ $K_{0}=\sigma ^{3}$; they read
explicitly as%
\begin{equation}
K_{1}=\left(
\begin{array}{cc}
0 & 1 \\
1 & 0%
\end{array}%
\right) ,\qquad K_{2}=\left(
\begin{array}{cc}
0 & 1 \\
-1 & 0%
\end{array}%
\right) ,\qquad K_{0}=\left(
\begin{array}{cc}
1 & 0 \\
0 & -1%
\end{array}%
\right)  \label{rho}
\end{equation}%
and they satisfy commutation relations with real structure constants as%
\begin{equation}
\left[ K_{1},K_{2}\right] =-2K_{0},\qquad \left[ K_{0},K_{1}\right]
=2K_{2},\qquad \left[ K_{0},K_{2}\right] =2K_{1}  \label{jsl}
\end{equation}%
By setting $H_{\alpha }=K_{0}$ and $E_{\pm \alpha }=(K_{1}\pm K_{2})/2$
\textrm{with adjoint conjugation} $\left( E_{\pm }\right) ^{\dagger }=E_{\mp
},$ we have%
\begin{equation}
\begin{tabular}{lllllll}
$\left[ H_{\alpha },E_{\pm \alpha }\right] $ & $=$ & $\pm 2E_{\pm \alpha }$
& $,\qquad $ & $\left[ E_{0},E_{\pm }\right] $ & $=$ & $\pm E_{\pm }$ \\
$\left[ E_{+\alpha },E_{-\alpha }\right] $ & $=$ & $H_{\alpha }$ & $,\qquad $
& $\left[ E_{+},E_{-}\right] $ & $=$ & $2E_{0}$%
\end{tabular}
\label{22}
\end{equation}%
with normalisation $H_{\alpha }=2E_{0}$. These three $E_{0}$ and $E_{\pm }$
are \textrm{related to} the\ non anomalous generators of the three Laurent
modes $L_{0}=E_{0}$ and $L_{\pm }=\lambda _{\pm }E_{\mp }$ ($\lambda
_{+}.\lambda _{-}=-1$) of the Virasoro algebra Vir$_{c}[\partial (AdS_{3})]$
living at the asymptote of the Anti-de Sitter geometry. \newline
In sum, the two real forms of $sl(2,\mathbb{C})$ are given by $\left(
\mathbf{i}\right) $ the compact real form $su\left( 2\right) $ with
generators $J_{a}^{su_{2}}$ realised in terms of the Pauli matrices like,%
\begin{equation}
J_{1}^{su_{2}}=\frac{i}{2}\sigma ^{1},\qquad J_{2}^{su_{2}}=\frac{i}{2}%
\sigma ^{2},\qquad J_{3}^{su_{2}}=\frac{i}{2}\sigma ^{3}
\end{equation}%
and $\left( \mathbf{ii}\right) $ the real split form $su(1,1)\sim sl(2,%
\mathbb{R})$ generated by $K_{a}^{sl_{2}}$ realised as in (\ref{rho}). They
are related to the Chevalley generators $\left( E_{+\alpha },E_{-\alpha
},H_{\alpha }\right) $ of $sl(2,\mathbb{C})$\ by bridge 3$\times $3 matrices
like $J_{a}^{su_{2}}=\mathcal{V}_{a}^{n\alpha }E_{n\alpha }$ and $%
J_{a}^{sl_{2}}=\mathcal{U}_{a}^{n\alpha }E_{n\alpha }$ with the bridging $%
\mathcal{V}^{su_{2}}$ and $\mathcal{U}^{sl_{2}}$\ learnt from
\begin{equation}
\begin{tabular}{c||c|c|c||c}
generators & $J_{1}$ & $J_{2}$ & $\quad J_{3}\qquad $ & $J_{a}$ \\
\hline\hline
$su\left( 2\right) $ & $i\left( E_{+\alpha }+E_{-\alpha }\right) $ & $%
E_{+\alpha }-E_{-\alpha }$ & $\quad iH_{\alpha }\qquad $ & $J_{a}^{su_{2}}=%
\mathcal{V}_{a}^{n\alpha }E_{n\alpha }$ \\ \hline
$sl(2,\mathbb{R})$ & $E_{+\alpha }+E_{-\alpha }$ & $E_{+\alpha }-E_{-\alpha
} $ & $\quad H_{\alpha }\qquad $ & $J_{a}^{sl_{2}}=\mathcal{U}_{a}^{n\alpha
}E_{n\alpha }$ \\ \hline\hline
\end{tabular}%
\end{equation}%
thus reading as%
\begin{equation}
\mathcal{V}^{su_{2}}=\frac{1}{2}\left(
\begin{array}{ccc}
i & i & 0 \\
1 & -1 & 0 \\
0 & 0 & i%
\end{array}%
\right) \qquad ,\qquad \mathcal{U}^{sl_{2}}=\frac{1}{2}\left(
\begin{array}{ccc}
1 & 1 & 0 \\
1 & -1 & 0 \\
0 & 0 & 1%
\end{array}%
\right)  \label{vu}
\end{equation}%
Notice that these two real forms of $sl(2,\mathbb{C})$ can be also
discriminated by their Killing forms $K\left( X,Y\right) =Tr(ad_{X}ad_{Y});$
which for $su(2)$ and $sl(2,\mathbb{R})$ read as follows%
\begin{equation}
K_{su(2)}=\left(
\begin{array}{ccc}
-8 & 0 & 0 \\
0 & -8 & 0 \\
0 & 0 & -8%
\end{array}%
\right) ,\qquad K_{sl(2,\mathbb{R})}=\left(
\begin{array}{ccc}
8 & 0 & 0 \\
0 & -8 & 0 \\
0 & 0 & 8%
\end{array}%
\right)  \label{ksu2}
\end{equation}%
For $K_{su(2)},$ all its eigenvalues are negative, meaning that the three
generators are compact, while $K_{sl(2,\mathbb{R})}$ has only one negative
eigenvalue corresponding to $J_{2};$ then, the two other generators of $sl(2,%
\mathbb{R})$ are non compact. This feature is nicely described by the Cartan
involution $\vartheta $ (with $\vartheta ^{2}=id$) given below with
illustration on the $sl(3,\mathbb{C})$ example. There, we show that $%
J_{2}\in \left. sl(2,\mathbb{R})\right\vert _{\vartheta =+1}$ while $J_{0}$
and $J_{1}\in \left. sl(2,\mathbb{R})\right\vert _{\vartheta =-1}.$

\subsubsection{The higher spin 3 algebra}

By higher spin 3 algebra, we mean the algebra generated by the non anomalous
generators $L_{0},$ $L_{\pm }$ and $W_{0},$ $W_{\pm },$ $W_{\pm 2}$ of the
conformal WA$_{3}$ symmetry living at the boundary of AdS$_{3}$ gravity with
gauge\ symmetry $SL(3,\mathbb{R})$. This is an eight dimensional algebra
which is isomorphic to the real split form sl(3,$\mathbb{R}$) of the complex
Lie algebra $\boldsymbol{A}_{2}$. In the principal embedding algorithm of
\cite{4A}, the sl(3,$\mathbb{R}$) is generated by monomials of the sl(2,$%
\mathbb{R}$) generators $J_{a}$ introduced before. This algebra has eight
generators; three given by $J_{a}$ and the other five denoted like $%
T_{\left( ab\right) }$ with the traceless condition $\eta ^{ab}T_{\left(
ab\right) }=0.$ The $T_{\left( ab\right) }$'s are given by quadratic
monomials in $J_{a}$ as follows
\begin{equation}
\begin{tabular}{lll}
$T_{11}$ & $=$ & $2J_{1}^{2}+\frac{2}{3}\boldsymbol{J}^{2}$ \\
$T_{22}$ & $=$ & $2J_{2}^{2}-\frac{2}{3}\boldsymbol{J}^{2}$ \\
$T_{33}$ & $=$ & $2J_{3}^{2}-\frac{2}{3}\boldsymbol{J}^{2}$%
\end{tabular}
\label{75}
\end{equation}%
with Casimir $\boldsymbol{J}^{2}=J_{a}\eta ^{ab}J_{b}$; and%
\begin{equation}
\begin{tabular}{lll}
$T_{12}$ & $=$ & $J_{1}J_{2}+J_{2}J_{1}$ \\
$T_{23}$ & $=$ & $J_{2}J_{3}+J_{3}J_{2}$ \\
$T_{13}$ & $=$ & $J_{1}J_{3}+J_{3}J_{1}$%
\end{tabular}%
\end{equation}%
The generic commutation relations read as follows%
\begin{equation}
\begin{tabular}{lll}
$\left[ J_{a},J_{b}\right] $ & $=$ & $\epsilon _{abc}J^{c}$ \\
$\left[ J_{a},T_{bc}\right] $ & $=$ & $\epsilon _{a(b}^{m}T_{c)m}$ \\
$\left[ T_{ab},T_{cd}\right] $ & $=$ & $\sigma \lbrack \eta _{a(c}\epsilon
_{d)bm}+\eta _{b(c}\epsilon _{d)am}]J^{m}$%
\end{tabular}%
\end{equation}%
with $\sigma =-1$ for sl(3,$\mathbb{R}$) and $\sigma =+1$ for su(2,1). We
can re-write this algebra in terms of the charged generators $L_{0},$ $%
L_{\pm }$ and the generators $W_{0},$ $W_{\pm },$ $W_{\pm 2}$ mentioned
before as%
\begin{equation}
\begin{tabular}{lll}
$W_{-2}$ & $=$ & $2\left( T_{11}-T_{12}\right) -T_{33}$ \\
$W_{+2}$ & $=$ & $2\left( T_{11}+T_{12}\right) -T_{33}$ \\
$W_{-1}$ & $=$ & $T_{13}-T_{23}$ \\
$W_{1}$ & $=$ & $T_{13}-T_{23}$ \\
$W_{0}$ & $=$ & $T_{33}$%
\end{tabular}%
\end{equation}%
and%
\begin{equation}
\begin{tabular}{lll}
$\left[ L_{i},L_{j}\right] $ & $=$ & $\left( i-j\right) L_{i+j}$ \\
$\left[ L_{i},W_{m}\right] $ & $=$ & $\left( 2i-m\right) W_{i+m}$ \\
$\left[ W_{m},W_{n}\right] $ & $=$ & $-\frac{1}{3}\left( m-n\right) \left(
+2m^{2}+2n^{2}-mn-8\right) L_{m+n}$%
\end{tabular}
\label{79}
\end{equation}%
where $i,j=0,\pm 1$ and $m,n=0,\pm 1,\pm 2.$ This description extends to $sl(%
\mathcal{N},\mathbb{R}).$

\subsection{Appendix B: real forms of complex $sl(3,\mathbb{C})$}

This appendix aims to describe useful aspects in the construction of real
forms of complex Lie algebras of Cartan classification \cite{SPIN} through
the example of $sl(3,\mathbb{C}).$ This is the Lie algebra of traceless
complex $3\times 3$ matrices ($M_{ij}$) with expansions $\sum_{a=1}^{8}%
\mathfrak{M}_{a}T_{ij}^{a}$. It is a representation of the Lie algebra $%
\boldsymbol{A}_{2}$ on the complex 3D space $\mathbb{C}^{3}$ with 8
Cartan-Weyl generators%
\begin{equation}
H_{\alpha _{1}},\quad E_{\pm \alpha _{1}},\quad E_{\pm \alpha _{2}},\quad
H_{\alpha _{2}},\quad E_{\pm \left( \alpha _{1}+\alpha _{2}\right) }
\end{equation}%
where $\alpha _{1}$ and $\alpha _{2}$ refer to the two simple roots of $%
\boldsymbol{A}_{2}$ and where $H_{\alpha _{3}}$ is just $H_{\alpha
_{1}}+H_{\alpha _{2}}$. The two simple roots generate the root system $\Phi
_{A_{2}}$ with cardinal $|\Phi _{A_{2}}|=6$ and elements as%
\begin{equation}
\pm \alpha _{1},\quad \pm \alpha _{2},\quad ,\quad \alpha _{3}=\pm \left(
\alpha _{1}+\alpha _{2}\right)  \label{s3}
\end{equation}%
Notice that by using $H_{\alpha _{3}}$ and $E_{\pm \alpha _{3}}$, one can
obtain the three $sl(2,\mathbb{C})$ subalgebras of $sl(3,\mathbb{C})$ useful
in the study of real forms. The commutation relations of $sl(3,\mathbb{C})$
are given by the commuting $\left[ H_{\alpha },H_{\beta }\right] =0,$ and%
\begin{equation}
\begin{tabular}{llllll}
$\left[ H_{\alpha },E_{\beta }\right] $ & $=$ & $+A_{\beta \alpha }E_{\beta
} $ &  &  &  \\
$\left[ H_{\alpha },E_{-\beta }\right] $ & $=$ & $-A_{\beta \alpha
}E_{-\beta }$ &  &  &  \\
$\left[ E_{\alpha },E_{\beta }\right] $ & $=$ & ${\small N}_{\alpha ,\beta }%
{\small E}_{\alpha +\beta }$ & $if$ & ${\small \alpha +\beta }$ & ${\small %
\in \Phi }$ \\
$\left[ E_{\alpha },E_{\beta }\right] $ & $=$ & $\delta _{\alpha +\beta
}H_{\alpha }$ & $if$ & ${\small \alpha +\beta }$ & ${\small =0}$ \\
$\left[ E_{\alpha },E_{\beta }\right] $ & $=$ & ${\small 0}$ \ \ \ \ \ \ \ \
\ \ \ \  & $if$ & ${\small \alpha +\beta }$ & ${\small \notin \Phi }$%
\end{tabular}
\label{714}
\end{equation}%
where ${\small N}_{\alpha ,\beta }$ are real structure constants and $%
A_{\beta \alpha }$ are the entries of the symmetric Cartan matrix
\begin{equation}
A_{\alpha \beta }=\left(
\begin{array}{ccc}
2 & -1 & 0 \\
-1 & 2 & -1 \\
0 & -1 & 2%
\end{array}%
\right)
\end{equation}%
It is known that the complex Lie algebra $sl(3,\mathbb{C})$ has, up to
automorphisms, three real forms represented by three different Tits-Satake
diagrams given by the Figure \textbf{\ref{STA}}. The three real forms can be
approached in various ways; in particular by using the so-called Cartan
involution $\vartheta $ acting on the roots $\alpha $ (\ref{s3}) and the
associated generators $E_{\pm \alpha },$ $H_{\alpha }$. Because $\vartheta
^{2}=1,$ the generators are characterised by the $\pm 1$ eigenvalues of the $%
\vartheta $ which act on the Killing form of $sl(3,\mathbb{C})$ ( generally
on Lie algebras g ) as follows%
\begin{equation}
\vartheta :K\left( X,Y\right) \rightarrow K_{\vartheta }\left( X,Y\right)
=K\left( X,\vartheta Y\right)
\end{equation}%
The signature of the eigenvalues of the representative matrix of $K\left(
X,\vartheta Y\right) $ permits to distinguish the real forms of $sl(3,%
\mathbb{C})$. Indeed, for real forms $\mathfrak{L}_{R}$ of a given complex
Lie algebra $\mathfrak{L}_{C}$ (here $\boldsymbol{A}_{2}$), the Cartan
involution $K_{\vartheta }\left( X,Y\right) $ of the Killing form of $%
\mathfrak{L}_{R}$\ should be negative definite as for $K_{su(2)}$ in (\ref%
{ksu2}). The three real forms of $sl(3,\mathbb{C})$ are described below:

\subsubsection{Compact real form $su\left( 3\right) $}

The compact real form $su\left( 3\right) $ of the Lie algebra $sl(3,\mathbb{C%
})$ is characterised by the following anti-hermitian Chevalley generators%
\begin{equation}
\begin{tabular}{lllllll}
$X_{\alpha }$ & $=$ & $i\left( E_{+\alpha }+E_{-\alpha }\right) $ & $,\qquad
$ & $\vartheta X_{\alpha }$ & $=$ & $X_{\alpha }$ \\
$Y_{\alpha }$ & $=$ & $E_{+\alpha }-E_{-\alpha }$ & $,\qquad $ & $\vartheta
Y_{\alpha }$ & $=$ & $Y_{\alpha }$ \\
$Z_{\alpha }$ & $=$ & $iH_{\alpha }$ & $,\qquad $ & $\vartheta Z_{\alpha }$
& $=$ & $Z_{\alpha }$%
\end{tabular}
\label{xyz}
\end{equation}%
They generate the $su\left( 2\right) $ subalgebras within $su\left( 3\right)
;$ and are invariant under the Cartan involution; i.e $\vartheta =id$. For a
given positive root $\alpha $, the bridge $3\times 3$ matrix $\mathcal{V}%
_{\alpha }^{su_{2}}$ between the basis $\left( X_{\alpha },Y_{\alpha
},Z_{\alpha }\right) $ and the Cartan basis $\left( E_{+\alpha },E_{-\alpha
},H_{\alpha }\right) $ is given by%
\begin{equation}
\mathcal{V}_{\alpha }^{su_{2}}=\frac{1}{2}\left(
\begin{array}{ccc}
i & i & 0 \\
1 & -1 & 0 \\
0 & 0 & i%
\end{array}%
\right)  \label{2ua}
\end{equation}%
Notice that because the $\vartheta $ is an involution ($\vartheta ^{2}=id$),
one uses its eigenvalues $\pm 1$ to decompose the real forms $\mathfrak{L}%
_{R}$ into two sectors: $\mathfrak{L}_{R}^{+}\equiv \mathfrak{t}$ generated
by compact generators as in (\ref{xyz}); and $\mathfrak{L}_{R}^{-}\equiv
\mathfrak{p}$ generated by non compact ones. So, we have the following $%
\vartheta $- decomposition%
\begin{equation}
\begin{tabular}{lllllll}
$\mathfrak{L}_{R}$ & $=$ & $\mathfrak{L}_{R}^{+}\oplus \mathfrak{L}_{R}^{-}$
& $,\qquad $ & $\vartheta (\mathfrak{t})$ & $=$ & $+\mathfrak{t}$ \\
& $=$ & $\mathfrak{t}\oplus \mathfrak{p}$ & $,\qquad $ & $\vartheta (%
\mathfrak{p})$ & $=$ & $-\mathfrak{p}$%
\end{tabular}%
\end{equation}%
As such for the compact real form $su(3)$, its generators are compact and
consequently sit into the $\mathfrak{t}$- sector such as
\begin{equation}
\begin{tabular}{|c|c|c|}
\hline
$su(3)$ generators & $\mathfrak{t}$ & $\mathfrak{p}$ \\ \hline
{\small Cartans (maximal in }$\mathfrak{t}${\small )} & $\left.
\begin{array}{c}
iH_{\alpha _{{\small 1}}} \\
iH_{\alpha _{{\small 2}}}%
\end{array}%
\right. $ & $\emptyset $ \\ \hline
{\small off diag- (maximal in }$\mathfrak{t}${\small )} & $\left.
\begin{array}{c}
E_{\alpha _{{\small 1}}}-E_{-\alpha _{{\small 1}}} \\
E_{\alpha _{{\small 2}}}-E_{-\alpha _{{\small 2}}} \\
E_{\alpha _{{\small 3}}}-E_{-\alpha _{{\small 3}}}%
\end{array}%
\right. ;$ $\
\begin{array}{c}
i\left( E_{\alpha _{{\small 1}}}+E_{-\alpha _{{\small 1}}}\right) \\
i\left( E_{\alpha _{{\small 2}}}+E_{-\alpha _{{\small 2}}}\right) \\
i\left( E_{\alpha _{{\small 3}}}+E_{-\alpha _{{\small 3}}}\right)%
\end{array}%
$ & $\emptyset $ \\ \hline
\end{tabular}
\label{V1}
\end{equation}%
The bridge 8$\times $8 matrix $\mathcal{V}^{su_{3}}$ between the basis ($%
X_{\alpha _{1}},Y_{\alpha _{1}},Z_{\alpha _{1}},X_{\alpha _{2}},Y_{\alpha
_{2}},Z_{\alpha _{2}},X_{\alpha _{3}},Y_{\alpha _{3}}$) and ($E_{\alpha _{%
{\small 1}}},E_{-\alpha _{{\small 1}}},H_{\alpha _{{\small 1}}},E_{\alpha _{%
{\small 2}}},E_{-\alpha _{{\small 2}}},H_{\alpha _{{\small 2}}},E_{\alpha _{%
{\small 3}}},E_{-\alpha _{{\small 3}}}$)%
\begin{equation}
\mathcal{V}^{su_{3}}=\left(
\begin{array}{ccc}
\mathcal{V}_{\alpha _{1}}^{su_{2}} & 0_{3\times 3} & 0_{3\times 2} \\
0_{3\times 3} & \mathcal{V}_{\alpha _{2}}^{su_{2}} & 0_{3\times 2} \\
0_{2\times 3} & 0_{2\times 3} & \boldsymbol{v}_{\alpha _{{\small 3}%
}}^{su_{2}}%
\end{array}%
\right)  \label{v8}
\end{equation}%
with $\mathcal{V}_{\alpha _{i}}^{su_{2}}$ as in (\ref{2ua}) and%
\begin{equation}
\boldsymbol{v}_{\alpha _{{\small 3}}}^{su_{2}}=\frac{1}{2}\left(
\begin{array}{cc}
i & i \\
1 & -1%
\end{array}%
\right)
\end{equation}

\subsubsection{Non compact real split form $sl(3,\mathbb{R})$}

The generators of the non compact real split form $sl(3,\mathbb{R})$ of the
complex Lie algebra $sl(3,\mathbb{C})$ can be constructed out of the
Chevalley basis $H_{\alpha },E_{\pm \alpha }$ within $sl(3,\mathbb{R})$. The
relation between the three generators ($X_{\alpha },Y_{\alpha },Z_{\alpha }$%
) of the $sl(2,\mathbb{R})$ contained into $sl(3,\mathbb{R})$ and the
Chevalley generators is given by the $\mathcal{U}^{sl_{2}}$ matrix as in eq(%
\ref{uv}). The action of the Cartan $\vartheta $- involution of $X_{\alpha
},Y_{\alpha },Z_{\alpha }$ is given by
\begin{equation}
\begin{tabular}{lllllll}
$X_{\alpha }$ & $=$ & $E_{+\alpha }+E_{-\alpha }$ & $,\qquad $ & $\vartheta
X_{\alpha }$ & $=$ & $-X_{\alpha }$ \\
$Y_{\alpha }$ & $=$ & $E_{+\alpha }-E_{-\alpha }$ & $,\qquad $ & $\vartheta
Y_{\alpha }$ & $=$ & $Y_{\alpha }$ \\
$Z_{\alpha }$ & $=$ & $H_{\alpha }$ & $,\qquad $ & $\vartheta Z_{\alpha }$ &
$=$ & $-Z_{\alpha }$%
\end{tabular}%
\end{equation}%
and on the Chevalley generators like $\theta \left( H_{\alpha }\right)
=-H_{\alpha }$ and $\theta \left( E_{\alpha }\right) =-E_{-\alpha }$; i.e:
\begin{equation}
\begin{tabular}{|l|l|l|l|}
\hline
real split \emph{sl(3,}$\mathbb{R}$\emph{)} & $\alpha _{i}$ & $H_{\alpha
_{i}}$ & $E_{+\alpha _{i}}$ \\ \hline
involution $\vartheta $ & $-\alpha _{i}$ & $-H_{\alpha _{i}}$ & $-E_{-\alpha
_{i}}$ \\ \hline
\end{tabular}%
\end{equation}%
It acts trivially on $Y_{\alpha }$ (compact generator as $\vartheta =1$) as
shown on the $K_{sl(2,\mathbb{R})}$ given by eq(\ref{22}), and non trivially
on $X_{\alpha }$ and $Z_{\alpha }$ (non compact generators as $\vartheta =-1$%
). Using the decomposition $\mathfrak{L}_{R}=\mathfrak{t}\oplus \mathfrak{p}%
, $ we have
\begin{equation}
\begin{tabular}{|c|c|c|}
\hline
$sl(3,\mathbb{R})$ & $\mathfrak{t}$ & $\mathfrak{p}$ \\ \hline
{\small commuting} & $-$ & $\left.
\begin{array}{c}
H_{\alpha _{{\small 1}}} \\
H_{\alpha _{{\small 2}}}%
\end{array}%
\right. $ \\ \hline
{\small root generators} & $\left.
\begin{array}{c}
E_{\alpha _{{\small 1}}}-E_{-\alpha _{{\small 1}}} \\
E_{\alpha _{{\small 2}}}-E_{-\alpha _{{\small 2}}} \\
E_{\alpha _{{\small 3}}}-E_{-\alpha _{{\small 3}}}%
\end{array}%
\right. $ & $\left.
\begin{array}{c}
E_{\alpha _{{\small 1}}}+E_{-\alpha _{{\small 1}}} \\
E_{\alpha _{{\small 2}}}+E_{-\alpha _{{\small 2}}} \\
E_{\alpha _{{\small 3}}}+E_{-\alpha _{{\small 3}}}%
\end{array}%
\right. $ \\ \hline
\end{tabular}
\label{U1}
\end{equation}%
The bridge 8$\times $8 matrix $\mathcal{U}^{sl_{3}}$ between the basis ($%
X_{\alpha _{1}},Y_{\alpha _{1}},Z_{\alpha _{1}},X_{\alpha _{2}},Y_{\alpha
_{2}},Z_{\alpha _{2}},X_{\alpha _{3}},Y_{\alpha _{3}}$) and ($E_{\alpha _{%
{\small 1}}},E_{-\alpha _{{\small 1}}},H_{\alpha _{{\small 1}}},E_{\alpha _{%
{\small 2}}},E_{-\alpha _{{\small 2}}},H_{\alpha _{{\small 2}}},E_{\alpha _{%
{\small 3}}},E_{-\alpha _{{\small 3}}}$)%
\begin{equation}
\mathcal{U}^{sl_{3}}=\left(
\begin{array}{ccc}
\mathcal{U}_{\alpha _{1}}^{su_{2}} & 0_{3\times 3} & 0_{3\times 2} \\
0_{3\times 3} & \mathcal{U}_{\alpha _{2}}^{su_{2}} & 0_{3\times 2} \\
0_{2\times 3} & 0_{2\times 3} & \boldsymbol{u}_{\alpha _{{\small 3}%
}}^{su_{2}}%
\end{array}%
\right)  \label{u8}
\end{equation}%
with $\mathcal{V}_{\alpha _{i}}^{su_{2}}$ as in (\ref{2ua}) and%
\begin{equation}
\boldsymbol{u}_{\alpha _{{\small 3}}}^{su_{2}}=\frac{1}{2}\left(
\begin{array}{cc}
1 & 1 \\
1 & -1%
\end{array}%
\right)
\end{equation}

\subsubsection{Non compact real form $su\left( 2,1\right) $}

The eight generators (${\small X}_{\alpha _{1}}{\small ,Y}_{\alpha _{1}}%
{\small ,Z}_{\alpha _{1}}{\small ,X}_{\alpha _{2}}{\small ,Y}_{\alpha _{2}}%
{\small ,Z}_{\alpha _{2}}{\small ,X}_{\alpha _{3}}{\small ,Y}_{\alpha _{3}}$%
) of the non compact real $su\left( 2,1\right) $ in terms of the Chevalley
generators (${\small E}_{\alpha _{{\small 1}}}{\small ,E}_{-\alpha _{{\small %
1}}}{\small ,H}_{\alpha _{{\small 1}}}{\small ,E}_{\alpha _{{\small 2}}}%
{\small ,E}_{-\alpha _{{\small 2}}}{\small ,H}_{\alpha _{{\small 2}}}{\small %
,E}_{\alpha _{{\small 3}}}{\small ,E}_{-\alpha _{{\small 3}}}$) read as
follows
\begin{equation}
\begin{tabular}{|c|c|c|}
\hline
$su_{{\small (2,1)}}$ & $\mathfrak{t}$ & $\mathfrak{p}$ \\ \hline
{\small Cartan} & $\left. i\left( {\small H}_{\alpha _{{\small 1}}}{\small -H%
}_{\alpha _{{\small 2}}}\right) \right. $ & ${\small H}_{\alpha _{{\small 1}%
}}{\small +H}_{\alpha _{{\small 2}}}$ \\ \hline
{\small root} & $\left.
\begin{array}{c}
{\small (E}_{\alpha _{{\small 1}}}{\small -E}_{-\alpha _{{\small 1}}}{\small %
)+}\left( E_{\alpha _{{\small 2}}}-E_{-\alpha _{{\small 2}}}\right) \\
{\small i(E}_{\alpha _{{\small 1}}}{\small +E}_{-\alpha _{{\small 1}}}%
{\small )-i(E}_{\alpha _{{\small 2}}}{\small +E}_{-\alpha _{{\small 2}}}%
{\small )} \\
{\small i(E}_{\alpha _{{\small 3}}}{\small +E}_{-\alpha _{{\small 3}}}%
{\small )}%
\end{array}%
\right. $ & $%
\begin{array}{c}
{\small (E}_{\alpha _{{\small 1}}}{\small +E}_{-\alpha _{{\small 1}}}{\small %
)+}\left( E_{\alpha _{{\small 2}}}+E_{-\alpha _{{\small 2}}}\right) \\
{\small i(E}_{\alpha _{{\small 1}}}{\small -E}_{-\alpha _{{\small 1}}}%
{\small )-i(E}_{\alpha _{{\small 2}}}{\small -E}_{-\alpha _{{\small 2}}}%
{\small )} \\
{\small E}_{\alpha _{{\small 3}}}{\small -E}_{-\alpha _{{\small 3}}}%
\end{array}%
$ \\ \hline
\end{tabular}
\label{W1}
\end{equation}%
The action of the Cartan involution $\vartheta $ on the simple roots and the
Chevalley generators is as follows:%
\begin{equation}
\begin{tabular}{|c|c|c|c|c|c|c|c|}
\hline
$su(2,1)$ & $\alpha _{1}$ & $\alpha _{2}$ & $H_{\alpha _{1}}$ & $H_{\alpha
_{2}}$ & $E_{+\alpha _{1}}$ & $E_{+\alpha _{2}}$ & $E_{\alpha _{1}+\alpha
_{2}}$ \\ \hline
involution $\vartheta $ & $-\alpha _{2}$ & $-\alpha _{1}$ & $-H_{\alpha
_{2}} $ & $-H_{\alpha _{1}}$ & $-E_{-\alpha _{2}}$ & $-E_{-\alpha _{1}}$ & $%
-E_{-\alpha _{1}-\alpha _{2}}$ \\ \hline
\end{tabular}%
\end{equation}%
The bridge 8$\times $8 matrix $\mathcal{W}^{su_{2,1}}$ is defined as $T_{%
\text{\textsc{a}}}^{su_{2}}=(\mathcal{W}^{su_{2,1}})_{\text{\textsc{a}}%
}^{n\alpha _{i}}E_{n\alpha _{i}}$ with the $T_{\text{\textsc{a}}}^{su_{2}}$s
referring to the generator basis ($X_{\alpha _{1}},Y_{\alpha _{1}},Z_{\alpha
_{1}},X_{\alpha _{2}},Y_{\alpha _{2}},Z_{\alpha _{2}},X_{\alpha
_{3}},Y_{\alpha _{3}}$) and the $E_{n\alpha _{i}}$'s designating the
Cartan-Weyl generator basis ($E_{\alpha _{{\small 1}}},E_{-\alpha _{{\small 1%
}}},H_{\alpha _{{\small 1}}},E_{\alpha _{{\small 2}}},E_{-\alpha _{{\small 2}%
}},H_{\alpha _{{\small 2}}},E_{\alpha _{{\small 3}}},E_{-\alpha _{{\small 3}%
}}$). It reads as follows%
\begin{equation}
\mathcal{W}^{su_{2,1}}=\frac{1}{2}\left(
\begin{array}{cccccccc}
1 & -1 & 0 & 1 & -1 & 0 & 0 & 0 \\
i & i & 0 & -i & -i & 0 & 0 & 0 \\
0 & 0 & i & 0 & 0 & -i & 0 & 0 \\
1 & 1 & 0 & 1 & 1 & 0 & 0 & 0 \\
i & -i & 0 & -i & i & 0 & 0 & 0 \\
0 & 0 & 1 & 0 & 0 & 1 & 0 & 0 \\
0 & 0 & 0 & 0 & 0 & 0 & i & i \\
0 & 0 & 0 & 0 & 0 & 0 & 1 & -1%
\end{array}%
\right)  \label{W21}
\end{equation}

\end{document}